\documentclass[preprint,aps,singlecolumn,superscriptaddress,preprintnumbers,nopacs,floatfix,amsmath,amssymb,overcite,citeauthorscript]{revtex4}

\usepackage[pdftex]{graphicx}
\usepackage{dcolumn}
\usepackage{bm}

\usepackage{amsmath}
\usepackage{graphicx}
\usepackage{amssymb}
\usepackage{mathrsfs}
\usepackage{bbm}
\usepackage{epsfig}

\newcommand{\la}[1]{\mbox{$\lefteqn{ \mbox{\,\, \tiny #1}}$} \label{#1}}
\def\thebibliography#1{{\section*{References}}\list
 {$^\arabic{enumi}$}{\settowidth\labelwidth{[#1]}\leftmargin\labelwidth
 \advance\leftmargin\labelsep
 \usecounter{enumi}}
 \def\newblock{\hskip .11em plus .33em minus .07em}
 \sloppy\clubpenalty4000\widowpenalty4000
 \sfcode`\.=1000\relax}

\setcitestyle{super}

\reversemarginpar

\newcommand{\be}{\begin{eqnarray}}
\newcommand{\ee}{\end{eqnarray}}

\begin{document}





\pagenumbering{arabic}
\renewcommand{\baselinestretch}{2}

%
\title{ On the r\^ole of thermal backbone fluctuations \\ in myoglobin ligand gate dynamics }

\author{Andrey Krokhotin}
\email{Andrei.Krokhotine@cern.ch}
\affiliation{Department of Physics and Astronomy and Science for Life Laboratory\\ Uppsala University,
P.O. Box 803, S-75108, Uppsala, Sweden}
\author{Antti J. Niemi$^{1}$}
\email{Antti.Niemi@physics.uu.se}
\affiliation{
Laboratoire de Mathematiques et Physique Theorique
CNRS UMR 6083, F\'ed\'eration Denis Poisson, Universit\'e de Tours,
Parc de Grandmont, F37200, Tours, France}
\affiliation {Department of Physics, Beijing Institute of Technology, Haidian District, Beijing 100081, P. R. China
           }
\author{Xubiao Peng$^{1}$}
\email{xubiaopeng@gmail.com}
\begin{abstract}
\noindent
We construct an energy function that describes the crystallographic structure of sperm whale myoglobin backbone.
As a model in our construction, we use the Protein Data Bank  entry 1ABS that has been 
measured at liquid helium temperature. Consequently the 
thermal B-factor fluctuations are very small, which is an advantage in our construction. 
The energy function that we utilize resembles that of the discrete non-linear Schr\"odinger 
equation. Likewise, ours supports solitons as local minimum energy configurations.  
We describe the 1ABS backbone in terms of solitons with a precision that deviates from 1ABS 
by an average root-mean-square distance, which  is less than the experimentally observed
Debye-Waller B-factor  fluctuation distance.  We  then subject the multisoliton solution to 
extensive numerical heating and cooling experiments, 
over a very wide range of temperatures.  We concentrate in particular to temperatures 
above 300K and below the $\Theta$-point unfolding temperature, which is around 348K. 
We confirm that the behavior of the multisoliton is fully consistent with Anfinsen's thermodynamic 
principle, up to very high temperatures. 
We observe that the structure responds to an increase of temperature consistently in a very similar manner. 
This enables us to characterize the onset of thermally induced conformational changes in terms of three 
distinct backbone ligand gates. 
One of the gates is made of the helix F and the helix E. 
This is a  pathway that is presumed to have a major r\^ole in ligand migration between 
the heme and the exterior. The two other gates are chosen similarly, when open they provide 
a direct access route for a ligand to reach the heme. We find that out of the three gates we investigate, 
the one which is formed by helices B and G is the most sensitive one 
to thermally induced conformational changes. Our approach provides a novel perspective to the important
problem of ligand migration. 
\end{abstract}

\maketitle
\section{Introduction} 

Myoglobin is a relatively small globular protein that has a central r\^ole in oxygen transport and storage
in muscle cells.  Its structure has been investigated very extensively, both for 
historical reasons \cite{myog} and as a tractable example of protein physiology. Myoglobin 
has become the paradigm specimen for exploring relations between protein
structure and function \cite{frau2003}.  

Myoglobin binds small non-polar ligands such as O$_2$, CO, and NO  in its interior,  where
they become attached to the iron atom of the heme. Various structural studies 
reveal that the natively folded myoglobin is very compact. In particular, there
does not appear to be any obvious channel for the 
ligands to enter, or exit, the interior. 
Consequently, under physiological conditions, myoglobin must undergo some kind of 
conformational deformations, for the ligands to reach the heme.  
These motions are known to have a thermal origin,  but their detailed 
character remains to be clarified.

Early experiments \cite{perutz} and theoretical approaches  \cite{karplus1} suggest that there is
one dominant, highly localized pathway,  which the ligand follows 
when it moves between the heme and the solvent. This pathway goes  through 
the distal histidine gate, which is made up of the His64 side-chain 
on the helix E. Thermal 
fluctuations may open and close this side-chain gate, for ligands to pass from solvent to heme and back. 
Subsequently, it was recognized that the pathway can not be unique. Several additional gates were
proposed  and investigated \cite{karplus2,huang}. It became also plausible, 
that ligand migration is due to an elaborate, collective structural motion that can involve 
both the backbone and several side-chains.  In addition of the distal cavity where the ligand becomes trapped
by the heme iron \cite{distal1,distal2}, four additional major folding defects were 
identified. These are now commonly  denoted as Xe1, Xe2, Xe3 and Xe4 ~ \cite{stilton}. 

At the moment, the understanding of ligand migration remains incomplete. On the one hand,
there are experiments supporting the original proposal \cite{perutz,karplus1}, that
a major pathway for ligand migration is located in the vicinity of
the distal histidine gate \cite{exp1,exp2}.  For example,  according to kinetic analysis,  as much as around 70-80$\%$ of 
ligands could enter and  exit  through this pathway \cite{histg}.  On the other hand,  
the energy landscape of myoglobin-ligand  interaction is known to be very complex \cite{frau75}. 
In particular, at physiological temperatures myoglobin most probably fluctuates between multiple conformational 
sub-states,  with many different folding intermediates.  Especially the four defects Xe1, Xe2, Xe3 and Xe4 are
presumed to participate in various different kind of ligand transport processes. 
For example, they can provide a route between the distal entry site and the proximal binding 
side of the heme.  These defects can also enable ligands to diffuse between the solvent 
and some other, normally inaccessible  sites within the myoglobin. As a consequence there  
are most likely many pathways for the ligand to enter and  exit, all of which
contribute to the binding process between the ligand and the heme. 

Several recent experimental and theoretical studies support the position that ligands utilize 
many different gates.  For example, experimental studies in  
\cite{exp1,exp2,path1,path2,path3,path4}  and molecular dynamics 
simulations in \cite{md1,md2,md3}  suggest, that 
CO can migrate from the distal heme pocket  to the Xe1 cavity, which is located on the proximal 
side of the heme. This migration takes place through a complex network of pathways, and involve 
several internal cavities. In the crystallographic x-ray structure of myoglobin, the pathways can not be 
identified. As a consequence, under {\it in vivo} conditions myoglobin must be highly dynamic, 
with ligand migration driven by a multitude of conformational fluctuations
including those that engage the backbone  \cite{md3}. 

In this article we aim to develop a novel theoretical approach to model the dynamical myoglobin. In particular, we 
propose some new ways for ligands to enter at exit. For this
we scrutinize the r\^ole of thermal backbone fluctuations, with the goal to 
identify backbone based ligand gates, and to describe in detail how they open and close. 
Our approach is based on an effective, coarse-grained free energy function. 
The general theoretical approach to protein folding we use, has been developed
in a series of articles \cite{oma,ulf,cherno,mart,nora,peng,dff,peng1,sha,marty,liwo}. The functional form of the effective energy 
that we utilize follows from purely geometrical arguments, in combination with 
the general concept of universality \cite{widom,kadanoff,wilson,fisher}; in the case of polymers see
also \cite{huggins,flory,degen,schafer,lim}. The relatively small number of parameters 
that appear in the energy function are specific to the given protein backbone, here myoglobin.
These parameters  are determined by solving the pertinent classical equation of motion. 
Fo this, one minimizes the root-mean-square distance (RMSD)
between the classical solution to the equation of motion that follows from the energy function, 
and the crystallographic x-ray backbone structure of myoglobin. 
Once the parameters have been determined, we have an explicit energy function that described 
the folded myoglobin backbone as its local energy minimum, in the limit of vanishing temperature.  
With this energy function at hand, various kind of energetic investigations and dynamical studies 
become possible. We can systematically study the effects of temperature to
thermal backbone fluctuations and, in the particular case of myoglobin considered here,
search for potential ligand gateways and scrutinize
how these gateways open and close. 

The static classical solution that we construct by solving the equations of motion,
describes the myoglobin backbone in the low temperature limit where all thermal fluctuations vanish. 
Therefore, we can introduce  a methodical  first principles approach to theoretically 
model the response of the backbone to a heat bath.  In our numerical investigations, we shall adiabatically increase the 
temperature from a vanishing value, by postulating that the ensuing thermally driven dynamical evolution follows
the Glauber protocol \cite{glauber,lebo,marti1,marti2};  we recall that this protocol determines a Markovian Monte Carlo 
evolution for a system that is off thermal equilibrium,  and for which the canonical Gibbsian probability 
distribution is the unique stationary equilibrium limit \cite{glauber,lebo,marti1,marti2}.

As the temperature slowly increases, the collapsed backbone starts fluctuating. The conformation
moves  around in the energy landscape, by swinging about the native state.  Those backbone ligand 
gates that have the lowest energy barrier, are the most likely to
be the first ones to start opening.  The opening and closing of the gates can be monitored by following the
temperature dependence in the amplitudes of local conformational fluctuations. 

As the temperature increases further, towards the $\Theta$-point regime, 
the backbone starts resembling a fully flexible random chain \cite{huggins,flory,degen,schafer}.
At this temperature range, we expect 
to observe both a clear transition in the radius of gyration, and a rapid increase in the local fluctuation amplitudes.  
Consequently, in our simulations, we  heat up the backbone configuration until we clearly pass a transition regime in 
each. This ensures  that we have also passed  the $\Theta$-point temperature, where the protein departs from 
the collapse phase and becomes a fully flexible  chain, eventually transiting to the self-avoiding random walk phase as the
temperature further increases.

When we conclude that we have reached a temperature value that is above the $\Theta$-point, 
we stop increasing the temperature. We allow the configuration to become fully thermalized  into its 
Gibbsian equilibrium state, which we know \cite{marti1,marti2} it approaches at an exponential rate. 
We then proceed to adiabatically cool the system. 
We follow how the protein collapses towards its native state, observing 
how the various backbone ligand gates close, one after another.  It is natural to expect, that  if there is a gate 
that fully exposes the heme to the solvent,  and which in addition is the first one to open and the last  one to close,
it is also the gate through which the
ligands most likely enter and exit.
  
Biologically, myoglobin is an important part of the oxygen diffusion network in many living orgasms. But for various experimental reasons, the 
myoglobin-ligand interaction is often studied using CO, instead of O$_2$.  As a consequence, there are several 
good quality crystallographic carbonmonoxony-myoglobin x-ray structures available in 
Protein Data Bank (PDB) \cite{pdb}, that we may utilize in our 
theoretical approach. Here we have chosen to base our construction on 
the configuration with PDB code 1ABS \cite{1abs}. It is
taken from a sperm whale (Physeter catodon). This  
structure has been measured at the very low liquid helium temperature value of around 20 Kelvin. 
As a consequence the thermal B-factors are very small.
This is the primary reason for us  to select 1ABS as the model, for which we  
construct the classical solution of our energy function.  Even though the resolution at 1.5 \AA ngst\"om, 
is not as good as we would like it to be; our method is fully 
capable for modeling the protein backbone at better resolutions. 

We start with a Methods section II. It outlines  our approach. In the sub-section II A we explain how to 
geometrically describe the protein backbone. In sub-section II B we introduce our energy function. In sub-section
II C we explain in detail, how to construct a multi-soliton solution that corresponds to the local minimum energy state of
our energy function, and how to determine the numerical values of the relatively few parameters, 
so that the energy function models
a given folded protein. In sub-section II D we comment on the r\^ole of the parameters, and estimate their number.
We then outline, in sub-section II E, how we implement our  algorithm numerically. 
In sub-section II F we introduce an explicit {\it Ansatz} to the multi-soliton, in terms of elementary functions.
In sub-section II G we comment on the precision that we aim for in our approach. 
In  sub-section II H of Methods section, we explain how we use the energy function to heat up the protein backbone. This is 
a process during which the protein is out of thermal equilibrium. Finally, in sub-section II I we analyze the effects of
temperature on the parameters in our model. 
 
We then proceed to describe our results, in Section III. In sub-section III A we present a soliton-based analysis of the myoglobin backbone
structure. This identifies the super-secondary helix-loop-helix structures, and in particular the number of solitons along the
backbone. In sub-section III B we do our best to estimate the effects of thermal fluctuations on the myoglobin background, using
available experimental structures.  In sub-section III C we discuss the r\^ole  of side-chains. We proceed to construct the
myoglobin backbone in terms of a multi-soliton solution to our energy function, in sub-section III D. In sub-section III E we
investigate the effects of heating and cooling. We study the phase structure, and in particular
identify the presence of $\Theta$-point transition. In sub-section III F we identify the three backbone ligand gates that we 
then investigate in detail their properties. In particular, we propose a novel mechanism, and likely pathway, for ligands to migrate between
the heme and the exterior of the myoglobin.  We conclude with a short summary, in section IV.


\section{Methods}

\subsection{Backbone geometry}
\label{sect:rgy}

The approach to describe the geometry of a protein backbone that we utilize,  has been 
developed  in \cite{oma,ulf,cherno,mart,nora,peng,dff,peng1,sha,marty,liwo}.  This formalism  aims to describe the protein 
geometry in terms of local energy minima of an energy function, that depends only
on the positions of the backbone C$_\alpha$ atoms. The ensuing
bond and torsion angles are the dynamical variables. 
These  angles are defined as follows:
We take $\mathbf r_i$ to be the coordinate sites of the C$_\alpha$ carbons. The 
index $i=1,...,N$ runs over all residues. In the case of the 1ABS myoglobin that is of interest here, we have $N=154$.
However, in our simulations  we do not include the apparently unstructured tails at the beginning and end
of the myoglobin chain. We doubt  that they are direct participants in ligand migration,  
even thought they can certainly affect the global fold. 
Mostly, since we are interested in ligand migration, 
we only consider the backbone segment that starts with the C$_\alpha$ atom with PDB index
8 and ends with the C$_\alpha$ atom with PDB index 149.
For each C$_\alpha$ carbon site $i$, we introduce the unit tangent vector 
\begin{equation}
\mathbf t_i = \frac{ {\bf r}_{i+1} - {\bf r}_i  }{ |  {\bf r}_{i+1} - {\bf r}_i | }
\label{t}
\end{equation}
the unit binormal vector
\begin{equation}
\mathbf b_i = \frac{ {\mathbf t}_{i-1} \times {\mathbf t}_i  }{  |  {\mathbf t}_{i-1} \times {\mathbf t}_i  | }
\label{b}
\end{equation}
and the unit normal vector 
\begin{equation}
\mathbf n_i = \mathbf b_i \times \mathbf t_i
\label{n}
\end{equation}
The orthogonal triplet ($\mathbf n_i, \mathbf b_i , \mathbf t_i$) is the discrete Frenet  
frame \cite{dff} at the position $\mathbf r_i$ of the backbone. 
The backbone bond angles are
\begin{equation}
\kappa_{i} \ \equiv \ \kappa_{i+1 , i} \ = \ \arccos \left( {\bf t}_{i+1} \cdot {\bf t}_i \right)
\label{bond}
\end{equation}
and the backbone torsion angles are
\begin{equation}
\tau_{i} \ \equiv \ \tau_{i+1,i} \ = \ {\rm sign}\{ \mathbf b_{i-1} \times \mathbf b_i \cdot \mathbf t_i \}
\cdot \arccos\left(  {\bf b}_{i+1} \cdot {\bf b}_i \right) 
\label{tors}
\end{equation}
If these angles are known, we can use  the discrete Frenet equation
\begin{equation}
\left( \begin{matrix} {\bf n}_{i+1} \\  {\bf b }_{i+1} \\ {\bf t}_{i+1} \end{matrix} \right)
= 
\left( \begin{matrix} \cos\kappa \cos \tau & \cos\kappa \sin\tau & -\sin\kappa \\
-\sin\tau & \cos\tau & 0 \\
\sin\kappa \cos\tau & \sin\kappa \sin\tau & \cos\kappa \end{matrix}\right)_{\hskip -0.1cm i+1 , i}
\left( \begin{matrix} {\bf n}_{i} \\  {\bf b }_{i} \\ {\bf t}_{i} \end{matrix} \right) 
\label{DFE2}
\end{equation}
to construct the frame at position $i+i$ 
from the frame at position $i$. Once we have the frames, we get the backbone using
\begin{equation}
\mathbf r_k = \sum_{i=0}^{k-1} |\mathbf r_{i+1} - \mathbf r_i | \cdot \mathbf t_i
\label{dffe}
\end{equation}
With no loss of generality we can set $\mathbf r_0 = 0$, and choose $\mathbf t_0$ so that it points 
along the positive $z$-axis.
Consequently, given an energy function that depends only on the bond and torsion angles, we can try
and relate its local minimal energy states to three dimensional protein backbone conformations using (\ref{DFE2})
and (\ref{dffe}).

When constructing the energy function, we shall combine a geometric line of arguments \cite{oma,ulf} with the
general concept of universality \cite{widom,kadanoff,wilson,fisher}. The geometric arguments are based on the following
observation:  We note that 
(\ref{dffe}) does not involve the vectors $\mathbf n_i$ and $\mathbf b_i$. Thus we may arbitrarily
rotate them, without affecting 
the backbone. We may even select a different
linear combination of these two vectors, at each backbone site $i$,
\begin{equation}
 \left( \begin{matrix}
{\bf n} \\ {\bf b} \\ {\bf t} \end{matrix} \right)_{\!i} \!
\rightarrow  \! 
 \left( \begin{matrix}
\cos \Delta_i & \sin \Delta_i & 0 \\
- \sin \Delta_i & \cos \Delta_i & 0 \\ 
0 & 0 & 1  \end{matrix} \right) \left( \begin{matrix}
{\bf n} \\ {\bf b} \\ {\bf t} \end{matrix} \right)_{\! i}
\label{discso2}
\end{equation}
Here the $\Delta_i$ are arbitrarily chosen local rotation angles. According to (\ref{dffe}), and since $\mathbf t_i$ remains
intact,  this local SO(2) transformation has no effect on the positioning of the C$_\alpha$ carbons. 

{\it A priori}, the fundamental range of the bond angle $\kappa_i$ is  $ [0,\pi]$. For the 
torsion angle the range is $\tau_i \in [-\pi, \pi)$. Consequently we may 
identify ($\kappa_i, \tau_i$) with the canonical 
latitude and longitude angles on the surface of a sphere. 
However, in the sequel we find it useful to extend the range
of $\kappa_i$ into $ [-\pi,\pi]$ $mod(2\pi)$, but with no change in the range of $\tau_i$. 
We compensate for this two-fold covering of the sphere, 
by introducing the following discrete symmetry \cite{cherno,dff}
\begin{equation}
\begin{matrix}
\ \ \ \ \ \ \ \ \ \kappa_{l} & \to  &  - \ \kappa_{l} \ \ \ \hskip 1.0cm  {\rm for \ \ all} \ \  l \geq i \\
\ \ \ \ \ \ \ \ \ \tau_{i }  & \to &  \hskip -2.5cm \tau_{i} - \pi 
\end{matrix}
\label{dsgau}
\end{equation}
This is a special case of (\ref{discso2}),  with
\[
\begin{matrix} 
\Delta_{l} = \pi \hskip 1.0cm {\rm for} \ \ l \geq i+1 \\
\Delta_{l} = 0 \hskip 1.0cm {\rm for} \ \ l <  i+1 
\end{matrix}
\]

We note that regular protein secondary 
structures  correspond to constant values of
$(\kappa_i, \tau_i)$.  For example standard $\alpha$-helix is
\begin{equation}
\alpha-{\rm helix:} \ \ \ \ \left\{ \begin{matrix} \kappa \approx \frac{\pi}{2}  \\ \tau \approx 1\end{matrix} \right.
\label{bc1}
\end{equation}
and standard $\beta$-strand is 
\begin{equation}
\beta-{\rm strand:} \ \ \ \ \left\{ \begin{matrix} \kappa \approx 1 \\ \tau \approx \pi \end{matrix}  \right.
\label{bc2}
\end{equation}
Similarly, we 
describe all the other regular secondary structures such as 3/10 helices, 
left-handed helices {\it etc.} with definite constant values of $\kappa_i$ and $\tau_i$.
Geometrically, a loop is thus defined to be 
any configuration that interpolates between the regular structures. In particular,
along a loop the values of ($\kappa_i, \tau_i$) are variable, from site to site.

Finally, in PDB structures with no {\it cis-}proline, such as 1ABS, the average distance 
between two C$_\alpha$ atoms is
\begin{equation}
 |\mathbf r_{i+1} - \mathbf r_i | = d  \approx  3.8 \, \dot{\mathrm A}
 \label{dist}
 \end{equation}
In (\ref{dffe}), we may then use the fixed bond length value (\ref{dist}).  During our dynamical simulations we
also impose the forbidden volume (steric) constraint
 \begin{equation}
 |\mathbf r_{i} - \mathbf r_k | \geq  3.8 \ \dot {\mathrm A}  \ \ \ \ {\rm for} \ \ |i-k| \geq 2 
\label{fvol}
\end{equation}
between the backbone C$_\alpha$ atoms.  Effectively, this also prevents
chain crossing.

%
%
%
%
%
%
%
%
%
%
%
%
%
%
%

%
%
%
%
%
%
%
%
%
%
%
%
%
%
%

\subsection{Backbone  energy }

Crystallographic proteins in PDB display a structural hierarchy \cite{cath,scop,peng1}. 
Folded proteins are composed of regular secondary structures such as $\alpha$-helices and $\beta$-strands,
together with the less regular loops that combine them together. 
Following  \cite{widom,kadanoff,wilson,fisher}, we propose that  
this hierarchy of structures should also be accounted for, when determining  the variables that describe 
protein dynamics. In particular,  our goal is to model the dynamics directly  in terms of  those variables that relate
only  to the super-secondary motifs, such as helix-loop-helix:
 
Molecular dynamics \cite{charmm,amber} aims to describe proteins  and their dynamics at the level of individual atoms
and their interactions. But as the  length scale increases, the shorter distance, atomic scale dynamical variables become 
gradually disengaged.  Therefore, at appropriate long distance scales the protein dynamics 
should become describable in terms of a relatively small set of effective variables. 
By adapting the Migdal-Kadanoff  \cite{migdal,kadan1,kadan2} block-spin transformation procedure 
and following the general concept of universality developed in \cite{widom,kadanoff,wilson,fisher} we may try and
systematically coarse grain the microscopic, atomic level energy function. In \cite{ulf,cherno,nora,dff}, see also \cite{oma},  
it has been argued that there is a universal, effective Landau-type free energy function that computes  
the overall fold geometry and describes the folding pathways of a protein, solely in terms of  
those variables that determine the positions of the central C$_\alpha$ atoms.
Since the fluctuations in the bond lengths are minimal, the leading order contribution to 
the energy involves can then only engage the bond and torsion angles of the C$_\alpha$  backbone. 
Therefore, in leading non-trivial order, the functional form of the energy function
can  be uniquely deduced  by symmetry considerations alone: 
{\it Any} backbone energy function that involves only the bond and torsion angles must 
remain invariant under the local SO(2) frame rotations 
(\ref{discso2}).  Consequently  one concludes  that
in a limit of  long distance scales, only the following SO(2) rotation invariant quantities can be present in  
the low temperature limit of the thermodynamic Helmholtz free energy (internal energy) \cite{ulf,cherno,nora,dff}
\begin{equation}
E = - \sum\limits_{i=1}^{N-1}  2\, \kappa_{i+1} \kappa_i  + \sum\limits_{i=1}^N
\biggl\{  2 \kappa_i^2 + q\cdot (\kappa_i^2 - m^2)^2  
\left.  + \frac{d}{2} \, \kappa_i^2 \tau_i^2   -  b  \kappa_i^2  \tau_i  - a \tau_i     +  
\frac{c}{2}  \tau^2_i 
\right\} 
\label{E1}
\end{equation}
The detailed derivation of  (\ref{E1}) can be found in  \cite{ulf}.  Here, it suffices to observe the following:
We  recognize in (\ref{E1}) a variant of the energy function of the discrete nonlinear Schr\"odinger  equation (DNLS),
together with its conserved quantities \cite{davy,scott,kevk,ludvig}: The first sum in combination with the three 
first terms in the second sum comprise the energy of  the standard  DNLS equation,  
when we express it  in terms of the discretized variable, following \cite{hasimoto}. Accordingly, this contribution relates 
to the description of  the motion of a filamental curve in the local induction approximation, in terms of
the  nonlinear Schr\"odinger equation.
The fourth ($b$) term is a conserved quantity in the DNLS hierarchy,  it is the  "momentum" of a filament. 
The fifth term ($a$) is the conserved "helicity", it enforces preference towards the right-handed chirality in the backbone.   
The last ($c$) term is the 
Proca mass that we include as a regulator, to ensure that the energy is bounded from below.  

For future reference, we shall propose the following interpretation of the parameter $q$ in (\ref{E1}):
In the limit of very large $q$, the third term in (\ref{E1}) yields the condition
\begin{equation}
\kappa^2_i \approx m^2
\label{qc1}
\end{equation}
on the bond angles. On the other hand, when $q$ is very small, we may ignore the third term. Due to  
the second term, the bond angle 
has then a propensity towards the value
\begin{equation}
\kappa_i \approx 0
\label{qc2}
\end{equation}
When we combine (\ref{qc1}), (\ref{qc2})  with (\ref{bc1}), (\ref{bc2}) we conclude that the strength of $q$ relates
to the stability of the regular secondary structures, such as $\alpha$-helices and $\beta$-strands. Their stability is due to
hydrogen bonds. Consequently the numerical value
of $q$ is a measure of the strength of these hydrogen bonds. Thus a large value of $q$ stabilizes structures such
as $\alpha$-helices and $\beta$-strands. But the smaller the numerical  value of $q$, the easier it is for hydrogen bonds 
in these structures to become broken.

Note that the energy function (\ref{E1}) does not  explain the detailed atomic level mechanisms, 
why a protein folds.  Instead, it models the protein backbone in terms of the coarse grained backbone variables,
in a manner which is based directly on the universal physical 
arguments of ~ \cite{widom,kadanoff,wilson,fisher}.  The universal character of these arguments
ensures that the energy  (\ref{E1}) emerges from {\it any} microscopic level Schr\"odinger 
operator that correctly describes the short distance interactions of all the atomic constitutes of the protein.   
The energy  (\ref{E1}) represents the long distance
{\it universality class } of the full microscopic level Schr\"odinger operator of the collapsed protein, {\it a priori}
in the limit of vanishing temperature. In particular, by the general arguments of \cite{widom,kadanoff,wilson,fisher} 
it emerges as the long distance limit from the classical molecular dynamics force fields \cite{charmm,amber} which
themselves are approximations to the full quantum mechanical atomic level Schr\"odinger equation.

Since the energy function  is a representative of the infrared (large distance)
universality class of the full atomic level Schr\"odinger equation,  we expect that thermal fluctuations around
its multi-soliton solution, when properly accounted for, 
correctly describe the finite temperature energy landscape in the neighborhood 
of the native fold of a give protein at or near thermal equilibrium.

Finally, we comment that there are also several other energy functions, both simplified and coarse-grained, to protein folding,
see {\it e.g.} \cite{hyeon} for a recent survey.  Examples range from the G\=o model and its 
various extensions and improvements \cite{go1,ciep,shak,meinke,polish}
to carefully crafted  Physics based energy functions such as UNRES  \cite{liwo2,liwo3,liwo4,liwo1}. Like ours, these  
approaches aim to describe the folding dynamics in terms of those degrees of freedom that are  
essential for describing  thermodynamically stable structures. 

Each of the simplified models has its own advantages and drawbacks. For example, in the case of models
of the G\=o type, the energy function is constructed  from the knowledge of all native contacts of all atoms,
in the protein of interest. These details are then carefully accounted for in building the energy function for the given
protein, to ensure that the experimental crystal structure is  the minimum energy configuration. 
As a consequence G\=o type models have a lacking predictive power, when it comes to the native fold. 
But they can still be applied  to study folding pathways \cite{shak,meinke,polish}. 
On the other hand, elaborate coarse grained energy functions such as UNRES \cite{liwo2,liwo3,liwo4,liwo1}
assume no {\it a priori} knowledge of the native structure.  As such, they are much closer to the molecular dynamics
force fields \cite{charmm,amber}, with a predictive power in 
determining how the folded protein should look like. 

When the  
number of details and parameters in simplified and coarse-grained force fields increases,
the numerical simulations become increasingly more complex and time consuming. 
At the same time, an increase in the number of parameters
inevitably leads to a decrease in the predictive power.

%
%
%
%
%
%
%
%
%
%
%
%
%
%
%

\subsection{Multi-soliton }

The energy function (\ref{E1}) has been derived using a very general line of reasoning. However, despite the generality
of the arguments,  it has been shown that the local minimum energy configurations of (\ref{E1}) 
are capable of describing practically all high resolution crystallographic protein backbones in PDB, 
with an accuracy that matches and even exceeds the experimental precision in x-ray crystallography. At least to  the extent,
that  this precision
can be related to the thermal Debye-Waller B-factor fluctuation distances in PDB data \cite{peng1}. 
This is a consequence of the remarkable property of the generalized DNLS energy function (\ref{E1}), 
that it supports solitons as local energy minima \cite{cherno,nora}.   

Soliton solutions to non-linear difference (differential) equations such as the one that follows from (\ref{E1})
by variational principle, are the archetype structural self-organizers not only in Physics, 
but elsewhere as well \cite{davy,scott,kevk,ludvig,sol1,sol2}. Generically, a soliton can be present  whenever 
non-linear interactions of the elementary constituents such as atoms along the protein chain, 
merge  into a localized collective excitation. This excitation is a soliton, whenever it is  stable against 
small perturbations and cannot easily decay, unwrap or disentangle. In particular, solitons can be 
very robust in preserving their form both under quantum mechanical and  thermal fluctuations. 

For example, solitons are being deployed for data transmission in transoceanic cables, they are utilized to
conduct electricity in organic polymers and they describe chemical energy transportation in proteins. 
Many phenomena from the formation of the morning glory cloud in the atmosphere to 
the Mei\ss ner effect in superconductivity and dislocations in liquid crystals are due to 
solitons. Solitons also model hadronic particles, cosmic strings and magnetic monopoles in high 
energy physics \cite{davy,scott,kevk,ludvig,sol1,sol2}.

We obtain the relevant soliton solution of (\ref{E1}) by 
starting with the $\tau$-equation of motion, 
\[ 
\frac{\partial E}{\partial \tau_i} =  d \kappa_i^2 \tau_i - b\kappa_i^2 - a+ c\tau_i = 0 
\]
From this we solve
\begin{equation}
\tau_{i} [\kappa] = \frac{a+ b\kappa_i^2}{c + d \kappa^2_{i} }  
\label{tauk} 
\end{equation}
There are four parameters in (\ref{tauk}). But one of them can be removed, as an overall scale.  
We can use this to choose  $a=-1.0$, in solving for the $\tau$-profile. 
For an  $\alpha$-helix (\ref{bc1}) we then have
\begin{equation}
\tau_{i} [\alpha] =  \frac{1 + b \kappa_i^2}{c + d \kappa^2_{i} }  \approx 1  \ \mod(2\pi) 
\label{tt11}
\end{equation}
and for a $\beta$-strand (\ref{bc2})
\begin{equation}
\tau_i [\beta] =  \frac{1 + b \kappa_i^2}{c + d \kappa^2_{i} }  \approx \pi \  \mod(2\pi) \ \ 
\label{tt22}
\end{equation}
We  use (\ref{tauk}) to eliminate the torsion angles from (\ref{E1}). This gives   
for the energy of bond angles 
\begin{equation}
E[\kappa] = - \sum\limits_{i=1}^{N-1}  2\, \kappa_{i+1} \kappa_{i}  + \sum\limits_{i=1}^N
 2 \kappa_{i}^2 + V[\kappa_i]  
\label{Ekappa}
\end{equation}
where
\begin{equation}
V[\kappa]  =  - \left( \frac{b c - a d}{d}  \right) \, \frac{1}{c+d\kappa^2} 
- \left( \frac{b^2 + 8 q m^2}{2b} \right)
\,\kappa^2 + q\, \kappa^4
\label{Vkappa}
\end{equation}
The first term is a generalization of the 
Vinetskii-Kukhtarev potential contribution \cite{vine}, introduced in the context of nonlinear waveguides.
In the case of proteins, it turns out that this  term has a numerical value which is small
in comparison to the second and the third term. The latter two have the 
conventional form of a symmetry breaking double-well potential.  Depending on 
the parameter values,  we are either in the broken symmetry phase 
where $\kappa$ and $\tau$  both acquire a non-vanishing and constant  
ground state value, or  in the symmetric  phase 
where $\kappa $ vanishes.  Regular protein structures such as 
helices  (\ref{bc1}) and strands (\ref{bc2}) correspond to different broken symmetry
ground states. Since  the numerical value of the first term in (\ref{Vkappa})
is  small, for an $\alpha$-helix (\ref{bc1})  we have the estimate
\begin{equation}
m \approx \frac{\pi}{2}
\label{malpha}
\end{equation}
and for a $\beta$-strand we have the estimate
\begin{equation}
m \approx 1.0
\label{mbeta}
\end{equation}
See also the discussion in connection of (\ref{qc1}), (\ref{qc2}). 

Loops are regions where ($\kappa_i, \tau_i$)  are variable  \cite{cherno,nora}. Loops correspond
to the so-called dark soliton solution of the generalized discrete nonlinear Schr\"odinger equation, that derives
from the energy (\ref{Ekappa}),
\begin{equation}
\kappa_{i+1} = 2\kappa_i - \kappa_{i-1} + \frac{ d V[\kappa]}{d\kappa_i^2} \kappa_i  \ \ \ \ \ (i=1,...,N)
\label{nlse}
\end{equation}
where we set $\kappa_0 = \kappa_{N+1}=0$.  
The equation (\ref{nlse}) is the {\it Master equation} from
which we compute the shape of a folded protein C$_\alpha$ backbone, in terms of the parameters provided these
are known.

Conversely,
given the protein structure, we can use the solution of (\ref{nlse}) to compute the parameters, by constructing
a  multi-soliton solution that matches the  experimental backbone structure.  For this we proceed as follows:
We introduce the iterative equation 
\begin{equation}
\kappa_i^{(n+1)} \! =  \kappa_i^{(n)} \! - \epsilon \left\{  \kappa_i^{(n)} V'[\kappa_i^{(n)}]  
- (\kappa^{(n)}_{i+1} - 2\kappa^{(n)}_i + \kappa^{(n)}_{i-1})\right\}
\label{ite}
\end{equation}
Here  $\{\kappa_i^{(n)}\}_{i\in N}$ denotes the $n^{th}$ iteration of an initial configuration  $\{\kappa_i^{(0)}\}_{i\in N}$ and $\epsilon$ is some 
sufficiently small but otherwise arbitrary numerical constant. We choose $\epsilon = 0.01$. 
The fixed point of this equation is clearly a solution of the
Master equation (\ref{nlse}); we note that the results of \cite{herr,nora} ensure the existence of a (dark) soliton solution that interpolates 
between the minima of the potential.

%
%
%
%
%
%
%
%
%
%
%
%
%
%
%

\subsection{Parameters}

The energy function (\ref{E1}) involves a number of parameters. Eventually, we would like
to compute their numerical values directly from the amino acid sequence. At the moment, this has
not yet been achieved. We hope that eventually the parameters could be evaluated directly from the sequence, 
for example by combining the present
approach with the UNRES energy function that utilizes a very similar set of variables \cite{liwo,liwo2,liwo3,liwo4,liwo1}.

{\it A priori} it seems that the number of parameters in (\ref{E1})
to describe an entire protein backbone, might be  quite large. However, since it describes the backbone 
in terms of  the soliton solution, the number of parameters turns out to be remarkably small: For each
super-secondary structure such as a helix-loop-helix,  the potential (\ref{Vkappa}) has  only
four independent parameter combinations. In addition of $q$ and $m$, there are only two 
independent parameter combinations that appear in the first term. Three of these four
parameters can be given the following  interpretations. Two of the parameters determine 
the values of $\kappa_i$ in the regular ground state structures that are adjacent to the soliton,
such as  (\ref{bc1}), (\ref{bc2}) {\it i.e.} the type of the helix 
that precedes and follows the soliton.  
The third parameter relates to the length of the loop. 
The fourth parameter can be included as one of the three independent parameters that determine the 
torsion profile (\ref{tauk}). It can be attributed to the length of the soliton, in terms of the torsion angle.
In addition, there is also the parameter that specifies the position of the soliton along the backbone. 

In the equation (\ref{tauk}) for the torsion angle, the overall scale cancels out. The two remaining 
parameter combinations in addition of the loop length, 
become determined by the values of $\tau_i$ 
in the ground states surrounding the soliton {\it i.e.} the type of 
the helix as in (\ref{bc1}), (\ref{bc2}); See (\ref{tt11}), (\ref{tt22}).

We conclude
that, for the backbone, the only loop specific parameters are those that determine
the lengths of the solitons. All additional  parameters in the energy function
determine the regular secondary structure content, such as (\ref{bc1}) and (\ref{bc2}). 
The profiles of all loops are completely determined by differently scaled versions of 
the {\it unique} dark solution solution to 
(\ref{nlse}). 

A typical super-secondary structure such as helix-loop-helix, involves around 15 amino acids.
Consequently, in order to describe the backbone geometry, one needs to specify 3$\times$15 = 45
coordinates. If we assume that the bond lengths are constant and given by (\ref{dist}), there are a
total of around 30 coordinates in the  typical super-secondary structure. These are the unknown variables that
we wish to determine. We have found that in the energy function (\ref{E1}) 
there are a total of six parameters (or seven if we include the dynamically determined position).
Consequently some 20 C$_\alpha$ coordinates per a super-secondary structure
must be determined by the functional form of the energy function. This is possible only if there is a firm
underlying universal physical principle that dictates the functional form of the energy function.  
This universal physical principle is also the reason, why proteins fold.

In \cite{peng1} we have found that over 92$\%$  crystallographic protein structures 
in PDB can be described in a modular fashion and with experimental B-factor precision, 
by combining together no more than 200 explicit soliton profiles. We propose that by
learning how to compute the parameter values directly from the sequence, 
the geometric shape of most
folded proteins can be constructed, simply by solving the Master equation (\ref{nlse}).

%
%
%
%
%
%
%
%
%
%
%
%
%
%
%

\subsection{Software } 

We have developed a software package  called {\it GaugeIt} to
tentatively identify the multi-soliton profile of a PDB structure, using the transformation (\ref{dsgau}). 
This software package is described at the www-address 
\begin{equation}
{\tt http://www.folding-protein.org/propro.html}
\label{www}
\end{equation}
{\it GaugeIt} reads the backbone C$_\alpha$ coordinates from the PDB and computes the bond
and torsion angles using (\ref{bond}) and (\ref{tors}), with the convention that all $\kappa_i $ are 
positive. We can then use {\it GaugeIt} to judiciously apply the transformation (\ref{dsgau}), 
to arrive at the proper soliton profile. 

We have also developed the software package {\it Propro}, that deploys (\ref{ite}) to determine
the parameters in (\ref{E1}) so that the ensuing multi-soliton profile constructed with {\it GaugeIt}
models the given PDB backbone structure.
This  software package is also described at the www-site (\ref{www}).
Once the parameter values are known, we can use the energy function (\ref{E1}) to study various dynamical 
properties of the protein.  The procedure ensures that the energy function indeed describes the desired static, 
crystallographic protein backbone conformation as a minimum energy configuration, in the limit of vanishing
temperature.

%
%
%
%
%
%
%
%
%
%
%
%
%
%
%

\subsection{Soliton Ansatz}

The explicit solution to the {\it Master equation} (\ref{nlse}) is not known to us, in terms of elementary functions. 
However, a general, mathematically rigorous existence proof can be presented to show that 
the soliton solution exists, and that the iterative equation (\ref{ite}) converges towards the 
soliton \cite{herr,nora}.

In \cite{cherno,peng} we have pointed out, that an excellent approximation can be obtained in
terms of hyperbolic functions, by discretizing the exact soliton solution to the continuum nonlinear Schr\"odinger
equation. This  approximative multi-soliton solution is a combination of (the index $r$ labels the super-secondary
structures)
\begin{equation}
\kappa_i  =  (-1)^{r+1} \frac{ 
(\mu_{r1} + 2\pi N_{r1})  \cdot e^{ \sigma_{r1} ( i-s_{r1})  } - (\mu_{r2} + 2\pi N_{r2}) \cdot e^{ - \sigma_{r2} ( i-s_{r2})}  }
{e^{ \sigma_{r1} ( i-s_{r1}) } +  e^{ - \sigma_{r2} ( i-s_{r2})}   }
\label{bond2}
\end{equation}
Here the $s_{r1}$ and $s_{r2}$ are parameters that determine the backbone site locations 
of the individual soliton centers in the multi-soliton, usually we set $s_{r1} = s_{r2}$. 
This determines the center of the fundamental  backbone loop \cite{cherno,peng} 
that we describe in terms of a single soliton solution. Long loops, with a complex structure, are described
by joining the profiles of (\ref{bond2}) together, one after another. 
The $\mu_{r1}, \ \mu_{r2} \in [0,\pi] $ are parameters, and their values are entirely determined by the adjacent helices and strands.
The $N_{r1}$ and $N_{r2}$ constitute the integer parts 
of $\mu_{r1} $ and $\mu_{r2}$ and for simplicity we shall take $N_{r1} = N_{r2} \equiv N_r$. This integer is 
like a covering number, it determines  how many times $\kappa_i$ covers  the fundamental 
domain $[0, \pi]$  when we traverse the soliton once. Therefore,
far away from the soliton centers we  have
\[
\kappa_i \ \to \left\{ \begin{matrix}  \mu_{r1} \  & \mod (2\pi) \ \ \ \ i >>  s_{r1} \\  -\mu_{r2} \  & \mod(2\pi) \ \ \ \ i << s_{r2}
\end{matrix} \right.
\]
For $\alpha$-helices and $\beta$-strands the $\mu_{r}$ values are given by (\ref{bc1}), (\ref{bc2}).
Negative values of $\kappa_i$ are related to the positive values by (\ref{dsgau}).
 
The Ansatz (\ref{bond2}) is a monotonic function. But in general the values of $\kappa_i \in [0,\pi]$ that we obtain
from PDB are not monotonic. When we encounter a site $i$ where $\kappa_i$ in the PDB data fails to be
monotonic, we can recover a monotonic structure by either adding or subtracting $2\pi$ to its value. In this manner
we can convert any 
sequence $\{ \kappa_i \}$ into a monotonic one, that we can then approximate by the Ansatz (\ref{bond2}).
Due to multi-valuedness of $\kappa_i$ as an angular variable in the three dimensional space, such
addition and subtraction of multiplets of $2\pi$ does not have any effect on the backbone geometry.

For $\mu_{r1} = \mu_{r2}$  and $\sigma_{r1} = \sigma_{r2}$ we have  the hyperbolic tangent. In this
case the two regular secondary structures before and after the loop are the same.
Moreover,  {\it only}  the (positive)  $\sigma_{r1}$ and  $\sigma_{r2}$ are 
intrinsically loop specific parameters. 
They specify the length of the loop.  Like the $\mu_{r}$,
they are combinations of the parameters in (\ref{E1}).

For the torsion angle, it turns out to be sufficient to introduce the following simplification  of (\ref{tauk}),
\begin{equation} 
\tau_i \ = \ \frac{ a_r }{1  + d_r \cdot \kappa_{i-1/2}^2}
\label{ti}
\end{equation}
That is, we  now set $b_\tau=0$ in (\ref{E1}), and we also set the overall
scale by selecting the Proca mass parameter 
$c_\tau = 1$.  We have also symmetrized the expression (\ref{ti}), 
by evaluating the value of the bond angle at the mid-point position $i-1/2$, which is  half-way between the  C$_\alpha$
sites.  We note that as a consequence, the torsion angle along the loop proper is  determined entirely by the bond angle:
There are only two parameters in (\ref{ti}). But each of the two regular 
secondary structures that are adjacent to the given  soliton have their own characteristic values 
of torsion, as exemplified in  (\ref{bc1}) and  (\ref{bc2}). Consequently
both of these two parameters are entirely determined by the regular secondary structures that are adjacent to the
soliton.  This leaves us with {\it no  loop specific parameter}  in (\ref{ti}) whatsoever, for the torsion angle. 
In this sense the {\it Ansatz} states, see subsection II.D,   that the 
geometric shape of {\it any} protein loop is essentially determined by the  adjacent 
regular secondary structures. The sole two loop specific parameters are $\sigma_{r1}$ and $
\sigma_{r2}$, but these parameters specify only the length of the loop.

\subsection{ On precision }

The interpretation of the protein backbone in terms of solitons can be used as a basis for a quantitative,
purely geometric secondary structure classification \cite{peng1}. This classification scheme can be developed
as a complement to existing schemes such as  CATH  \cite{cath} and SCOP \cite{scop}.

As a criterion for identifying the soliton structure, in our approach, we shall use the fluctuation  distance that we compute
from the experimental B-factors, in terms of  the  Debye-Waller relation
\begin{equation}
\sqrt{ <\mathbf x^2 >}  \ = \  \sqrt{ \frac{B}{8\pi^2}}  
\label{x}
\end{equation}
Here $B$ are the experimentally measured temperature factors, as given in PDB. 
If the RMSD distance between a soliton profile and an experimentally measured 
putative secondary structure is less than the average B-factor
fluctuation distance of the latter, it is meaningful to identify the two. On the other hand, if the RMSD distance is larger than the
average fluctuation distance, 
there is an experimentally meaningful and observable difference between the two. As a consequence,
their identification is not well grounded.  For a given
crystallographic structure, we then need to find the smallest number of soliton configurations, either by solving the
{\it Master equation} (\ref{nlse}), (\ref{tauk}) or by utilizing the {\it Ansatz} (\ref{bond2}), (\ref{ti}), that describe
the backbone with RMSD distance accuracy that is better than B-factor fluctuation distance accuracy. 

Obviously, this soliton identification 
method can be truly valuable only in the case of crystallographic structures that have been
measured  with very high resolution
and with very low B-factors. We remind, that we have chosen 1ABS as our model of myoglobin, due to its very low B-factor values.
Even though the resolution is not quite as small as we would like it to be.

The advantage of the {\it Ansatz} (\ref{bond2}) over the {\it Master equation} (\ref{E1}),  for classification
purposes,  is in its simplicity.  The (present) disadvantage 
is the lack of an energy function: We do not (yet) 
know the explicit relationship between the parameters in (\ref{bond2})  and  those in the energy function
(\ref{E1}). Consequently we do not know at the moment, how to perform energetic studies in terms of the 
{\it Ansatz}. But we can still utilize it as an effective tool, to classify a given protein backbone in terms of solitons, and to inspect
the fine structure of its geometry, as a complement \cite{peng1} to existing methods  \cite{cath,scop}  
In particular, we can try to identify any potential geometric anomaly in the backbone, those sites where the backbone
deviates from a {\it perfect soliton crystal}.

In \cite{peng1},  it has been shown 
using the {\it Ansatz} (\ref{bond2}), that over 92$\%$ of high resolution PDB configurations can be constructed in 
terms of 200 explicit {\it Ansatz} profiles. This makes a strong case that  the solitons of the DNLS equation are  indeed 
the correct modular building blocks of folded proteins \cite{peng1}.

\subsection{Nonequilibrium dynamics}

We shall be interested in the dynamics of the multi-soliton configuration, during repeated 
heating and cooling processed. For this we need a proper framework of non-equilibrium 
statistical mechanics. We shall approach the issue of constructing the non-equilibrium
dynamics using  same kind of principle of universality, that we have used in deriving the energy function
(\ref{E1}).  

We shall average over all very short time scale oscillations, vibrations and other 
tiny fluctuations in the positions individual atoms that, as such, are basically irrelevant to the 
way how the folding progresses over those time scales that are 
biologically relevant. The general concept of universality \cite{widom,kadanoff,wilson,fisher}
proposes us to introduce a non-equilibrium dynamics which is based on a symmetric Markovian evolution towards 
the Gibbsian probability measure, that describes the backbone at thermal equilibrium.  For this,  we adopt the following
coarse grained heat bath probability measure that we  implement using a Monte Carlo procedure,  
to describe the heating and cooling of the backbone, \cite{glauber,lebo,marti1,marti2} 
\begin{equation}
\mathcal P = \frac{x}{1+x} \ \  \ \ {\rm with}  \   \ \ \ x =     \exp\{ - \frac{ \Delta E}{kT} \}  
\label{P}
\end{equation}
Here $\Delta E$ is the energy difference between consecutive MC time steps,  that we 
compute from (\ref{E1}). The scale of units in the temperature factor 
$kT$  depend on the overall normalization of the energy
function (\ref{E1}); note  in particular that we have chosen  
the numerical value of $2$ for the normalization factor in the nearest neighbor interaction.
To determine the unit, we need in addition a renormalization condition. For this we need to perform a proper
experimental measurement(s), and compare the properties of our model to those 
of the protein that it describes, at that temperature.
One suitable renormalization point could be,   to try and identify  the  experimentally
measured $\Theta$-point temperature with the temperature value of the rapid transition in the
specific energy that takes place  in our model \cite{cherno2},  between the collapsed phase and the fully flexible chain (random walk)
phase.
In the next sub-section we show how to utilize  the determination of the $\Theta$-point  value as
a renormalization point for determining the temperature scale, in the case of myoglobin.

The MC implementation of
the probability distribution (\ref{P}) determines the standard  Glauber protocol \cite{glauber,lebo,marti1,marti2}. 
A time evolution of a system that follows Glauber dynamics has the characteristic property, that 
it approaches the Gibbsian thermal equilibrium state at an  exponential rate \cite{marti1,marti2}. 
This ensures that during our adiabatic heating and cooling cycles our 
backbone remains, at least locally, very close to a Bolzmannian equilibrium conformation. 
This is obviously a very reasonable property in the case of {\it any} protein under {\it in vivo} conditions. 
Normally, there are no large and abrupt temperature fluctuations in living cells.

During the course of our simulations we monitor the state of the configuration by keeping watch on
the temperature dependence of different quantitative measures.  In particular, the radius of gyration 
is a widely used order parameter, in the context of 
polymer physics. Asymptotically, for large values of monomers $N$ (number 
of residues in the case of proteins)  the value of the radius of gyration $R_g$  increases according to  \cite{lim},
\begin{equation}
R_g \ = \ \sqrt{ \, \frac{1}{2N^2}  \sum_{i,j} ( {\bf r}_i 
- {\bf r}_j )^2 \,} \  \approx \  R_0 N^{\nu} ( 1 + \beta_1 N^{-\Delta_1} + ... ) 
\label{nu}
\end{equation} 
Here ${\bf r}_i$ ($i=1,2,...,N$) are the locations of the $N$
monomers (C$_\alpha$ atoms). The critical exponents $\nu$ and $\Delta_1$ are universal quantities.
But the form factor $R_0$ that characterizes the effective distance between the monomers, 
and  the amplitude $\beta_1$ that parametrizes finite size corrections, are not universal.
These two parameters are in principle computable. They contain all the effects of temperature and 
chemical microstructure, and all the atomic level details of the polymer.
For a linear polymer such as a protein, the  compactness index has the following mean field
values \cite{degen,schafer}: 
\begin{equation}
\nu  \ = \ \left\{ \!\! \!\! \begin{matrix} 3/5 \ \   {\rm SARW} \\ 
\! \! \! 1/2 \ \ \  \ {\rm RW}
 \\
 \ \ \   \ 1/3 \ \  {\rm collapsed}
 \end{matrix} \right.
 \label{nus}
 \end{equation}
As a function of temperature, the collapsed phase occurs at low temperatures (bad solvent) while the self-avoiding 
random walk (SARW) phase describes the high temperature (good solvent) behavior of polymers. 
The random walk (RW) phase takes place when a polymer is in 
its $\Theta$-temperature regime.  This is the transition regime that 
separates the  SARW phase from the collapsed phase.  

%
%
%
%
%
%
%
%
%
%
%
%
%
%
%
%
%
%
%
%
%
%

\subsection{Temperature renormalization}

During the heating and cooling simulations,  we do not change the parameter values in 
(\ref{E1}) but keep them fixed.
In particular, in the probability distribution (\ref{P}) we  normalize 
the nearest-neighbor coupling contribution in (\ref{E1}) as follows,
\begin{equation}
- \frac{2}{kT} \sum\limits_{i=1}^N  \kappa_{i+1} \kappa_i   
\label{norms}
\end{equation}
This implies that
the temperature factor $kT$ depends on the physical temperature $t$ in a non-trivial fashion. That is,  we really have
\begin{equation}
 \frac{2}{kT} \ \to \  \frac{J(t)}{k_B t}
\label{JT}
\end{equation}
where $J(t)$ is the strength of the nearest neighbor coupling at temperature $t$ (with $k_B$ the Boltzmann constant). Its numerical value depends
on the temperature in a manner that is governed by the 
standard renormalization group equation  \cite{widom,kadanoff,wilson,fisher,schafer}
\begin{equation}
t \, \frac{ d J}{dt} = \beta_J (J; b,c,m,d,q,e) \ \sim \ \beta_J(J) + \dots
\label{RGEJ}
\end{equation}
For simplicity, we shall assume 
that to leading order the dependence of $\beta_J$ on the other couplings can be ignored.

In general, the parameters and as a consequence their $\beta$-functions, depend also
on the properties of the solvent. In particular, a
change in the solvent properties can be compensated by a change in
temperature scale.

In the low temperature limit we may expand the nearest neighbor coupling as follows, 
\begin{equation}
J(t) \ \approx \  J_0 - J_1 t^\alpha + \dots \ \ \ \ \ \ {\rm as} \ t \to 0
\label{J0}
\end{equation}
Here the value of $J_0 $ can not vanish. 
The critical exponent $\alpha$ controls the low temperature behavior of $J(t)$. 
The asymptotic expansion (\ref{J0}) corresponds to a $\beta$-function (\ref{RGEJ}) that in the $t\to 0$ limit
approaches
\[
\beta_J (J) = \alpha( J - J_0) + \dots 
\]
Consequently, at low temperatures
\begin{equation}
kT \ \approx \ \frac{2 k_B}{ J_0} \, t
\label{Tlow}
\end{equation}

In terms of the temperature factor, (\ref{RGEJ}) translates into
\begin{equation}
t  \frac{ d }{dt}\left( \frac{1}{kT} \right) = - \frac{1}{k T} + \frac{1}{2 k_B t}\,  \beta_J\!\left( \frac{2 k_B t}{kT} \right)
\label{betaT}
\end{equation}
We search for an approximative  solution in the collapsed phase, where the temperature $t $ is below 
the critical $\Theta$-point temperature $t_{\theta}$ for the transition between the collapsed phase and
the random walk phase.   This is the physical temperature value that corresponds to the unfolding transition temperature factor value
$kT_\Theta$, in our dimensionless units.

We introduce
\[
 \beta_J\!\left( \frac{2 k_B t}{kT} \right)  \ = \  \frac{2 k_B t}{kT}  +  F\left( \frac{2 k_B t}{kT} \right)
\]
We define
\[
y = \frac{1}{kT}  
\]
\[
x = \frac{1}{2k_B t}
\]
The equation (\ref{betaT}) then becomes
\[
\frac{dy}{dx} =  - F(\frac{y}{x})
\]
The solution is
\[
\ln ( {\lambda} \, x ) = - \int^{\frac{y}{x}} \frac{du}{F(u) + u }
\]
Here $\lambda$ is an integration constant. For simplicity, we shall assume that 
the leading non-linear corrections are logarithmic, as this is often the case \cite{widom,kadanoff,wilson,fisher,schafer}. 
As a consequence, to the leading order
\[
F(u) = (\eta-1) \, u + \alpha  \, u \! \ln u  + \dots 
\]
However, we note that in general higher order corrections are present.
We re-introduce the original variables and choose
\[
\eta = - \alpha \ln J_0
\]
This gives for the temperature factor
\begin{equation}
kT  \ \approx \ \frac{2}{J_0}\,  k_B t \, \exp\{ \frac{J_1}{J_0} t^\alpha \}  
\label{kTt}
\end{equation}
\[
\approx \frac{2}{J_0}\,   k_B t  +  \frac{2J_1}{J_0^2} \, k_B t^{\alpha+1} + ... \ \ \ \ {\rm as } \ \ t \to 0 \ \ \ \ (\alpha >0)
\]
where we choose the integration constant so that 
in the low temperature limit we recover (\ref{Tlow}). 

Note that for the value of the nearest neighbor coupling, the result (\ref{kTt}) gives
\[
J(t) \ \approx  \ J_0 \exp\{ -\frac{J_1}{J_0} t^\alpha \} 
\]
As a consequence, the coupling between bond angles becomes weak at an exponential rate, as the temperature 
approaches the transition temperature $t_{\theta}$ between the collapsed phase and the random walk phase.

Similarly, all the other couplings that appear in (\ref{E1}) are also temperature dependent, each with
their own renormalization group equations. 
For example, the quartic $\kappa_i$ self-coupling 
$q$ in (\ref{E1}) flows according to a renormalization group 
equation that has the standard  functional form
\begin{equation}
t \, \frac{ dq } {dt} \ = \  \beta_q (q) 
\label{betac}
\end{equation}
where for simplicity,  we again assume that in the leading order $\beta_q$ depends only on $q$. 
In (\ref{qc1}), (\ref{qc2}) we have argued that $q$ can be interpreted as a measure of the strength 
of hydrogen 
bonds. The hydrogen bonds  are presumed to 
become vanishingly weak when the protein unfolds. This takes place when the system reaches 
the transition temperature $t_{\theta} $ between
the collapsed and random walk phases. Thus we expect that, asymptotically, 
\[
q(t)  \  \to  \ q_\theta \ | t - t_{\theta} |^ {\gamma_q} \ \ \ \ \ {\rm as}  \ \ t \to t_{\theta}  \ \ {\rm from ~ below}
\]
Here $\gamma_q$ is a critical exponent that characterizes the vanishing of the strength of 
hydrogen bonds.  More broadly, 
we may transplant here $t_\theta \to t_H$, the  temperature value at which all
hydrogen bonds disappear also in the solvent. It may have a value which is higher than $t_\theta$.

Above $t>t_{\theta}$, when the hydrogen bonds become vanishingly weak, we 
expect that effectively $q\approx 0$ in (\ref{E1}). 
On the other hand,  we expect that as the temperature decreases the value of $q(t)$ increases,
so that in the low temperature limit  we have
\[
q(t)  \  \to  \ q_0 - q_1 t^{\gamma_0}  + \dots \  \ \ \ \ {\rm as} \ \ t \to 0  
\]
Thus 
\[
\beta_q (q) \ \approx \ \gamma_0 ( q - q_0) + \mathcal O[(q-q_0)^2]
\]
Here $q_0$ is close to the value we obtain from PDB, when we compute the parameters in (\ref{E1}) from
the crystallographic low temperature structure.

%
%
%
%
%
%
%
%
%
%
%
%
%
%
%
%
%
%
%
%
%
%

\section{Results And Discussion}

\subsection{Soliton Ansatz and classification of 1ABS backbone}

We now proceed to investigate 
the myoglobin backbone with PDB entry 1ABS \cite{1abs},  first in terms of the soliton {\it Ansatz}. The goal is
to identify  and classify the secondary structure assignment, in terms of solitons. In the subsequent sub-sections
we shall  analyze  the dynamics of myoglobin, by explicitly constructing 
the energy function (\ref{E1}) using the  1ABS backbone.

The index $i$ for the bond and torsion angles of myoglobin 1ABS takes values $i=3,...,151$;  the definition
of $\kappa_i$ involves three C$_\alpha$ sites while for $\tau_i$ we need four C$_\alpha$ 
sites.  
 In Figure 1 (top)
 we display the
$\kappa_i$ and $\tau_i$ profiles along the 1ABS  backbone. 
In this Figure, we use the standard differential
geometric convention that  a bond angle $\kappa_i$ should be  non-negative. 
We adopt a purely geometric soliton based approach to secondary structure classification:  
We use relations such as (\ref{bc1}), (\ref{bc2})  to identify
structures such as  $\alpha$-helices and $\beta$-strands. These are the
regions where $(\kappa_i, \tau_i)$ have the definite constant values given by (\ref{bc1}), (\ref{bc2}).
Loops are identified simply as those regions where  $(\kappa_i, \tau_i)$ are variable, loops connect the 
regular secondary structure regions to each other.  Each loop is either a single soliton or a composite of several single
solitons, each with the profile given by our {\it Ansatz}. Solitons can be so close to each other that they partially overlap,
and we use the B-factor fluctuation distance (\ref{x}) as a criterion,
to determine the soliton content of a loop.

At visual level, the soliton identification of secondary structures becomes very precise when we utilize 
the symmetry (\ref{dsgau}) 
together with the freedom to define both angles modulo $2\pi$. In the case of 1ABS, 
we apply the software {\it GaugeIt} described at (\ref{www}) 
to  convert  Figure 1 (top) into Figure 1 (bottom).  In the Figure 1 (bottom) the  
individual loops  that are not really visible in the Figure 1 (top), have 
become commuted  into regions where $\kappa_i$ changes its sign as it interpolates between the regular secondary structures.
In  Figure 1 (bottom) we are able to immediately visualize  
the regular secondary structures, and  loops that correspond to single solitons; In particular,  the center of a single soliton
loop can be unambiguously located to 
the point where the linear polygon that interpolates between the $\kappa_i$ changes its sign. 
Such points of vanishing curvature play a special r\^ole in the geometry of differentiable plane curves, they are 
the {\it inflection points}. The existence of an inflection point is a $\mathbb Z_2$  invariant  topological property of a plane curve. 
In particular,  an  isolated  inflection point can not be made or removed by any
continuous local deformation of the curve. Inflection points in planar curves can only 
be made or removed in pairs, or, by translating them individually through
the endpoints of the curve. This is the kind of stability that is the hallmark of a topological (kink) soliton. 

The top and bottom of Figure 1 describe
the same intrinsic backbone geometry. From the bottom Figure 1 we conclude that in terms of $\kappa_i$  we may {\it putatively}
interpret the myoglobin backbone as a space polygon, with {\it eleven} helices that are separated by {\it ten} single soliton
loops. These numbers 
are unambiguously determined by the number of inflection points in the space curve, 
and accordingly we start by dividing the backbone  into ten soliton profiles. These profiles are identified in Table 1. 

We remind that our geometry based identification of the loops and helices along
the 1ABS backbone does not necessarily need to coincide 
with the one based on inspection of hydrogen bonds.  In particular, according to the common classification, see for example \cite{dssp},
the soliton pair  3 and 4, the soliton pair  6 and 7, and the soliton pair  9 and 10,  are all interpreted as a single loop. 

We remind that we are primarily  interested in the backbone gates that ligands can use to 
enter and depart the interior of the myoglobin.  Consequently, in Table 1, we have limited our attention 
to those residues that are located 
between the sites 14 and 136. Furthermore, we observe 
that from our geometric point of view, the PDB data reveals  that,   in 1ABS,
there are four different types of solitons. Those that connect two $\alpha$ helices, those that connect an $\alpha$-helix 
with a $3/10$-helix or {\it vice versa},  and finally, those that connect two $3/10$-helices.

We  proceed to describe each of the 10 solitons in Table 1  in terms of the {\it Ansatz} 
profile. For the $\kappa_i$ we introduce (\ref{bond2}), with
$r=1,...,10$ labeling  the ten helix-loop-helix motifs. We remind that
the $ \mu_{r1}$ and  $\mu_{r2}$ specify the asymptotic values of $\kappa_i$.  
Thus these parameters are entirely 
determined by the nature of the adjacent regular secondary structures 
like in (\ref{bc1}) and  (\ref{bc2}), they are not intrinsic to the loop. 
The parameter $s_r$ determines the location of the corresponding soliton, {\it i.e.} the putative position of an
inflection point where $\kappa_i$ vanishes. Note that since $\kappa_i$ for a given soliton
depends only on the difference $i-s_r$ the soliton profile is translation invariant,
the profile of $\kappa_i$ is not influenced by the value of $s_r$, except for a translation along the backbone. 
As we have argued in sub-section II D
this leaves us with {\it only two} loop specific parameter, the $\sigma_{r1}$ and $\sigma_{r2}$. 
They quantify the length of the soliton, both before and
after the  inflection point. 

We have determined the parameters in (\ref{bond2}), (\ref{ti})  for all 
the solitons in 1ABS, using a standard Monte Carlo based Metropolis algorithm \cite{metro}
to minimize the RMSD distance between the space polygon that is described by our 
{\it Ansatz}, and  the $C_\alpha$ backbone of 1ABS.  
In Table II we present the parameter values, together with the 
RMSD distances between the given soliton and the corresponding loop in 1ABS. 
Please keep in mind that in the case of $a_r$, $d_r$ and also in the case of the $\mu_{r1}, \ \mu_{r2}$, the large 
numerical values are somewhat misleading,  since both $\kappa_i$ and $\tau_i$ are defined modulo $2\pi$. We observe that the
solitons are highly symmetric, the differences between the values $\sigma_{r1}$ and $\sigma_{r2}$, and between the
values $m_{r1}$ and $m_{r2}$ are remarkably small. In fact, we may identify these parameters with only a slight loss
in accuracy of our description.

According to Table II, the RMSD distances between the individual soliton profiles and the ensuing loops of 1ABS
are very small. For a more detailed comparison, we compute the average experimental  fluctuation distance 
for each individual C$_\alpha$ atom in the 1ABS, and compare it to the difference between 1ABS backbone and our soliton profiles.
We compute the fluctuation  distance using  the  Debye-Waller relation (\ref{x}).
In Figure 2  we show how the difference between our soliton and the Debye-Waller fluctuation distance (\ref{x})
varies from site to site,  for each of the ten solitons in the case of 1ABS. 
This Figure also shows an estimate of the 
zero-point fluctuation regime around the soliton, as a grey area. According to  \cite{kln} 
the extent of this regime in PDB data is around 0.15 \.A.  As visible in Figure 2, for most of the sites our solitons give an accuracy that is fully comparable to, 
and even clearly exceeds the experimental B-factor precision. This justifies our interpretation of
1ABS as a 10 soliton state. 

But in the vicinity of two sites, around 76 and around
123, the soliton visibly exceeds the B-factor fluctuation distance, by about $0.4 \ - \ 0.5$ \.Angstr\"om. 
The first site 76 where we observe such a large deviation, is located near the end of the helical structure E. 
This site is
close to the loop that connects the  helices E and F, that  together form the ÓVÓ-shaped 
pocket where the heme group is located. 
Consequently this deviation between the soliton and the PDB structure
could be due to the presence of heme: We propose that the heme deforms the shapes of the adjacent helices, bending them
away from ideality, and this is detected by our Ansatz profile, as a deviation from the perfect soliton crystal structure.

The second site, number 123, where our Ansatz profile detects a deviation, is located in 
the loop that separates the helical structures
G and H. These structures are also close to the heme pocket. 
They are located on the opposite site of heme, from the E and F helical structures. 
From Table I we observe that according to our classification, the site 123 is in the very short boundary region 
between two solitons. Geometrically the site is in an $\alpha$-helical position, even though the hydrogen bond 
patterns place it in a loop \cite{dssp}. We suspect that the deviation from the {\it Ansatz} 
that we observe around site 123, reflects the presence of an 
interaction between the two solitons, numbers 9 and 10 in Table I, 
that are located {\it very} close to each other. 

%
%
%
%
%
%
%
%
%
%
%
%
%
%
%
%
%
%
%
%
%
%

\subsection{Thermal fluctuations and PDB structures}

Even though the accuracy is not always equally good, 
NMR spectroscopy is better suited than x-ray crystallography for analyzing the presence of multiple 
conformational sub-states and the thermal effects on protein structure. Unfortunately, at the moment, 
the NMR data on myoglobin is very sparse. We only find one NMR structure \cite{1myf}, with PDB code 
1MYF. Like 1ABS,  it is from sperm whale. The measurements have been made at $308 \ K$. 

There are twelve structures in the NMR data. We have used them to estimate the 
backbone fluctuations. For this we have computed the average values 
of the C$_\alpha$ coordinates,  and the one standard deviation
fluctuation distance around the averages. This gives some estimation 
on the range of fluctuations, that we can try to 
compare with the B-factor fluctuations distances in Figure 2. The results are given
in Figure 3, that we have  modeled according to Figure 2. In particular, we have similarly 
divided the backbone into ten segments and then plotted the one-sigma NMR
fluctuation distance for each site along the backbone. 
%
%
%
%
%
%
%
%
%
%
%
%
%
%
%
%
%
%
%
%
%
We record that the fluctuation amplitudes in 1MYF have the largest values in the segment, which 
located between sites 80-86. This segment is also visible as a slight anomaly, in the soliton profile of Figure 2.  
There also appears to be a clear  increase in the fluctuation amplitudes around sites 48-58, in comparison to Figure 2.
However, unlike the soliton profile in  Figure 2, in the region between sites 72-78 the NRM data detects  almost no
fluctuations, as shown in Figure 3.  The vicinity of site 121 shows the presence of a fluctuation, in both cases. 

On the basis of the very limited number of NMR measurements on myoglobin, there is only one PDB entry, 
and in particular since the level of accuracy at which
we aim to scrutinize the myoglobin structure is very high (below 1 \AA ngstr\"om), 
we can not draw conclusions beyond these observations.
We hope, that in the future there will be more high precision NMR analysis available, over a wide temperature range
and in particular above 300 K.

We can also try to estimate, from available crystallographic x-ray 
data, how thermal fluctuations might influence the myoglobin backbone
structure. For this we compare the RMSD distances over the positions of the ten solitons in 
two PDB structures, 1ABS and  1MBC \cite{1mbc}.
Both are carbonmonoxony-myoglobins from sperm whale. The resolutions are equal {\it i.e.} 1.5 \AA, and the 
R-values are also comparable (0.207 {\it vs.} 0.171).  But the B-factors of 1ABS are clearly smaller;
 the data of 1ABS has been measured at 20K while 
the data of 1MBC has been measured at 260K. Consequently there might be differences in the structures, that we can try to 
interpret in terms of thermal fluctuations. 

In Table III we present the RMSD distances 
between 1ABS and 1MBC, over our ten soliton structures. 
We find that the RMSD distances between the 1ABS and 1MBC are mostly either smaller or 
quite comparable to the average Debye-Waller B-factor fluctuation distances along the 1ABS backbone.
The only exception is in the region that corresponds to our soliton number 9. Here,
the RMSD distance between 1ABS and 1MBC is almost twice as large as the average
B-factor fluctuation distance in 1ABS. We note that this is the soliton that 
terminates around site 123, where our {\it Ansatz} analysis of 1ABS also detects an anomaly.
Excluding this anomaly, we conclude that up to temperatures at least as high as 260K, thermal backbone
fluctuations are not experimentally visible in myoglobin, to the extent crystallographic 
B-factors measure these fluctuations.

Experimentally,  not very much seems to be known about the $\Theta$-point transition in myoglobin. 
We note the following experimental observations, obtained by circular dichroism spectroscopy  \cite{theta1,theta2}:
It appears that the helical structures of myoglobin (horse heart) are stable all the way up to around 
348K.  At this temperature,  around  $67\%$ of the original low temperature $\alpha$-helical structures are still retained.
But when the temperature increases beyond 348K, there is a rapid decrease so that at 363K only
around $24\%$ of the $\alpha$-helical structures remain. After this, there is a slower decrease in the amount of $\alpha$-helices
so that at around  400K some $14\%$ of the original helical structure remains.   As a consequence, the thermal denaturation of myoglobin 
can be considered to have a critical temperature somewhere around 348-353K. It was also observed that, upon cooling 
from temperatures below 348K, the protein recovered its  original low temperature helical structure. But for temperatures above
353K, this was no longer the case.
On the other hand, sei whale myoglobin \cite{theta2}, which 
is assumed to be very similar to the myoglobin of sperm whale, starts to denature already  at 
temperatures as low as 293K \cite{theta2}. There are two steps, with the major denaturation taking place at around 337K.  
For bluefish tuna, the temperature values are again slightly different \cite{theta2} and  now there appears to be three steps. 

%
%
%
%
%
%
%
%
%
%
%
%
%
%
%
%
%

\subsection{Side-chains and backbone slaving}

The energy function (\ref{E1}) computes the protein geometry entirely in terms of the backbone C$_\alpha$ carbons. 
In particular, there are no terms that {\it explicitly} describe side-chains and their interactions. In \cite{marty} an extension has been
proposed, to account for side-chain interactions. There, it was observed that the side-chain contributions are relatively small, and can
be described by the backbone: 
Since the energy function (\ref{E1}) is based on the universality arguments \cite{widom,kadanoff,wilson,fisher}, one
can argue that the effect of side-chains is accounted for by a renormalization of the couplings in (\ref{E1}). 

In the case of myoglobin, we can use the existing PDB data to try and estimate the extent to which the side chain 
conformation is  slaved to the backbone. For this we compare the directional angular distribution of the side-chain 
C$_\beta$ and C$_\gamma$ carbons in relation to the backbone, at different temperatures.  We again 
compare 1ABS and 1MBC. This gives us {\it some} impression, how thermal fluctuations affect the side-chain
orientations over the temperature range 20K-260K. 
For each of these two myoglobins, we compute the angular distribution of the side-chain C$_\beta$ and C$_\gamma$ atoms
along the entire backbone, as they are seen at the position of the corresponding C$_\alpha$ atoms. 
In Figure 4 we show the angular distribution of the C$_\beta$ atoms for 1ABS, and in Figure 5 we show this distribution for
1MBC. 

%
%
%

%
%
%
%
In the case of 1ABS, all the side-chain  C$_\beta$ orientations are in the expected region, as shown in Figure 4. 
For the most part, there are no observable thermal effects in the orientations of the C$_\beta$ in 1MBC either. The only exceptions are the 
side-chains with PDB indices 98, 122, 123 and 152 where we find that the directions of the C$_\beta$ atoms are slightly 
outside of the expected region  {\it i.e.} the grey background in Figure 5. This may  be an indication of thermal sensitivity, at these 
side-chain sites. 
But it can as 
well be due to an experimental  refinement procedure, that tend to target the side-chain atoms. 
We propose that these exceptions we have identified,
could be scrutinized experimentally. We remind that in the case of 1ABS (see Figures 2 and 3) we have already observed certain 
anomaly around site 121.

At the level of the C$_\gamma$ carbons, in the case of 1ABS all the directions are again within the expected regions, as shown in  Figure 6.

%
%
%
%
%
%
 In the case of 1MBC, the C$_\gamma$ fluctuations around the expected regions are also minor, as shown in Figure 7.

%
%
%
%
%
%
We conclude from Figures 4-7, that at least within the temperature range 20-260K,  the side-chain directions appear to be strongly slaved to the
backbone, and their effects can thus be accounted for by a renormalization  in the parameters in (\ref{E1}), 
in determining the backbone geometry. However,
we record that there is a slight anomaly that we observe in 1MBC C$_\beta$ atoms, at sites 122 and 123.
This is the same region where we have previously observed slight backbone anomaly.

\subsection{1ABS backbone as a multi-soliton}
\label{sect:phase}
%
%
%
%
%
%
%
%
%
%

We proceed to the construction of the multi-soliton solution to the {\it Master equation} (\ref{nlse}) that
describes the 1ABS backbone. We aim to  determine the parameter values, for which the 
1ABS backbone is a local energy minimum of (\ref{E1}). Once these parameters are known, we  can
perform  
a dynamical heating and cooling analysis of myoglobin.
 
We construct the multi-soliton for the sites with PDB index between
N=8-149. That is, we do not include the flexible tails at the ends of the backbone. These tails could be included, 
but with added complexity, and it appears to us, without much additional insight to the issues that we wish to address here.

We start by recalling  the backbone bond and torsion angle spectrum in terms of the putative multi-soliton profile,
shown in Figure 1 (bottom). 
We use our program package {\it Propro}  described at (\ref{www}), to solve the {\it Master equation}  (\ref{nlse}) 
for the ensuing parameter values in (\ref{E1}).
In Table IV we give parameter values for the most accurate multi-soliton
profile that we have found. It describes the 1ABS backbone with 0.78 \AA ~ RMSD accuracy. 

When we assume that all the bond lengths have the constant 
value  (\ref{dist}),  we have 282 C$_\alpha$ angular coordinate values that we need to determine from the energy function (\ref{E1})
in order to construct the backbone from (\ref{DFE2}), (\ref{dffe}).
Since  there are 
a total of 80 parameters in Table IV,  a total of 202 coordinates 
remain to be determined by the multisoliton that minimizes the energy  function. 
Therefore, these 202 unknowns are the {\it predictions} of the model, they directly probe 
the physical principles on which (\ref{E1}) has been built. 

%
%
%
%
%
%
%
%
%
%
%
%
%
%
%
%
%
%
%
%
%
%
%
%
%
%
%
%
We observe that the $\sim$ 0.8 \AA ~ RMSD accuracy of the multi-soliton solution  is not as good as
the accuracy that we obtain for each individual loop, using the {\it Ansatz} (\ref{bond2}), (\ref{ti}). But now, for technical simplicity,
we have also 
restricted the values of $\kappa_i$ strictly into the range $[-\pi,\pi]$. Furthermore, we describe the full chain using one single
multi-soliton solution to the {\it Master equation} and as a consequence we have an energy function that we can use
in dynamical considerations.  If instead we solve the {\it Master equation} for each of the individual solitons independently, we obtain accuracies that match and even exceed those of the {\it Ansatz}. We have done this individual soliton construction, 
and in this way confirmed our
interpretation of the backbone in terms of ten solitons.  However, this construction does not give us a {\it single} 
energy function that 
describes the {\it full} backbone, which is a complication for the energetic and dynamical issues that we wish to address.
We note that it is likely  
that a multi-soliton solution with better accuracy can be found, but the present one is sufficient for our purposes. The
convergence of our numerical algorithm becomes slow and somewhat time-consuming on the MacPro desktop that we use,
and for this reason we have simply stopped the numerical simulation at the point we reach 0.78\AA.

In Figure 8 we compare, site-wise, the precision of the multi-soliton
%
%
%
%
%
profile with the PDB structure 1ABS. Again, the 15 pico-meter gray-scaled region around the multi-soliton profile 
corresponds to the regime of zero point fluctuations, see \cite{kln}. 
The red line describes the B-factor fluctuation distances in the PDB data of 1ABS.
We have computed these using the Debye-Waller relation (\ref{x}). 

We proceed to a detailed comparison between the 1ABS myoglobin backbone and the multi-soliton soliton
of (\ref{E1}), using the
parameters given in Table I. We first note that conceptually, the multi-soliton describes a {\it single} 
structure in the limit of vanishing temperature.
In particular, it does not account for any 
kind of conformational fluctuations that are due to thermal, lattice imperfection, or any other kind of conformational
sub-state effects. 
On the other hand, the PDB configuration 1ABS is more like an average over a subset of 
different conformational sub-states. The experimentally measured 1ABS crystal should 
not be interpreted as a single static structure, but rather as an average over 
a large number of possible instantaneous structures.  

From Figure 8 we observe that the distance between the multi-soliton solution
and the C$_\alpha$  carbon backbone of 1ABS has its largest values mainly in two 
regimes. These are located {\it roughly} between the sites 35-45,  and between  the sites 79-98. 
The first regime corresponds to the single soliton that models the loop between helix B and helix C.
The second regime corresponds to the location of the helix F which is part of the "V"-shaped pocket of helices
E and F, where the heme group is located. In particular, the helix F includes the proximal histidine at site 93,
which is bonded to the iron ion of the heme.  Note that in addition, we again detect the anomaly at around site 121.
Finally, when we compare Figure 3 and Figure 8, we observe that  there is also clear
correlation between regions, where fluctuations are enhanced.

In order to understand the origin of these deviations from a perfect multi-soliton crystal, 
we  check for the presence of potential 
structural 
disorders in 1ABS using  {\it Molprobity} \cite{molprobity}. 
In Figure 8, along the top at the level of the 2.0 \AA ~ line, we have marked with blue those regions where 
according to {\it Molprobity}  there are potential clashes. The  {\it Molprobity} clash score
of 1ABS is 20.32 which puts it in the 10th percentile among structures with comparable resolution,
100 $\%$ being the best.   The regions of potential structural clashes correlate with those
regions, where our multi-soliton profile has the largest deviations from the 1ABS backbone.
Except the vicinity of the site 123, which is unproblematic according to {\it Molprobity}.

We first consider the difference between the multi-soliton and the 1ABS backbone 
around the sites 79-98, that was also identified by the {\it Ansatz} as a potentially 
troublesome one. The difference appears to be largely due to a deformation of the
helix F. It could be caused  by a  bond between the proximal histidine at site 93 and the
heme iron. This might introduce a strain which modifies the backbone. The effect of the heme is 
not accounted for by our energy function, in the present form. Consequently we propose the
histidine-heme interaction
to be the likely explanation for  the relatively large deviation between our multi-soliton 
profile and the 1ABS backbone, at this point. 

We proceed to consider the difference between the multi-soliton and the 1ABS backbone around the sites 35-45.
These sites are also located very close to the heme. For example, the distance between the C$_\alpha$ carbon
at site 45 (Arg) and the hem oxygen 154 is 4.84 \AA, and the C$_\alpha$ of Phe at site 43 is even closer to the heme. 
This proximity between the backbone and the heme is reflected in the {\it Molprobity} clash at site 45 (between
C$_\delta$ and 154 HEM). We conclude that there could be strain in the backbone structure which is 
due to the heme,  and this could explain the difference between 1ABS backbone and the multi-soliton configuration
in this regime. 

Finally, we note that in Figure 8, there is also the previously observed anomaly at site 121 (glycine). 
At this point, we have no explanation for the anomaly, except that it was also observed both by using the {\it Ansatz} 
and by comparing the experimental PDB structures 1ABS and 1MBC, both for the backbone and side-chains.  
We also note that glycine is flexible
and that  this region is on the exterior of the protein. This leaves the hydrophobic 
phenylalanine at nearby site 123 exposed to the solvent. Consequently relatively strong fluctuations 
between several local conformational sub-states are possible, and the difference between 1ABS and 1MBC could
indicate for the presence of thermal sensitivity in these fluctuations.  Since the site 123 and its immediate vicinity,
come up so persistently as a slight anomaly in our analyses,  an experiment should be 
performed to understand whether something of interest indeed takes place.

%
%
%
%
%
%
%
%
%
%
%
%
%
%
%
%

%
%
%
%
%
%
%
%
%
%
%
%
%
%
%
%

\subsection{ Heating, Cooling and the $\Theta$ point transition}
\label{sect:phase}
%
%
%
%
%
%
%
%
%
%

At the present, there is still an incomplete understanding how small non-polar 
ligands such as O$_2$, CO and NO, move between the external solvent and the heme 
inside the myoglobin. 
From the available static crystallographic PDB structures one can not  identify any obvious ligand pathway.
It is likely, that  the process involves thermally driven large scale conformational motions.   
Thermal fluctuations can open and close gates through which the ligands migrate, and these
gates are not necessarily visible in the crystallized low temperature structures.    

We have performed extensive heating and cooling simulations using the energy function (\ref{E1}),
with the 1ABS specific parameter values that are given in Table I. The goal is to locate and
identify thermally driven {\it backbone} ligand gates, and in particular to study how they open 
and close when the myoglobin is subjected to repeated heatings and coolings.  
The reason why we concentrate on {\it backbone} gates
is that, as argued in sub-section III C, 
the side-chain conformations appear to be slaved to the backbone ones. 
Consequently a gate that opens in the side-chains, should make its presence known at the backbone level.

We describe the non-equilibrium heating and cooling 
processes statistically,  in terms of
Glauber dynamics (\ref{P}) \cite{glauber,lebo,marti1,marti2}. The protocol has been described in sub-section II G.
We start our simulations at a low temperature value, with the multi-soliton configuration that models the 
1ABS. We note that conceptually, as a classical solution, 
the multi-soliton configuration is properly defined in the limit of vanishing temperature, where fluctuations
are tiny.

For the numerical value of the low temperature, in terms of the dimensionless unit
that is fixed by our choice of  the overall energy  scale in (\ref{E1}), we have chosen
\[
kT_L = 10^{-16}
\]
But we have confirmed that substantially smaller values can also be introduced, without affecting {\it any} of our results or
conclusions. 
We select the numerical high temperature value to be 
\[
kT_H =  10^{-13}
\]
In Figure 9 we show how these dimensionless temperature 
units can be converted into Kelvin scale. This Figure 
has been obtained by applying the renormalization group flow described in sub-section II I., to relate the two temperature factors
$kT$ and $k_Bt$. The conversion relation is given by (\ref{kTt}),  with explicit parameter values
\begin{equation}
kT = 1.6181 \cdot 10^{-9} k_B \, t \, \exp\{ 0.05506 \, t \}
\label{Ttot}
\end{equation}
where we use CGS units so that $k_B = 1.381 \times 10^{-16} \, erg/K$. 
The detailed determination of the parameters in this conversion relation is postponed to 
subsection III E.
%
%
%
%

We note that under {\it in vivo} conditions myoglobin always interacts with water. This interaction is 
essential for maintaining the collapsed phase. In our approach we account for the solvent (water) 
implicitely, in terms of the parameter values in (\ref{E1}). 
In particular, as such our model does not directly take into
account the highly complex phase properties of water at sub-freezing temperatures  \cite{mos,frau,dow1,ringe,wat2,wat1}. 
Nor does it
account for the complications that appear when the temperature raises above the boiling point of  
water. We simply do the best we can and trust that, to the extent the crystallographic low temperature structure of myoglobin 
relates to its biologically active native fold, our approach also describes the thermal dynamics 
of the myoglobin backbone under {\it in vivo} conditions.

We start the simulation at $kT_L$. The heating takes place  at a very slow but
exponential rate of increase, during 5 million Monte Carlo steps.  According to Figure 9, in terms of physical temperature $t$,
this corresponds to an adiabatically slow nearly-linear rate of increase. The system is modeled by the standard Glauber 
protocol \cite{glauber,lebo} which tends to a Gibbsian equilibrium distribution, also at an exponential rate \cite{marti1,marti2}. 
When we arrive at the high temperature $kT_H = 10^{-13}$, we fully thermalize the system by keeping it at this temperature value for another 5 million steps. 
We then proceed to cool it back down to $kT_L$, at the same rate as we heated it up,
during  5 million steps. Each complete heating-cooling cycle takes about 3 minutes of wall-clock time when we use  a single 
processor in an ordinary personal computer (MacPro). Consequently we are able to  
collect very good statistics. In particular, we have confirmed that our results and conclusions are not sensitive to
the rate of heating and cooling. 

During the simulations, we monitor the state of the myoglobin backbone 
by following the evolution of both the radius of gyration $R_g$, and the RMSD
distance $R_{rmsd}$ between the simulated configuration and the folded 1ABS structure. 

In Figure 10 
we show the evolution of the radius of gyration, and in Figure 11  
we show the evolution 
of the RMSD distance to 1ABS,   as a function of steps during 100 repeated 
heating and cooling cycles; Notice that in these Figures we have converted the temperature into 
Kelvin scale, using the diagram in Figure 9. 

We observe the
following:
At low temperatures, with temperature factors
\[
kT < 10^{-15}
\]
the radius of gyration is essentially constant $R_g \approx 14.6 $ and subject to 
very small thermal fluctuations. Between
\[
10^{-15} < kT < 10^{-14}
\]
there is a regime when the radius of gyration increases at an accelerating rate in the number of steps. 
The increase in $R_g$ continues until we reach a temperature near $kT_H$. For temperatures where
the temperature factor is
in the range
\[
10^{-14} < kT < kT_H = 10^{-13}
\]
the rate of increase decelerates, and when we reach the vicinity of the temperature 
$kT_H$ we observe no further increase
in $R_g$.  This proposes that the system has entered the random walk ($\Theta$-point) phase. 
When we decrease the temperature, the evolution of $R_g$ becomes 
inverted. At the end of the cycle, when the temperature reaches $kT_L$, 
the configuration returns back to the folded state 
in terms of the radius of gyration.  This establishes
that our energy function, and in particular the multi-soliton configuration we have constructed, is consistent with 
Anfinsen's thermodynamic principle \cite{anf},  according to which a protein should return to its original shape
if heated and then cooled. We also conclude that the native state which is described by the multi-soliton solution
is  the unique and stable minimum energy state of of the Helmholtz free energy, to the 
extent that our heating and cooling simulations probe the surrounding energy landscape. 

We have confirmed Anfinsen's thermodynamic principle and verified that the transition near $kT_H$ is indeed 
the $\Theta$-point transition between collapsed phase and random walk
phase, by heating the configuration to substantially higher temperature factor values. We have found that
the principle remains valid for temperatures all the way to 
\[
kT = 10^{-8} 
\]
In Figure 12 we show how the RMS distance between the heated configuration and 
1ABS changes as a function of temperature, during heating and cooling cycles between 
$kT_L = 10^{-16}$ and $kT = 10^{-8}$. We observe two clear transitions that are consistent with the
transitions between collapsed and random walk phases, and between random walk and self-avoiding random
walk phases according to (\ref{nus}). When heated to temperatures above  $kT = 10^{-8}$, we find that the system does not
always return to the native state of 1ABS, indicating that there is a limit to Anfinsen's principle above which the cooled configuration
can become misfolded.
%
%
%
%
%
%
%

In Figure 13 we show  
%
%
%
%
%
the {\it average} values of  $R_g$, evaluated at several different temperature values 
over 100 runs.  Both  during the heating period when $0 < x < 7.5 $, and during
the cooling period when $ 7.5 < x < 15 $ ($x$ is number of MC steps in millions), 
we can describe the data with a very good precision by
\begin{equation}
R_g (x) \ \approx  \  a \cdot \tanh\{ b (x -c)  \} + d
\label{Rg1}
\end{equation}
The corresponding parameter values are listed in Table V.
In Figure 13 we also display the derivative of (\ref{Rg1}).
Putatively, we can try and use the maximum of the derivative to identify the $\Theta$-point 
transition temperature in our model. For this, we tacitly assume that the transition 
temperature at $T_c \approx 348 \ K$ 
reported in  \cite{theta1}, \cite{theta2} corresponds 
to the $\Theta$-point.  By identifying this with the 
maximum of the derivative of $R_g$, we 
conclude that during the heating cycle the $\Theta$-point temperature relates to our dimensionless temperature values
as follows,
\begin{equation}
T_g^{h} \approx 1.63 \cdot 10^{-14}  \approx 348 \ K
\label{renT}
\end{equation}
We utilize this value, to determine on of the two parameters in (\ref{Ttot}).
During the cooling cycle, we find 
\[
T_g^{c} \approx 1.71 \cdot 10^{-14}  \approx 349 \ K
\]
We observe that there appears to be very slight asymmetry present, during the heating and cooling cycle.
This kind of asymmetry has also been reported  experimentally \cite{theta1}.

The RMSD distance between the simulated configuration and the 1ABS backbone depends on temperature 
in a very similar manner. In Figure 14  we show 
the comparison between 
simulation, and the corresponding approximation (\ref{Rg1}), 
\begin{equation}
R_{rmsd} \ \approx \ a \cdot \tanh\{ b (x-c) \} + d
\label{rmsd1}
\end{equation}
The parameter values are listed in Table V
for the heating period $0<x< 7.5$  and for the cooling period $7.5<x<15$.
The Figure 14 also 
shows the derivative of $R_{rmsd}(x)$. In parallel with the radius of gyration, we use the maximum of the derivative
to estimate the peak rate of change in the transition temperature. 
During the heating period the increase in $R_{rmsd}$ peaks  at
\[
T_{rmsd}^{h} \approx 1.35 \cdot 10^{-14} \ \approx \  344 \ K
\]
During the cooling period the peak is located at a slightly higher temperature value,
\[
T_{rmsd}^{h} \approx 1.45 \ 10^{-14} \ \approx \  346 \ K
\]

%
%
%
%
%
%
%
%
%
%
%
%
%
%
%
%

%
%
%
%
%
%
%
%
%
%
%
%
%
%
%
%

\subsection{ Backbone ligand gate dynamics and temperature scale determination}
\label{sect:scale}

In Figures 10-14 of the previous sub-section we have displayed the temperature scale in terms of Kelvin scale,
instead of the dimensionless scale that we use in our simulations. 
The conversion between the dimensionless temperature and the Kelvin scale temperature, using the diagram 
shown in Figure 9, is 
made by using the approximative solution (\ref{kTt}) of the renormalization group flow equation (\ref{betaT}).
One of the two parameters can be fixed using the experimental estimate (\ref{renT}) of the $\Theta$-point.
The other parameter is determined by considering {\it backbone ligand gate dynamics}, which we shall now
proceed to investigate.


By visually investigating the shape of the backbone during the heating and cooling, we find that, qualitatively,
the thermal fluctuations follow a very similar pattern. The backbone  becomes deformed and unfolds in more 
or less the same manner, as the temperature increases. This repeats itself during the cooling.
The onset of the unfolding transition
can be described in terms of the backbone ligand gates. We have identified three particularly 
interesting gates that we call Gate 1,2 and 3, and we define them as follows:

The  Gate 1 
is defined as the area between the following two backbone segments: The first segment starts 
from PDB site 37 (Pro) and ends at PDB site 44 (Asp). The second, opposite, segment  
starts at 96 (Lys) and ends at 103 (Tyr). The opening of this gate takes place with the distance between
the two segments increasing, and the open gate exposes the heme to the solvent.
Figure 15 shows the location of this gate on the 1ABS backbone.

The  Gate 2 is located between the helical structures E and F, as shown in Figure 16.
In order to compare this gate, that extends over the entire length of both helices E and F, with Gate 1 
that is composed of segments with only eight residues, we have selected two segments in Gate 2.  
Each of them consists similarly of eight amino acids. The first segment, located in the 
helical structure E, starts with site 61 (Leu) and ends with site 68 (Val). The second segment, located 
in the helical structure F opposite to the first segment, starts with site 89 (Leu)  and ends with site 96 (Lys).
We have intentionally selected these two segments to be far from the loop that connects the helices
E and F.  This is because in our simulations, we have observed that the amplitudes 
of the thermal fluctuations in the segment distances tend to increase, 
the further away the segment is located from the connecting loop: The opening and closing of the gate resembles the 
opening and closing of scissors,
with blades formed by helices E and F,  connected by  the loop between these two helices. 
Note that the first segment along helix E, includes both the distal histidine at site 64 and the valine at the end site 68. This valine
is also inside the heme pocket, and it is presumed to have an important role in CO {\it vs.} O$_2$ discrimination. 
Similarly, the opposite segment in the helical structure F includes the proximal histidine at site 93. 

We remind that in our simulations, we do not take into account the side-chains. 
The side-chains are for sure very important. In particular, one can expect that the interactions 
between the proximal and distal histidines and the heme
help in stabilizing the structure formed by helices E and F.  However, as we have argued in sub-section III C,
everything in our analysis points towards a strong master-slave coupling 
between the backbone and the side-chain conformations. Consequently, we have all the reasons 
to expect, that it is sufficient to consider the backbone ligand gates only. There is no experimental data suggesting otherwise.

Finally, the Gate 3  is shown in Figure 17. 
It is  located between the helical structures B and G. Again, in order to compare this 
relatively long gate with Gate 1 which is relatively short,
we select two segments that each consist of eight amino acids.
The segment in the helical structure B starts at  site 25 (Gly) and 
ends at site 32 (Leu). The segment in helix G starts at
site 106 (Phe)  and ends at site 113 (His). 

During the heating and cooling cycle of the myoglobin, we monitor the sizes of these three gates. For this
we  compute the distance $d_i \ (i=1,2,3)$ between the respective segments, as a functions of temperature. 
The distance $d_i$ between the segments,  for each of the three gates, is defined as follows:
\begin{equation}
d_i = \sqrt{ \  \sum\limits_{n=1}^8 \ (\mathbf x_n - \mathbf y_n )^2 \ \ }
\label{di}
\end{equation}
Here $\mathbf x_n$ are the eight coordinates in the first segment, and $\mathbf y_n$ are the corresponding 
coordinates in the second segment, in the Gate $i=1,2,3$. Note that the two segments in each 
of the three gates, are spatially oriented in an anti-parallel manner with respect to PDB indexing.
Consequently,  in computing (\ref{di}), we invert the indexing in one of the two segments  
with respect to the PDB indexing. 

We start by investigating, what does the existing data in PDB reveal on the temperature dependence of 
the three gates.  For this we compute the following three gate ratios
\begin{equation}
\frac{\rm Gate 1}{\rm Gate 2}  = \frac{d_1}{d_2} \ \ \ \ \&  \ \ \ \  \ \frac{\rm Gate 3}{\rm Gate 2}  = \frac{d_3}{d_2}
\ \ \ \ \&  \ \ \ \  \ \frac{\rm Gate 3}{\rm Gate 1}  = \frac{d_3}{d_1}
\label{gr}
\end{equation}
for all present PDB myoglobins, that have been measured with resolution 2.0 \AA~ or better.  We display the results is Figures 18-20,
separately for each of the three gates.  
In each Figure 18-20 we observe that there are substantial fluctuations in the gate ratios,  in the PDB data that has been 
taken at around $100$K. This reflects the fact that the majority of PDB data has been collected at this temperature value,
overall the gate ratios show no temperature dependence for $T< 300K$. 

We have computed the temperature dependence of the gate ratios using the energy function (\ref{E1}), with the 1ABS parameters in 
Table IV. The results are shown in Figure 21-23.
We have found that the first gate to open as the temperature increases,
and the last one to close as temperature decreases, is the Gate 3. The Gate 2 is the one to open last, and the one to close first. 
In the low temperature limit the Gate 3 is about half the size of the Gate 2. But its size exceeds that of the Gate 2 in the
segment separation distance (\ref{di}) at temperature
\[
kT^{c}_{23} \approx 10^{-14} \ \sim  \ 340 \, K
\]
The transition is very rapid. This is in line with the general results of reference \cite{cherno2}:  
When the temperature reaches the $\Theta$-point value $\sim 348 \, K$,  the Gate 3 is about twice as large as the Gate 2. 

The Gate 1 also opens much faster than Gate 2, but slower than Gate 3. It also closes slower than Gate 2, but faster than 
Gate 3. In the low temperature limit the Gate 1 is about half as wide as Gate 2, but becomes wider than Gate 2
when the temperature reaches similarly a value
\[
kT^{c}_{12} \approx 10^{-14} 
\]
However, the Gate 1 does not become quite as wide as Gate 3. This is shown in Figure 23.

We are now in a position to determine the second parameter in (\ref{kTt})  to arrive at  (\ref{Ttot}), that we have displayed 
in Figure 9; we remind that one of the parameters is determined in (\ref{renT}).
For this we proceed as follows:
When we compare Figures 18-20 we conclude that experimentally, the gate ratios do not display any observable
temperature dependence whenever $t<300 \, K$. Consequently, the lowest possible value of the temperature factor $kT$ where
Figures 21-23 can display any change in the gate ratios, should correspond to a temperature which is above $300\, K$.  
When we read off the lowest possible $kT$ value where we have an observable effect in Figures 21-23, we conclude that, necessarily,
\[
k T \approx 10^{-15} >   k_B \, 300 \, K 
\]
This gives a lower bound. When we adopt this lower bound value as our estimate for the gate opening temperature we obtain
the second parameter value in (\ref{Ttot}) and arrive at Figure 9. In reality the actual gate opening temperature can be higher, but at the moment
there is no experimental basis for choosing a higher value; the single existing NMR data\cite{1myf} of 1MYF, even though it has been
taken at the slightly higher temperature value of $308 \, K$ does not suffice for us to  improve our estimate.  Thus we can only estimate a lower bound.

We note that qualitatively, a higher gate opening temperature has no effect to our conclusions, and quantitatively 
the differences are also  minor; the only effect is a sharpening of the $\Theta$-transition
onset.

We proceed to inspect the effects of our results in Figures 21-23 to ligand migration: Our 
results show, that to the extent backbone thermal fluctuations have a r\^ole in ligand
migration, the Gate 3 between the helical structures B and G can be very important.  This gate opens very much 
like a baseball glove, as we increase temperature.   The Gate 1 might also play a r\^ole, but probably a lesser one 
than Gate 3. On the other hand, the V-shaped Gate 2 between helices E and F seems to be quite sturdy, 
it does not seem to open as much as the other two gates. The presence of the distal and proximal histidines in
Gate 2 and their attractive interactions with heme, might have an additional stabilizing effect that is not accounted by our model. 
Consequently we do not see how the thermal backbone fluctuations that take place in the Gate 2, could 
play a major r\^ole in ligand migration. At least, to the extent that  backbone fluctuations are relevant.  
 
Finally, in Figure 24  we show representative thermal fluctuations in the Gate 3, 
during six consecutive time-steps as we heat the system to the vicinity of the representative physiologically
relevant value $kT \sim 37 ^oC$. 
The first frame (1) is the crystallographic structure 
The subsequent frames show the gate structure, as time evolves. It appears that there are oscillations between a closed gate
position which corresponds to the crystallographic state, and a thermally excited, much more open gate: The gate keeps on
opening and closing under thermal fluctuations.  In the open position, the gate exposes the distal 
cavity where the CO molecule is located, to the solvent. The relevance of this oscillatory behavior, and the temperature
dependence  of its amplitude, to ligand migration remains to be clarified.

\section{Summary}
\label{sect:summary}
In summary, we have develop a general approach to describe protein folding and unfolding dynamics in terms of
an effective energy function. The energy function models a given protein backbone in terms of a multisoliton,
that locally minimizes the energy. This facilitates a wide range of energetic studies, for example investigations how the
protein responds to changes in the temperature. As a concrete example, we have constructed the multisoliton 
that describes myoglobin, using the Protein
Data Bank  structure 1ABS as our crystallographic model. The multisoliton approximates the PDB configuration
with a sub-\AA ngstr\"om accuracy. We have applied the energy function to study in detail the response of the myoglobin
structure to heating and cooling cycles, from low temperature values where the backbone is in the collapsed
phase to high temperature values where the backbone is in the random walk phase, and even in the self-avoiding random walk phase.
By repeated heating and cooling simulations, we have found that our description of myoglobin is fully consistent
with Anfinsen's thermodynamic principle. Furthermore, we have applied the model 
to investigate the backbone ligand gate dynamics, how thermal
fluctuations can expose the heme to the exterior and allow ligands to move in and out. We have identified three different
backbone ligand gates, that we have scrutinized in detail. He have found a gate, that appears to be the first to open
and last to close, when the myoglobin is in an environment where temperature fluctuates. This gate is located between
helices B and G. Its opening appears to expose the heme, providing a direct passage for the ligand to enter and exist. 
The mechanism and the pathway that we have identified, appears to be novel in the present context.

\vskip 0.5cm

\section*{Acknowledgements}

\vskip 0.3cm
We thank M. Lundgren for help with simulations and for discussions, L.  Kamerlin for discussions, and G. Petsko for communications and
proposing us to use 1ABS as a model of myoglobin in our investigations. We thank support 
from CNRS PEPS grant, Region Centre Rech\-erche d$^{\prime}$Initiative Academique grant,
Sino-French Cai Yuanpei Exchange Program, and Qian Ren Grant at BIT.

\vskip 2.0cm

\newpage

\section*{Figure legends.}

\newcounter{fig}

\setcitestyle{plain}

\begin{list}
{\bf Fig. \arabic{fig}.}{\usecounter{fig}}

    \item  (color online) Top: The $\kappa_i$  (black) and $\tau_i$ (red) profiles of 1ABS using 
    the standard differential geometric convention that bond angles
    are positive. Bottom:  The soliton structure becomes visible in the $\kappa_i$ profile once we implement the transformations 
    (\ref{dsgau}).
        \la{fig1}

    \item (color online)  Comparison of the fluctuation distance  (\ref{x}) (red line) to the soliton  (black line) along the 1ABS backbone.
    The distances are measured from the PDB data of the C$_\alpha$ atoms. The shaded region around the soliton 
    describes the 0.15 \.A zero point fluctuation regime.
    \la{fig2}

    \item (color online) 
    Comparison of the  one standard deviation fluctuation distance from the average C$_\alpha$ coordinate,
    along the 1MYF backbone. Following Figure 2, for ease of comparison, the backbone has been divided equally
    into ten segments. 
    \la{fig3}

    \item (Color online) Angular distribution of the C$_\beta$ directions (red dots) along the 1ABS backbone in the Frenet frames, as seen
    from the C$_\alpha$ carbons that are located at the origin {\it i.e.} at the center of the sphere. The (grey) background is constructed
    from all PDB proteins that have been measured with resolution 2.0 \AA~ or better. We have also identified the regions of $\alpha$-helices,
    $\beta$-strands, loops and left-handed $\alpha$-helices.
    \la{fig4}
    
    \item (Color online) Angular distribution of the C$_\beta$ directions (red dots) along the 1MBC backbone in the Frenet frames, as seen
    from the C$_\alpha$ carbons that are located at the origin {\it i.e.} at the center of the sphere. As in Figure 4, the (grey) background is constructed
    from all PDB proteins that have been measured with resolution 2.0 \AA~ or better. We have identified the sites that display observable, 
    apparently thermal effects in their orientations, by their PDB site number.
    \la{fig5}
 
    \item (Color online) Angular distribution of the C$_\gamma$ directions (red dots) along the 1ABS backbone in the Frenet frames, as seen
    from the C$_\alpha$ carbons that are located at the origin {\it i.e.} at the center of the sphere. As in Figure 4, the (grey) background is constructed
    from all PDB proteins that have been measured with resolution 2.0 \AA~ or better. We have identified the $\alpha$-helical and $\beta$-stranded
    regions, they are connectd by loop regions. Also identified are the {\it gauche $\pm$} (g$\pm$) and {\it trans} (t) regions.     \la{fig6}

    \item (Color online) Angular distribution of the C$_\gamma$ directions (red dots) along the 1MBC backbone in the Frenet frames, as seen
    from the C$_\alpha$ carbons that are located at the origin {\it i.e.} at the center of the sphere. As in Figure 4, the (grey) background is constructed
    from all PDB proteins that have been measured with resolution 2.0 \AA~ or better.   The sites identified as deviations in Figure 5 are now within the
    expected regions. There are slight deviations at sites 49 and 141 that have been marked.
    \la{fig7}
    
    \item (Color online) Comparison of the RMSD distance between the 1ABS configuration and the 
    multi-soliton solution,  with the Debye-Waller B-factor fluctuation distance around the 1ABS backbone. The blue marking
    at top, along 2.0 \AA~ line, denotes sites where {\it Molprobity} detects imperfections.
    \la{fig8}

    \item (Color online) Conversion diagram between Kelvin scale physical temperature $t$ and the dimensionless 
    temperature factor $kT$ used in our simulations. The details  are explained in subsection III E.    
    \la{fig9}

    \item (Color online) The evolution of the radius of gyration during 100 repeated heating and cooling cycles. 
    The (blue) line is the average, and the surrounding  (orange) 
    shaded area  describes the one standard deviation extent of fluctuations. Along the top
    axis, we have converted the temperature into Kelvin scale, using the conversion diagram
    between dimensionless and Kelvin scales in Figure 9.
    \la{fig10}

 \item (Color online) The evolution of the RMSD distance between the 1ABS backbone and the simulated configuration during
    100 repeated heating and cooling cycles. The (blue) line denotes the average,
    and the shaded (orange) area around it is the extent of one standard deviation fluctuations. Along the top
    axis, we have converted the temperature into Kelvin scale, using the conversion diagram
    between dimensionless and Kelvin scales in Figure 9.    
    \la{fig11}

 \item (Color online) The evolution of the RMSD distance between the 1ABS backbone and the simulated configuration during
    100 repeated heating and cooling cycles, to very high temperature values. The (blue) line denotes the average,
    and the shaded area around it is the extent of one standard deviation fluctuations. Along the top
    axis, we have converted the temperature into Kelvin scale, using the conversion diagram
    between dimensionless and Kelvin scales  in Figure 9. This conversion is for indicative purposes only.
    The validity of the conversion relation in Figure 9 has been derived using properties of myoglobin at temperatures
    not exceeding the  $\Theta$-point value. Consequently, when extended above the $\Theta$-point transition,  the conversion
    relation is  not very reliable.      
    \la{fig12}

 \item (Color online) The (red line) fitting of (\ref{Rg1}) to the average values of (blue dots) 
    $R_g$, over the heating and cooling periods. The grey area around the (red) fitting line is one standard
    deviation estimate. Note: The difference between these three is so small that it is barely observable 
    in the Figure. Also shown is the
    derivative of (\ref{Rg1}) (light blue line). Along the top
    axis, we have converted the temperature into Kelvin scale, using the procedure that is described in subsection III E; the conversion
    between dimensionless and Kelvin scales are as in Figure 9.
     \la{fig13}

 \item (Color online) The fitting of (\ref{Rg1}) to the average values $R_{rmsd}$, over the heating and cooling periods. 
    Also shown is the derivative of $R_{rmsd}(x)$.   Along the top
    axis, we have converted the temperature into Kelvin scale, using the procedure that is described in subsection III E; the conversion
    between dimensionless and Kelvin scale temperature values are shown in Figure 9.
    \la{fig14}

 \item (Color online) Stereographic cross-eyed view of the Gate 1, defined as 
    the area between a segment that starts 
from PDB site 37 (Pro) and ends at PDB site 44 (Asp), and a segment that  
starts at 96 (Lys) and ends at 103 (Tyr). We also show the location of the heme (orange), the proximal
histidine (93), the valine (68), the distal histidine (64) (all green) and the CO (black ellipsoid)   
  \la{fig15}

 \item (Color online)  Stereographic cross-eyed view of the Gate 2, defined as 
    the area between a (red) segment that starts 
from PDB site 61 (Leu) and ends at PDB site 68 (Val), and a (blue) segment that  
starts at 89 (Leu) and ends at 96 (Lys). We also show the location of the heme, the proximal
histidine (93), the valine (68) and the distal histidine (64).  Also shown are the heme (orange), the Gate 1 (green and labelled), 
and the CO (black ellipsoid)  
    \la{fig16}

 \item (Color online) Stereographic cross-eyed view of the Gate 3, defined as 
    the area between a (green) segment that starts 
from PDB site 25 (Gly) and ends at PDB site 32 (Leu), and a (green) segment that  
starts at 106 (Phe) and ends at 113 (His). Also shown are  the location of the heme, the proximal
histidine (93) (blue), the valine (68) and the distal histidine (64) (both red).  We also show the  heme (orange)
and the CO (black ellipsoid).  
    \la{fig17}

 \item (Color online) Gate ratio (\ref{gr}) between the gate 1 in Figure 15 and gate 2 in Figure 16. The dotted (red) lines are
    intended to guide eye only. 
    \la{fig18}

 \item (Color online) Gate ratio (\ref{gr}) between the gate 3 in Figure 17 and gate 2 in Figure 16. The dotted (red) lines are
    intended to guide eye only.
        \la{fig19}

 \item (Color online) Gate ratio (\ref{gr}) between the gate 3 in Figure 17 and gate 1 in Figure 15. The dotted (red) lines are
    intended to guide eye only.
    \la{fig20}

 \item (Color online) The temperature dependence of the ratio between gates 1 and 2 during our heating and cooling cycle.
     \la{fig21}

 \item (Color online) The temperature dependence of the ratio between gates 3 and 2 during our heating and cooling cycle.
    \la{fig22}

 \item (Color online) The temperature dependence of the ratio between gates 3 and 1 during our heating and cooling cycle. 
    \la{fig23}

 \item (Color online) Thermal fluctuations of Gate 3, near the physiologically relevant value $kT \sim 37 ^oC$. The first frame is the initial crystallographic
    structure,  and the remaining ones are snap-shots with increasing time. There appears to be oscillations
    between a state which is close to the crystallographic Gate 3, and a wider, thermally excited gate position.
    \la{fig24}

\end{list}

\newpage

\begin{figure}
  \begin{center}
    \resizebox{13.cm}{!}{\includegraphics[]{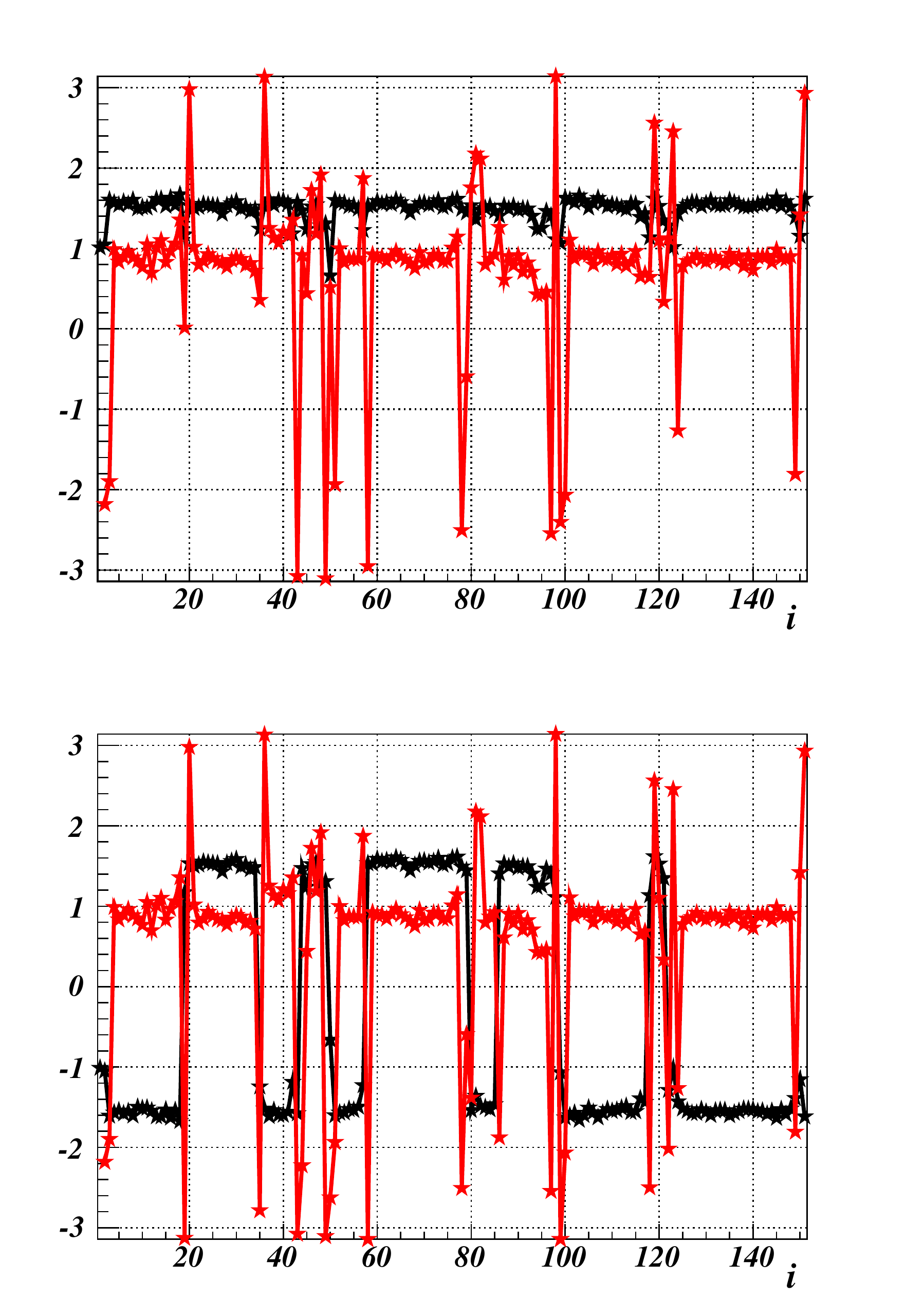}}
  \end{center}
\end{figure}

\vfill

\hfill{\large\sf Figure  \ref{fig1}}

\clearpage


\begin{figure}
  \begin{center}
    \resizebox{15.5cm}{!}{\includegraphics[]{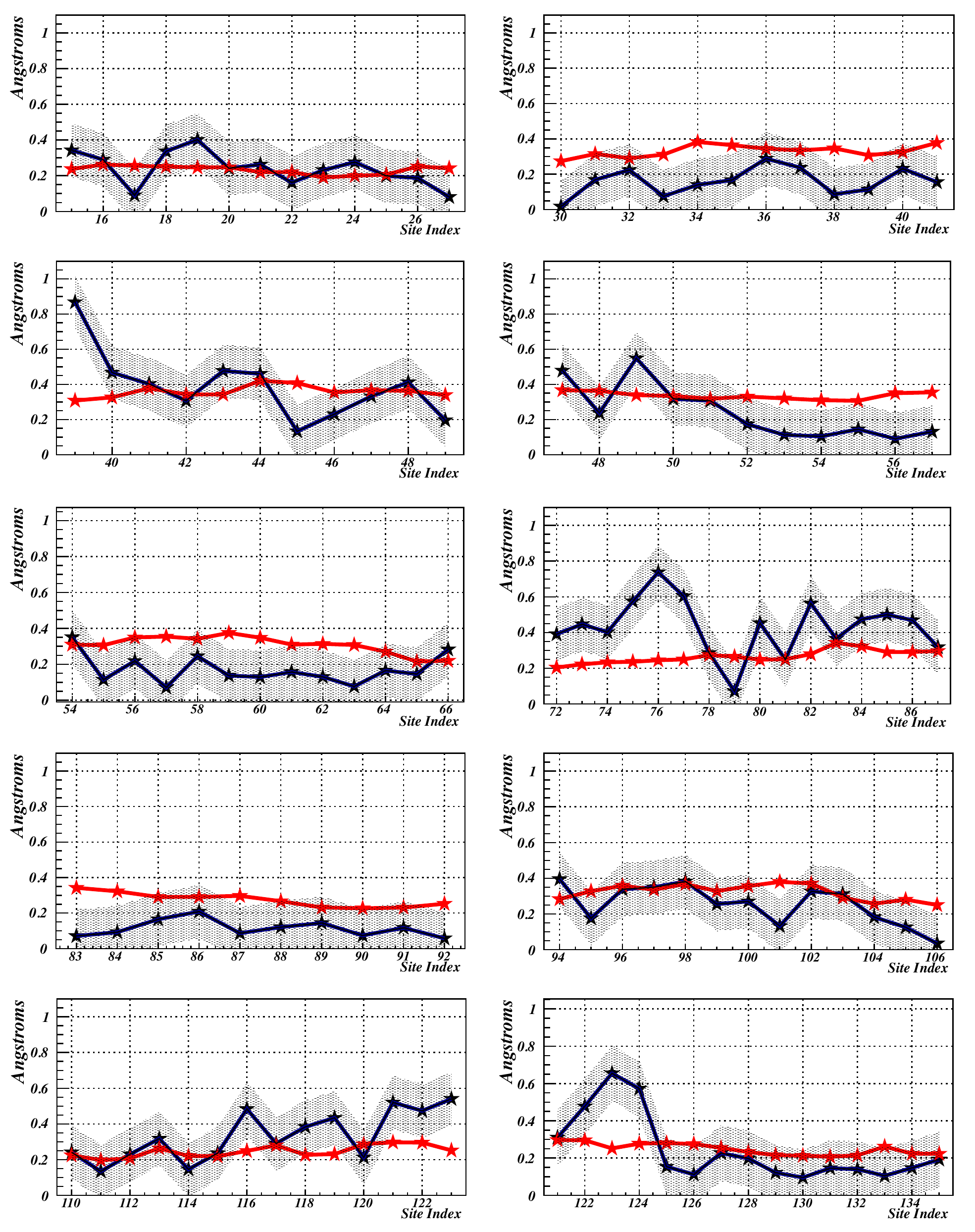}}
  \end{center}
\end{figure}

\vfill

\hfill{\large\sf Figure  \ref{fig2}}

\clearpage


\begin{figure}
  \begin{center}
    \resizebox{14.5cm}{!}{\includegraphics[]{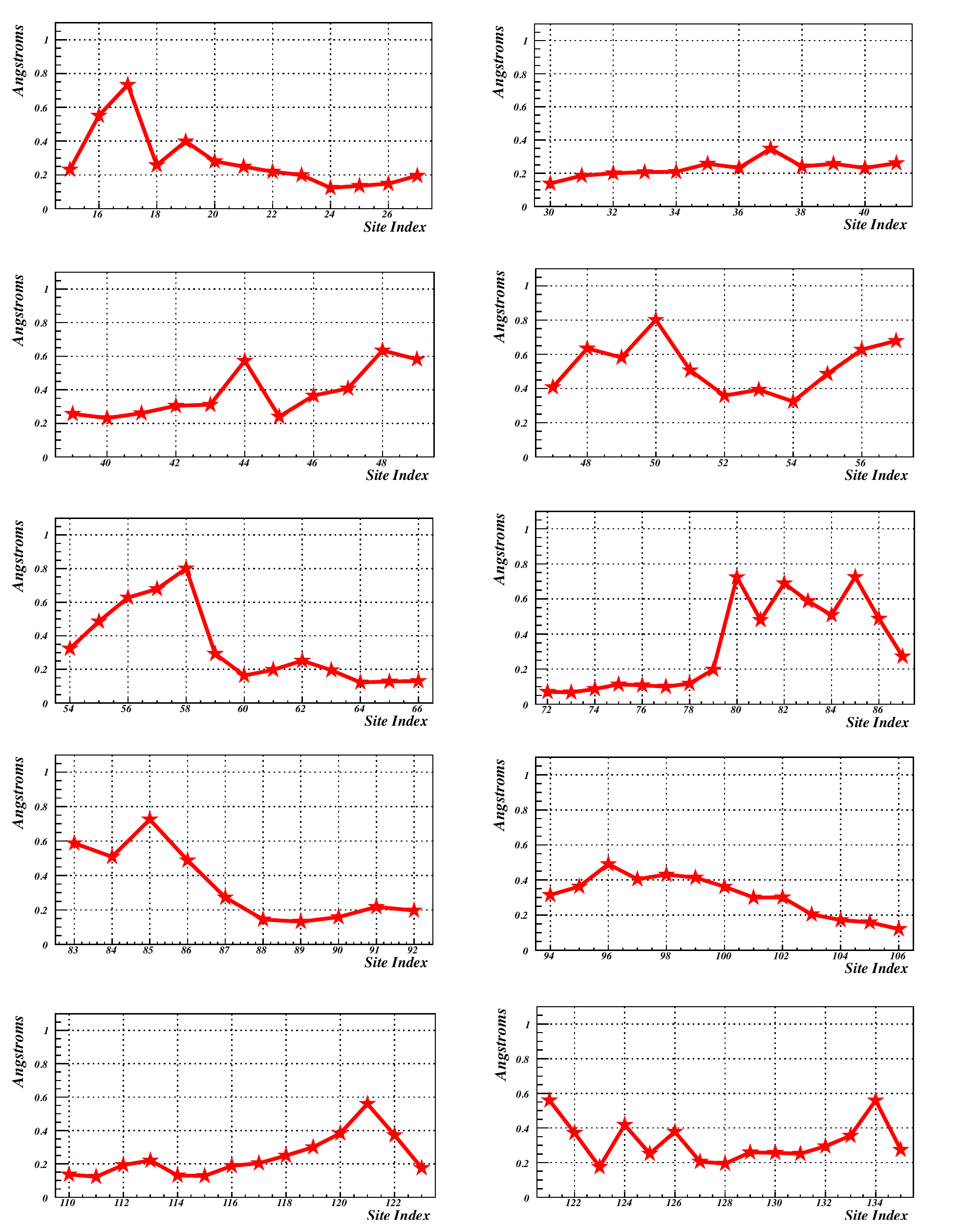}}
  \end{center}
\end{figure}

\vfill\hfill{\large\sf Figure  \ref{fig3}}

\clearpage

\begin{figure}
  \begin{center}
    \resizebox{16.5cm}{!}{\includegraphics[]{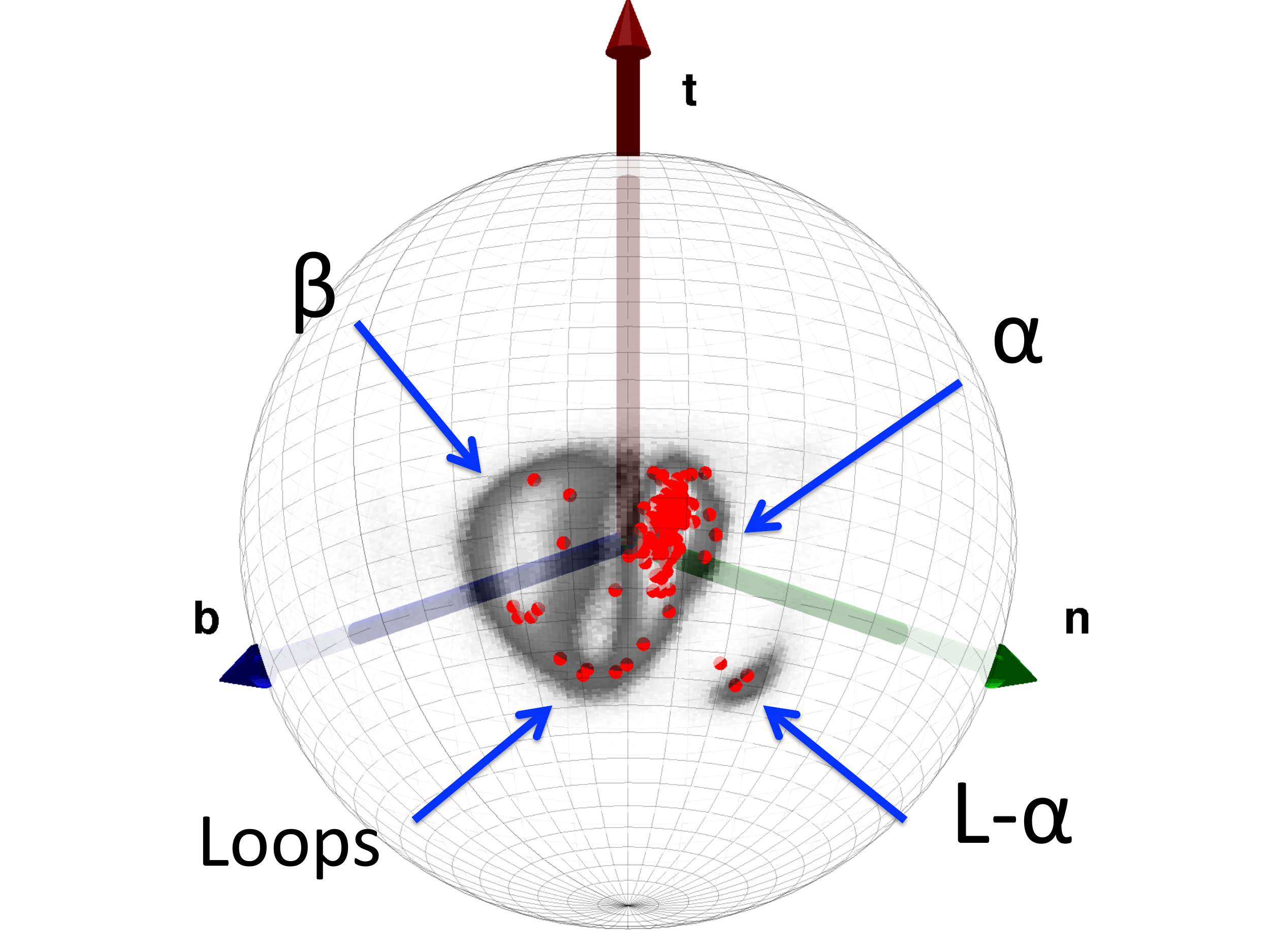}}
  \end{center}
\end{figure}

\vfill

\hfill{\large\sf Figure  \ref{fig4}}

\clearpage

\begin{figure}
  \begin{center}
    \resizebox{16.5cm}{!}{\includegraphics[]{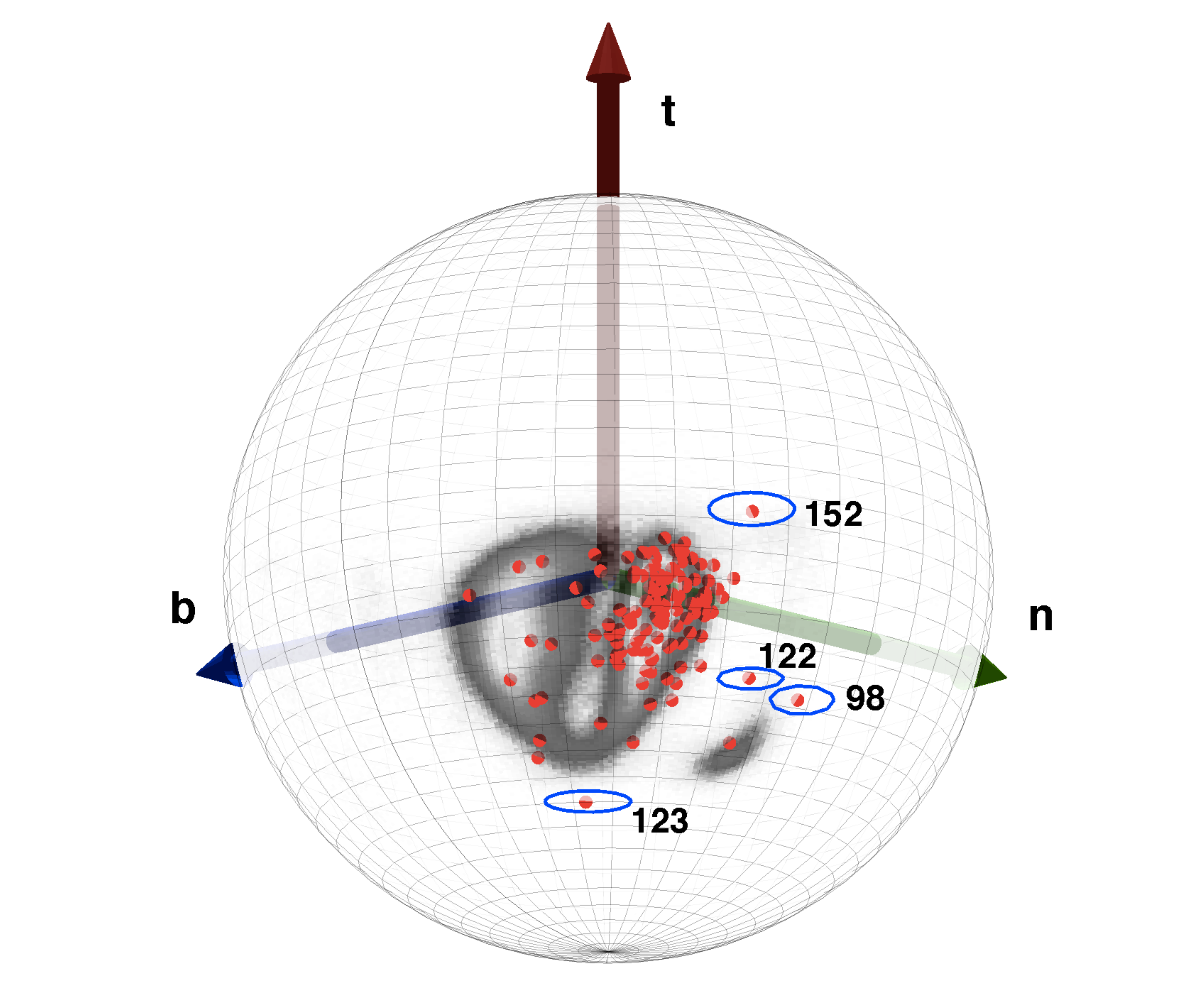}}
  \end{center}
\end{figure}

\vfill

\hfill{\large\sf Figure  \ref{fig5}}

\clearpage

\begin{figure}
  \begin{center}
    \resizebox{16.5cm}{!}{\includegraphics[]{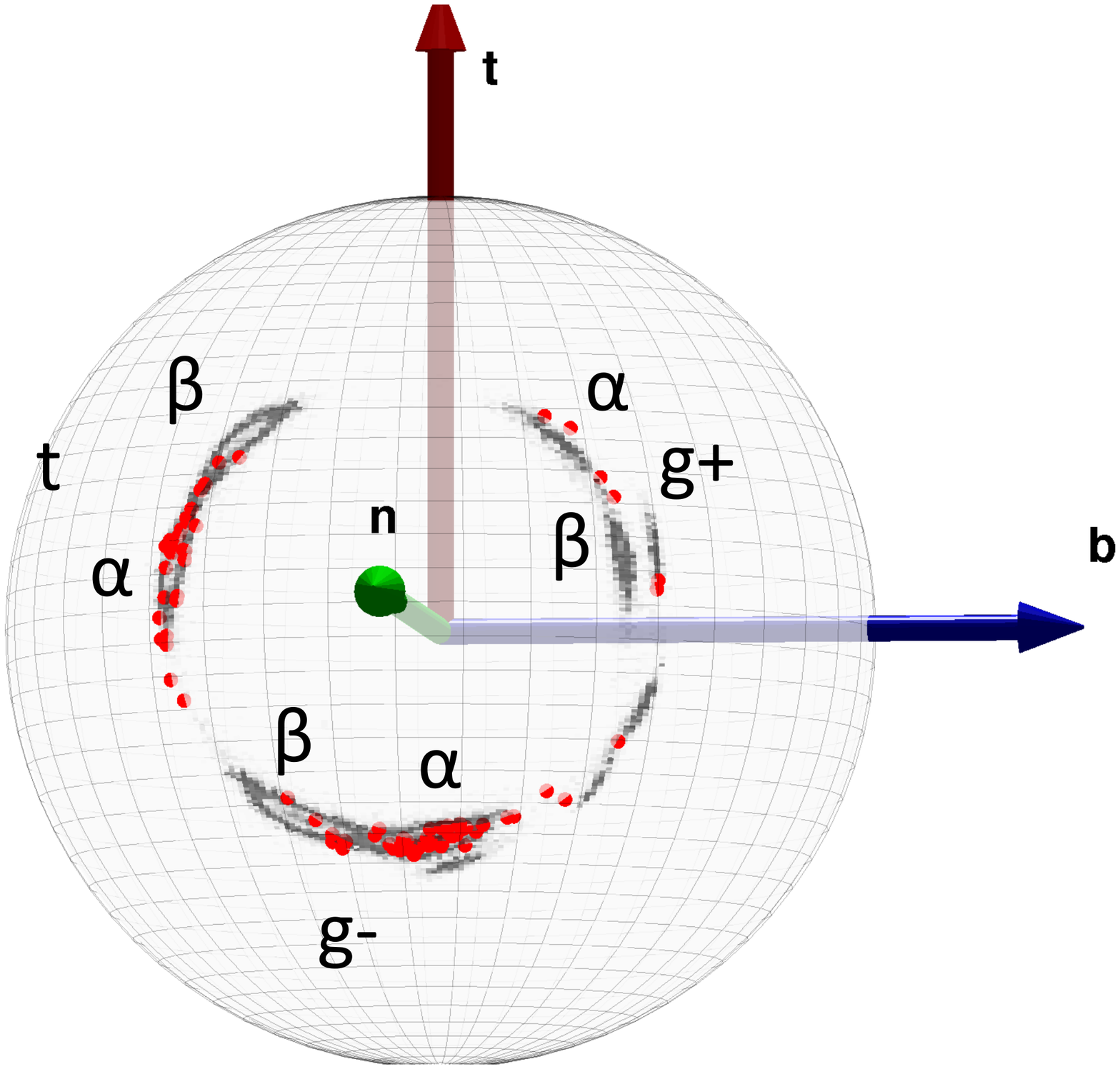}}
  \end{center}
\end{figure}

\vfill

\hfill{\large\sf Figure  \ref{fig6}}

\clearpage

\begin{figure}
  \begin{center}
    \resizebox{16.5cm}{!}{\includegraphics[]{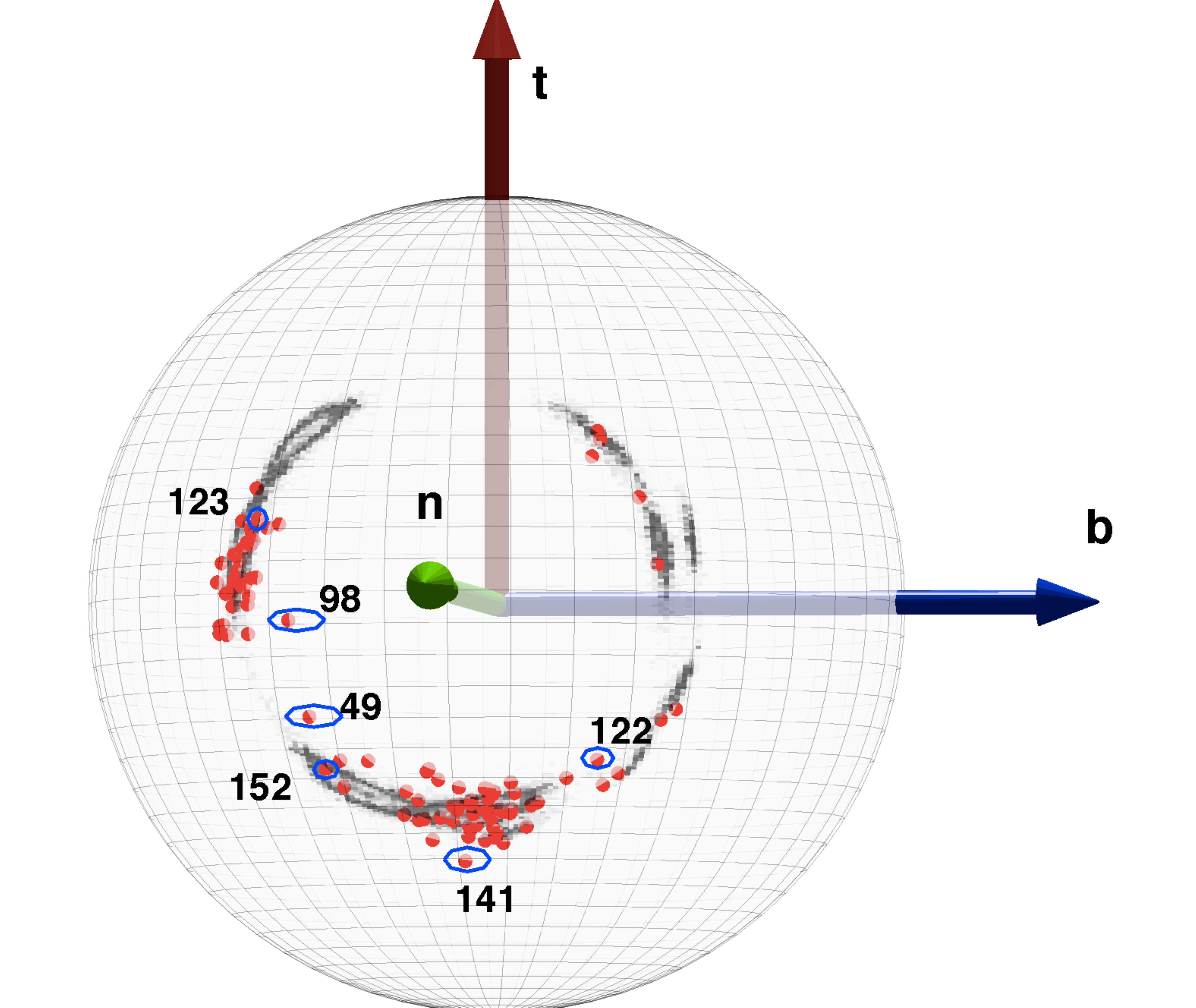}}
  \end{center}
\end{figure}

\vfill

\hfill{\large\sf Figure  \ref{fig7}}

\clearpage

\begin{figure}
  \begin{center}
    \resizebox{16.5cm}{!}{\includegraphics[]{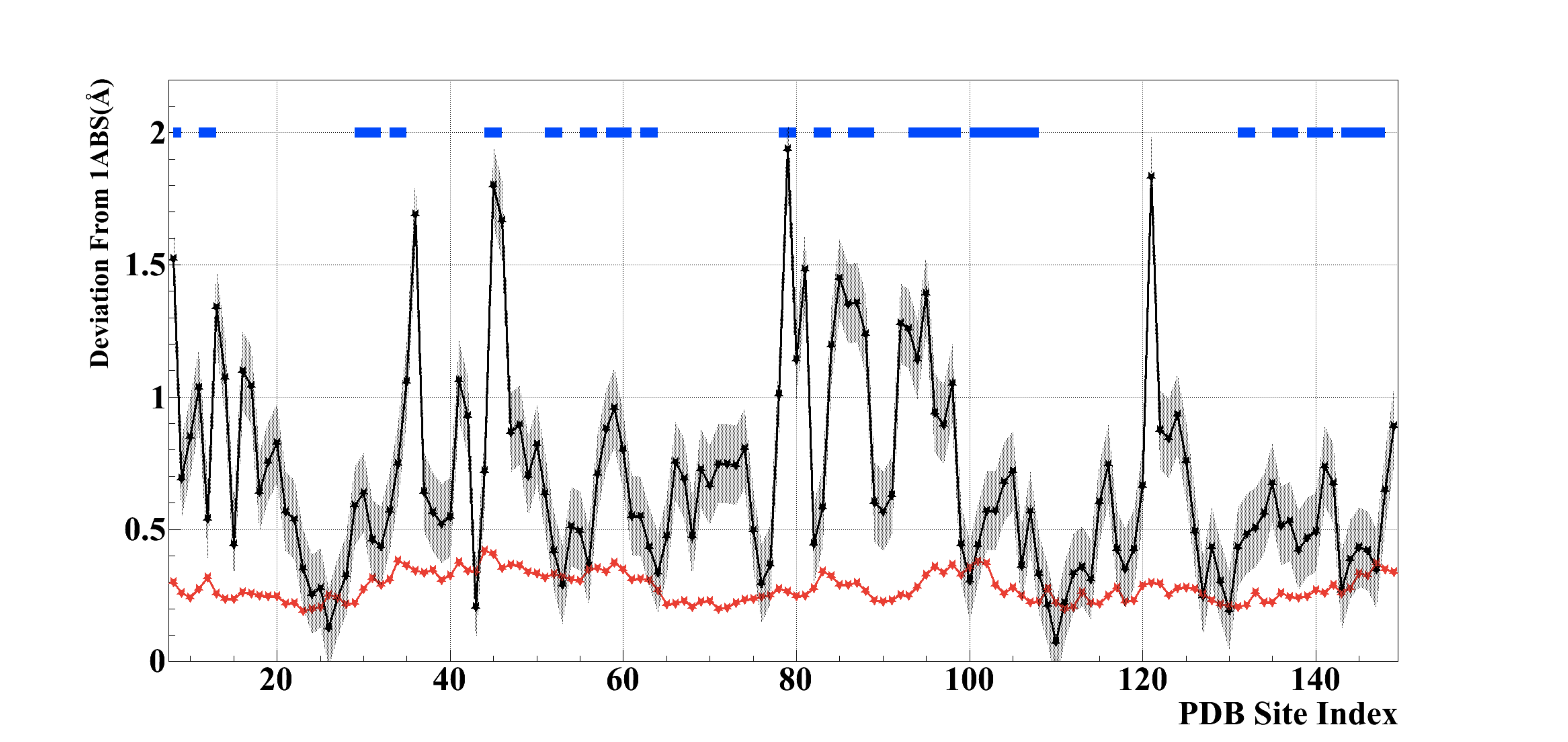}}
  \end{center}
\end{figure}

\vfill

\hfill{\large\sf Figure  \ref{fig8}}

\clearpage

\begin{figure}
  \begin{center}
    \resizebox{16.5cm}{!}{\includegraphics[]{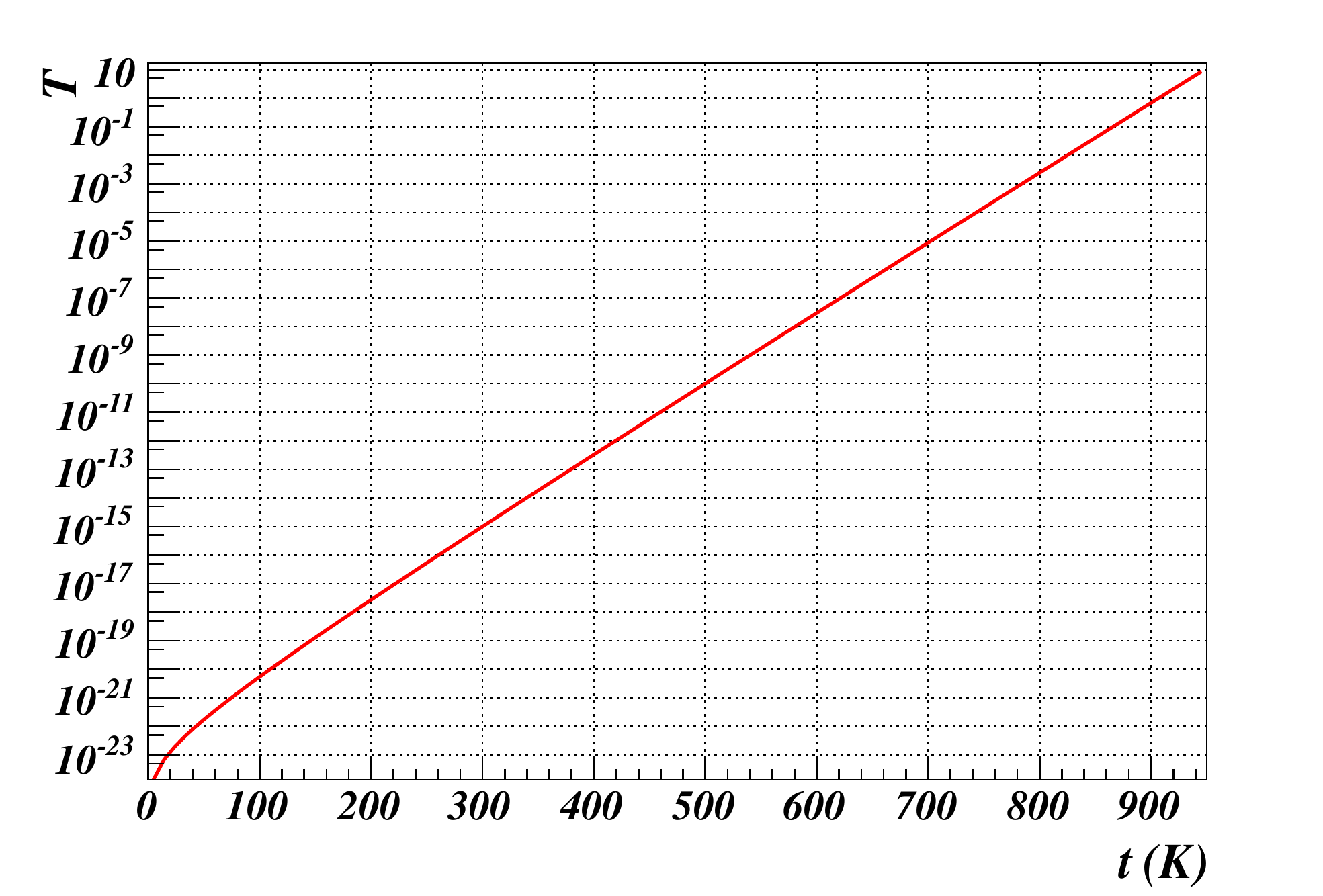}}
  \end{center}
\end{figure}

\vfill

\hfill{\large\sf Figure  \ref{fig9}}

\clearpage

\begin{figure}
  \begin{center}
    \resizebox{16.5cm}{!}{\includegraphics[]{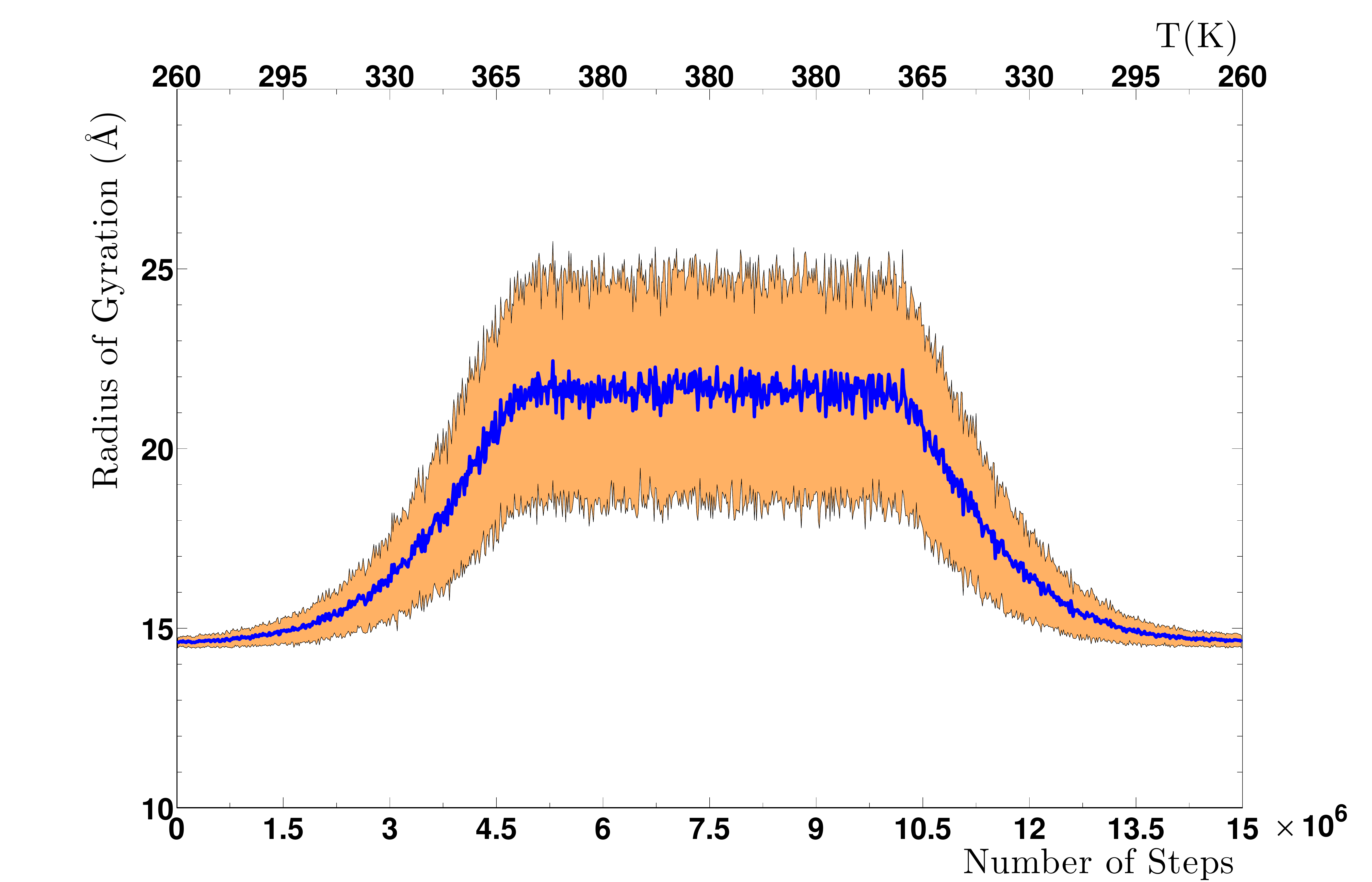}}
  \end{center}
\end{figure}

\vfill

\hfill{\large\sf Figure  \ref{fig10}}

\clearpage

\begin{figure}
  \begin{center}
    \resizebox{16.5cm}{!}{\includegraphics[]{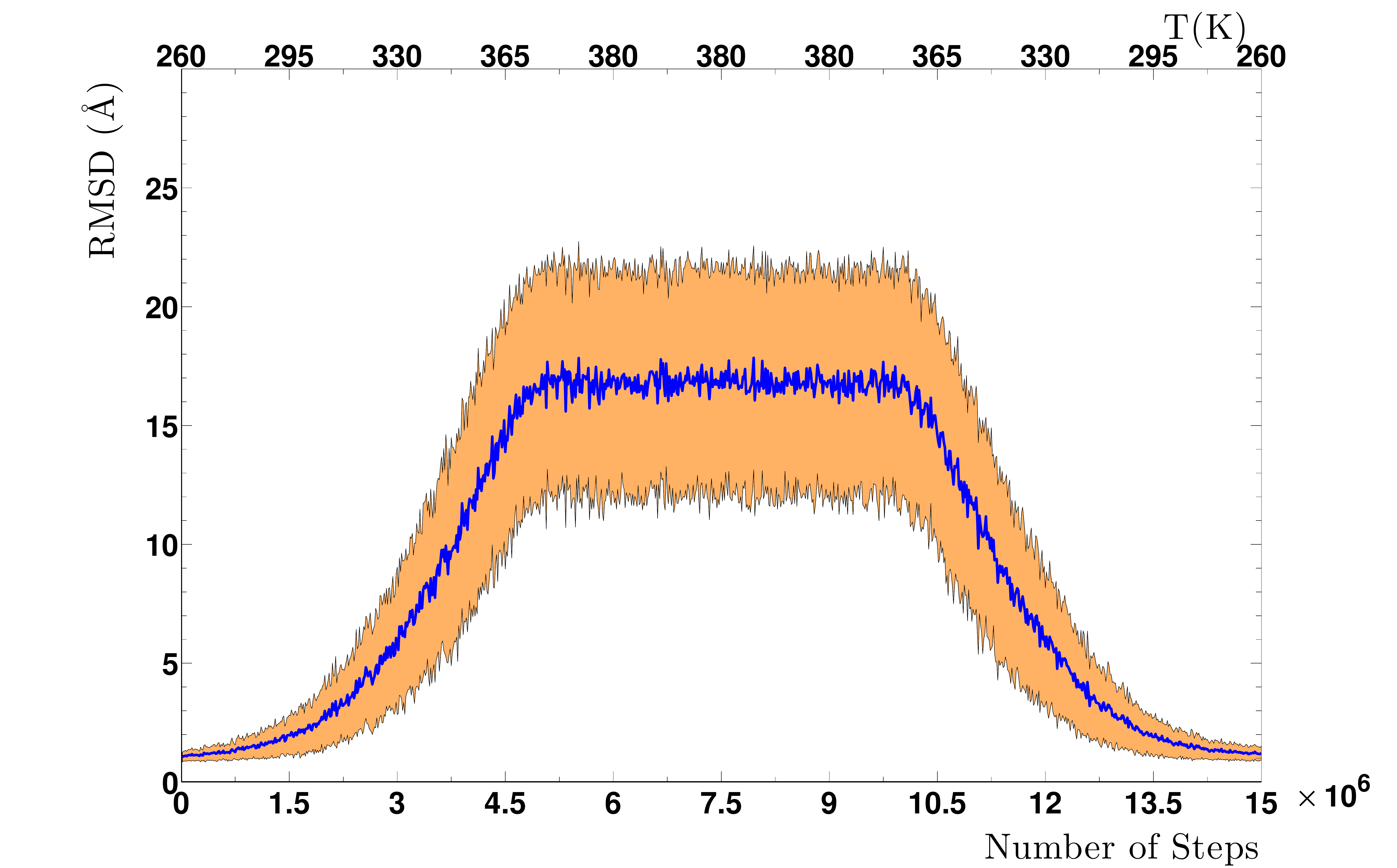}}
  \end{center}
\end{figure}

\vfill

\hfill{\large\sf Figure  \ref{fig11}}

\clearpage

\begin{figure}
  \begin{center}
    \resizebox{16.5cm}{!}{\includegraphics[]{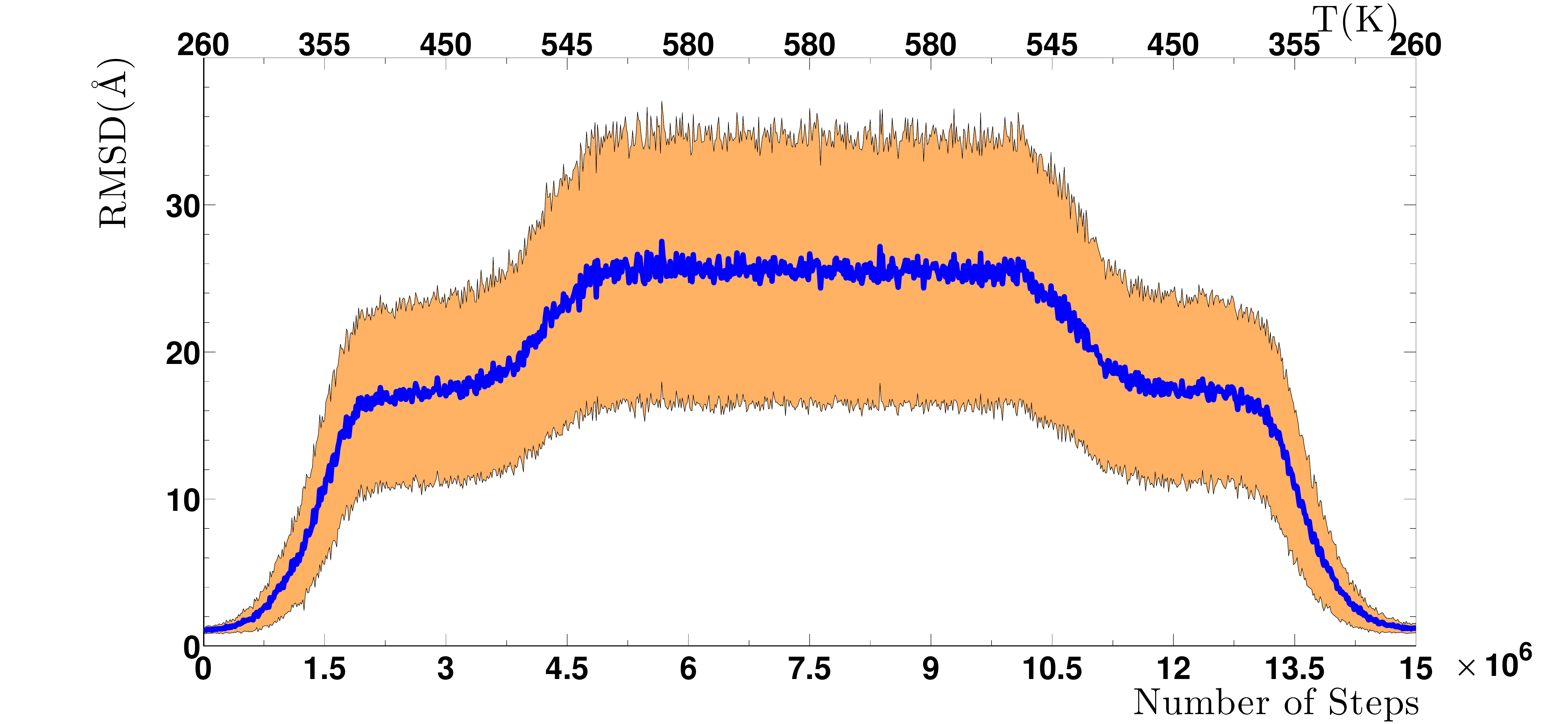}}
  \end{center}
\end{figure}

\vfill

\hfill{\large\sf Figure  \ref{fig12}}

\clearpage

\begin{figure}
  \begin{center}
    \resizebox{16.5cm}{!}{\includegraphics[]{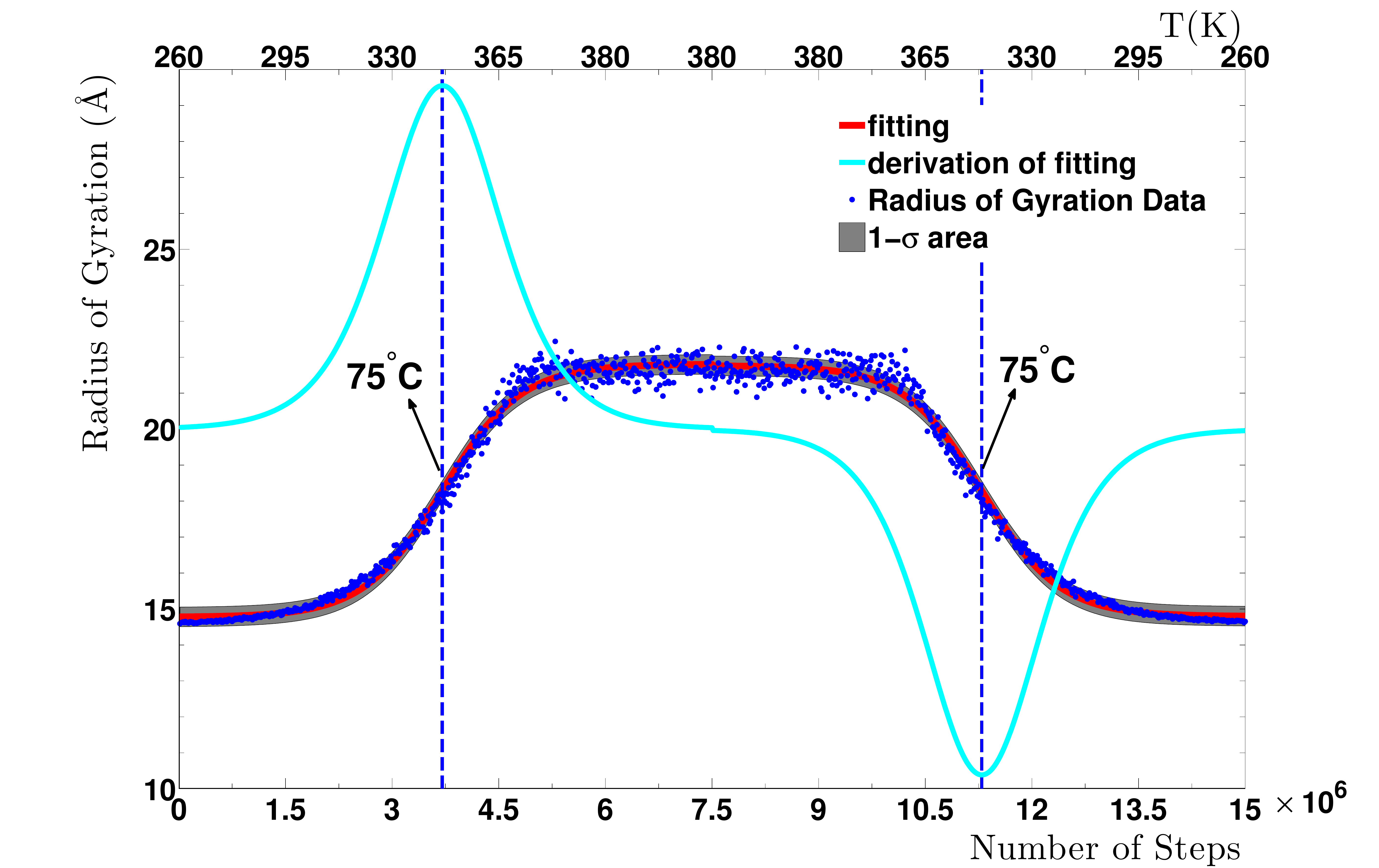}}
  \end{center}
\end{figure}

\vfill

\hfill{\large\sf Figure  \ref{fig13}}

\clearpage

\begin{figure}
  \begin{center}
    \resizebox{16.5cm}{!}{\includegraphics[]{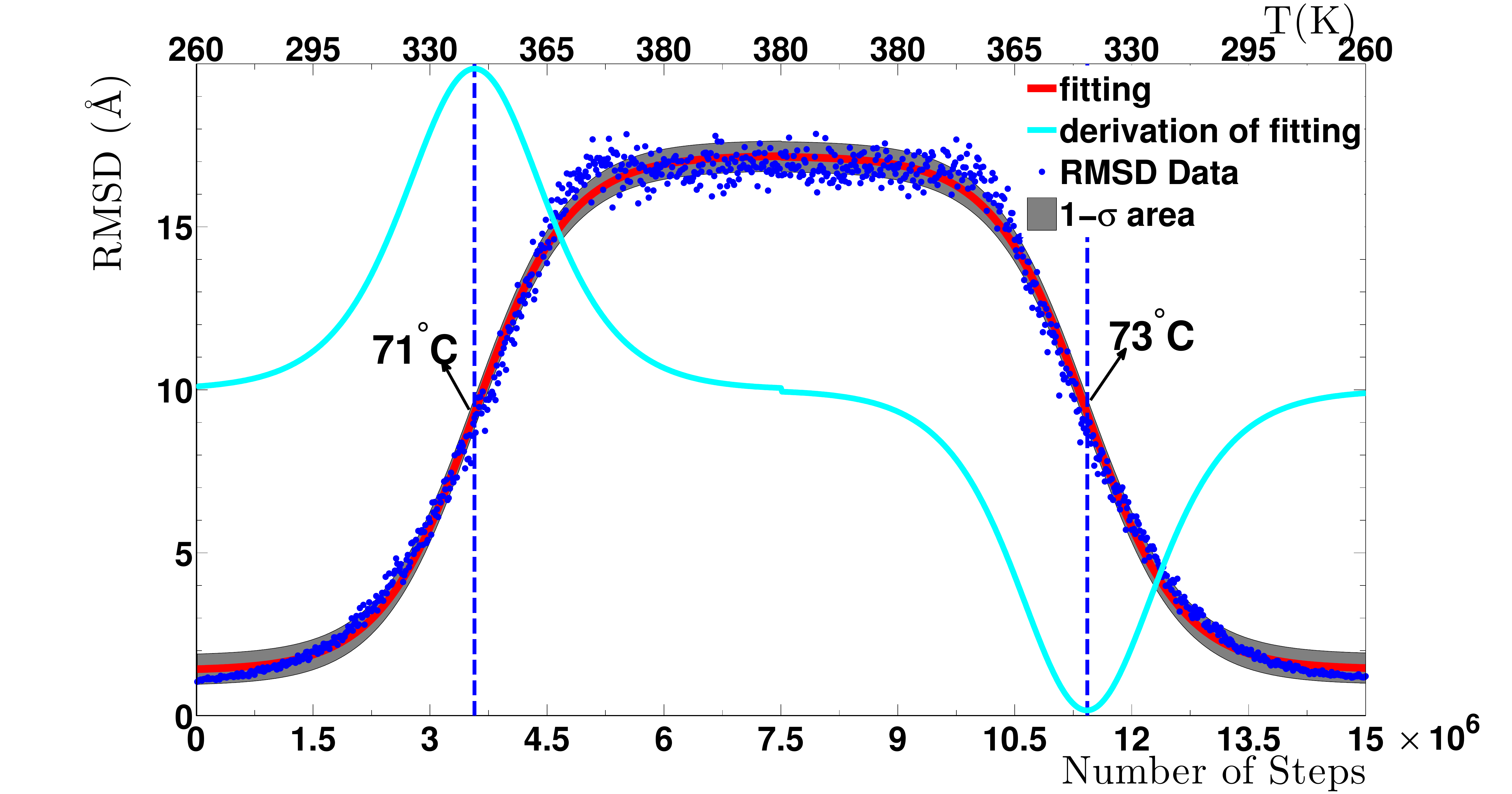}}
  \end{center}
\end{figure}

\vfill

\hfill{\large\sf Figure  \ref{fig14}}

\clearpage

\begin{figure}
  \begin{center}
    \resizebox{16.5cm}{!}{\includegraphics[]{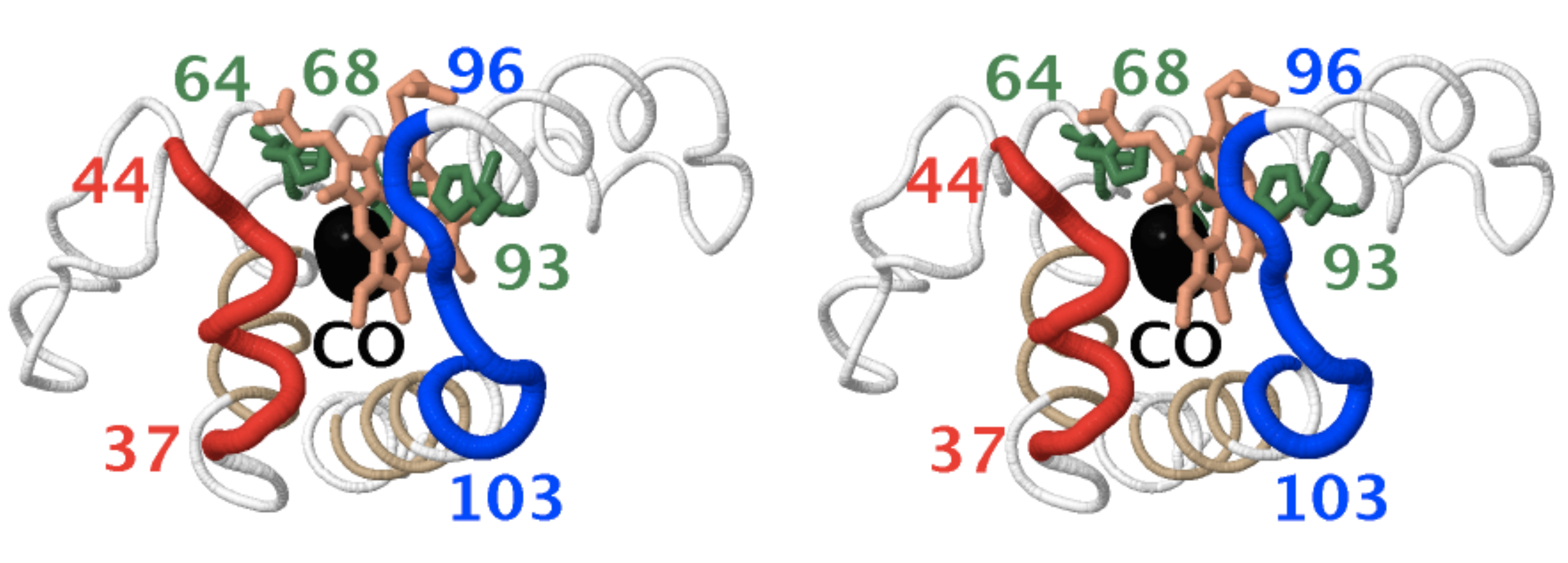}}
  \end{center}
\end{figure}

\vfill

\hfill{\large\sf Figure  \ref{fig15}}

\clearpage

\begin{figure}
  \begin{center}
    \resizebox{16.5cm}{!}{\includegraphics[]{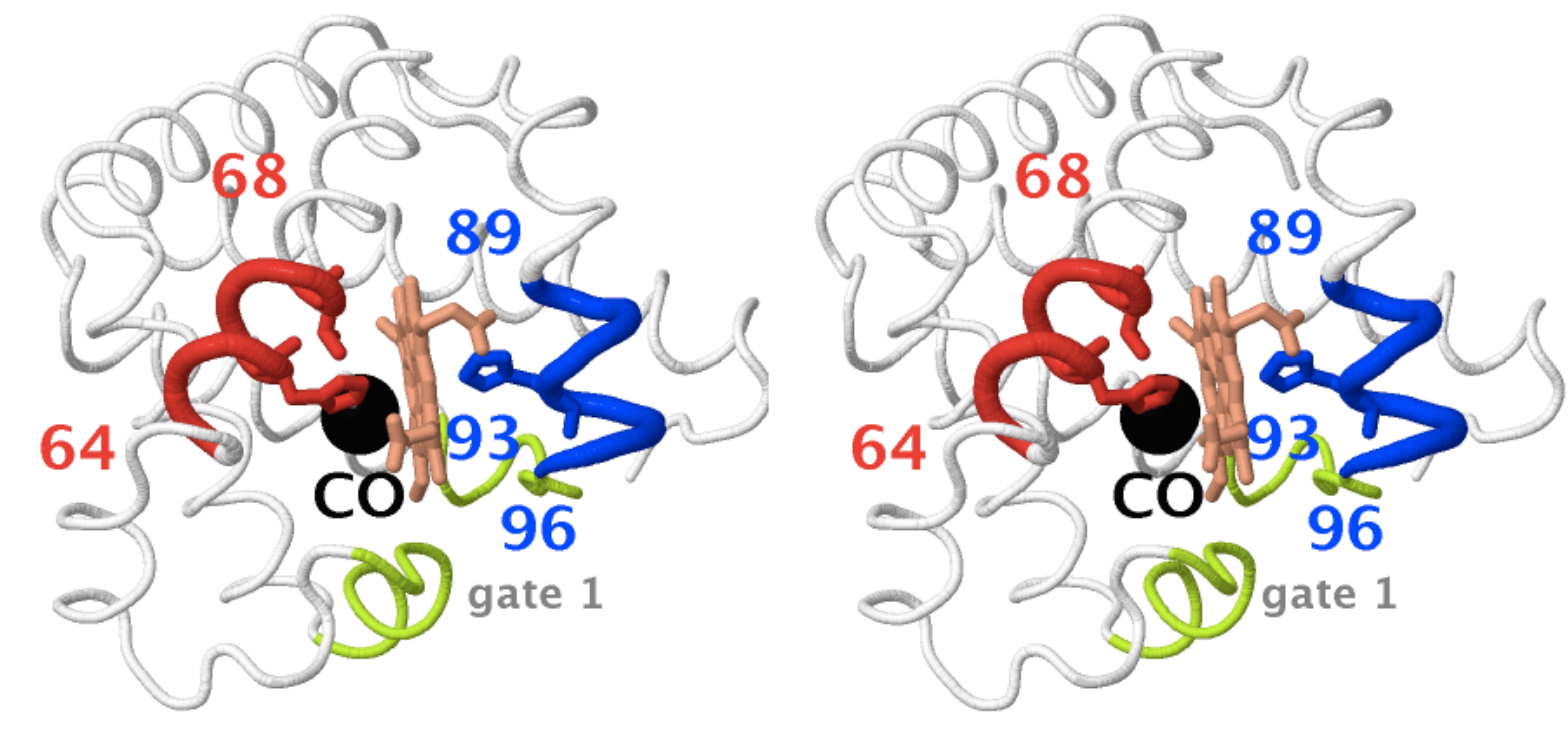}}
  \end{center}
\end{figure}

\vfill

\hfill{\large\sf Figure  \ref{fig16}}

\clearpage

\begin{figure}
  \begin{center}
    \resizebox{16.5cm}{!}{\includegraphics[]{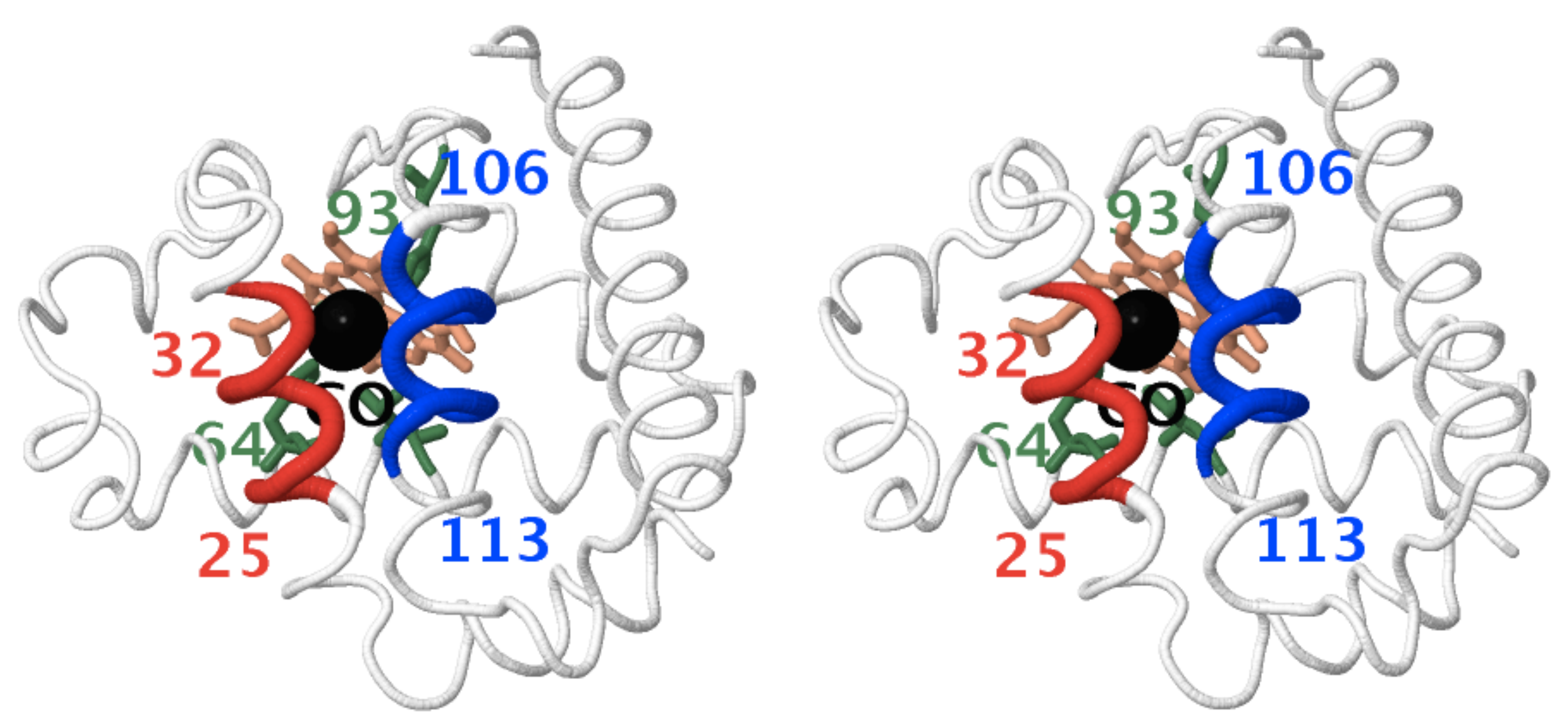}}
  \end{center}
\end{figure}

\vfill

\hfill{\large\sf Figure  \ref{fig17}}

\clearpage

\begin{figure}
  \begin{center}
    \resizebox{16.5cm}{!}{\includegraphics[]{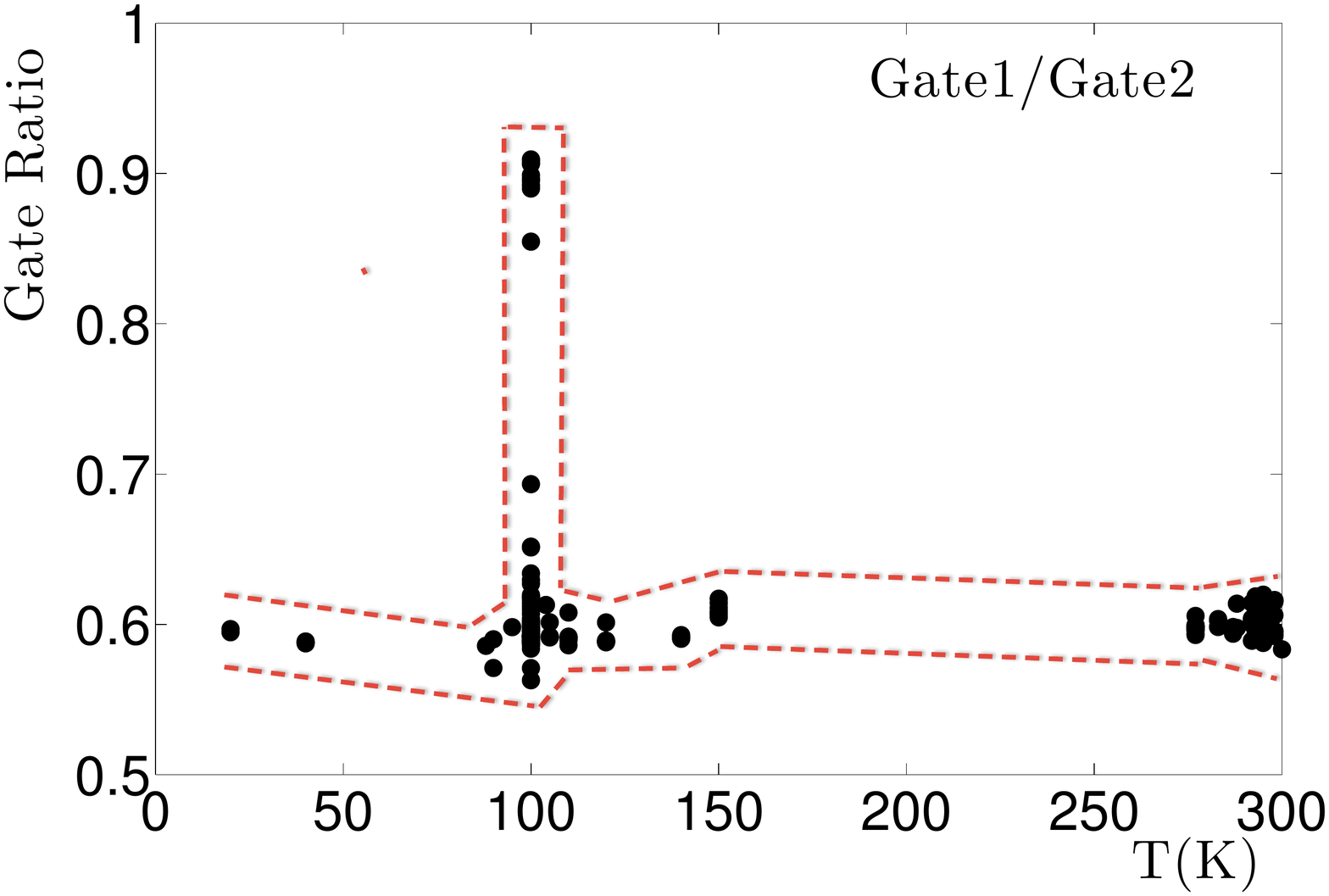}}
  \end{center}
\end{figure}

\vfill

\hfill{\large\sf Figure  \ref{fig18}}

\clearpage

\begin{figure}
  \begin{center}
    \resizebox{16.5cm}{!}{\includegraphics[]{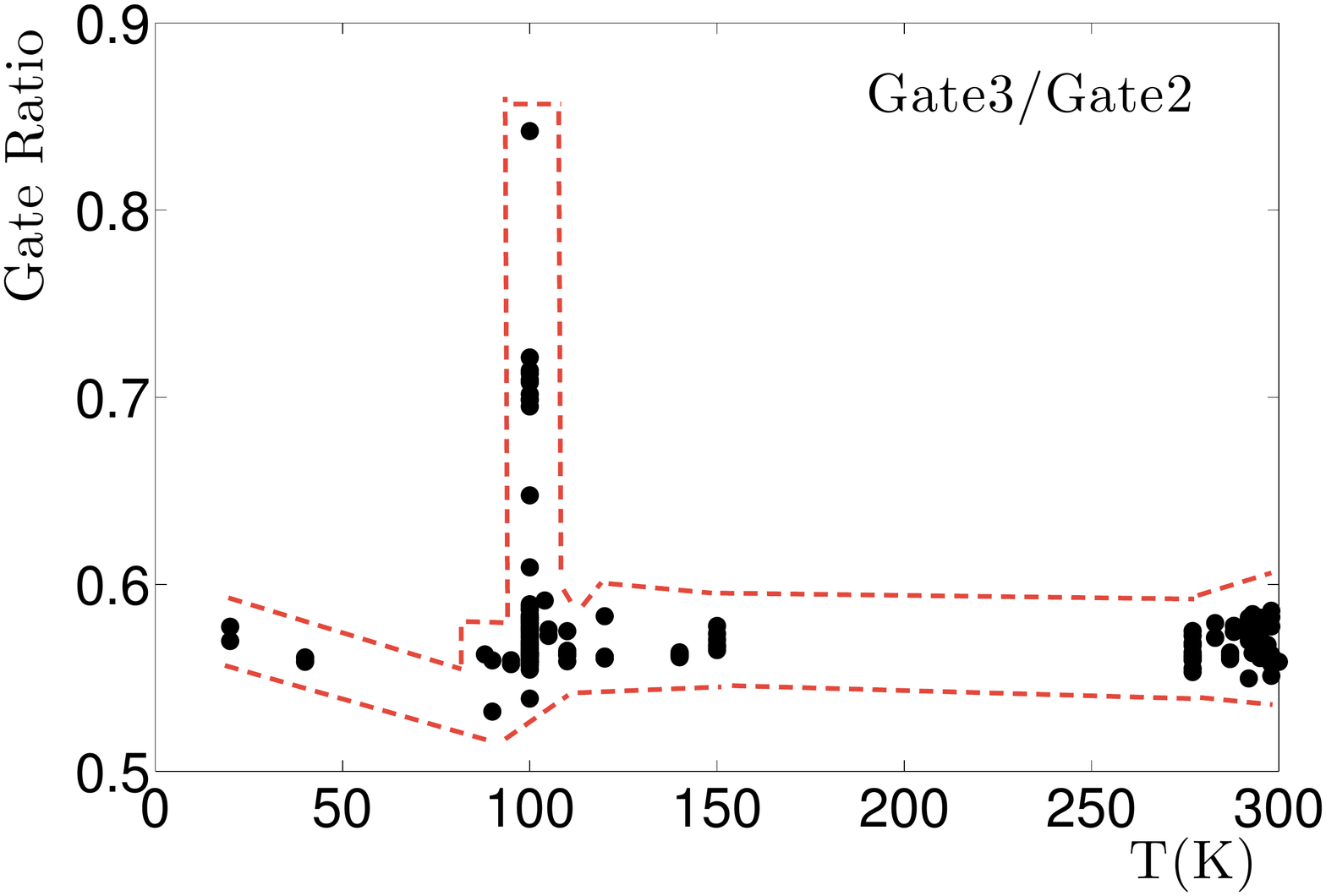}}
  \end{center}
\end{figure}

\vfill

\hfill{\large\sf Figure  \ref{fig19}}

\clearpage

\begin{figure}
  \begin{center}
    \resizebox{16.5cm}{!}{\includegraphics[]{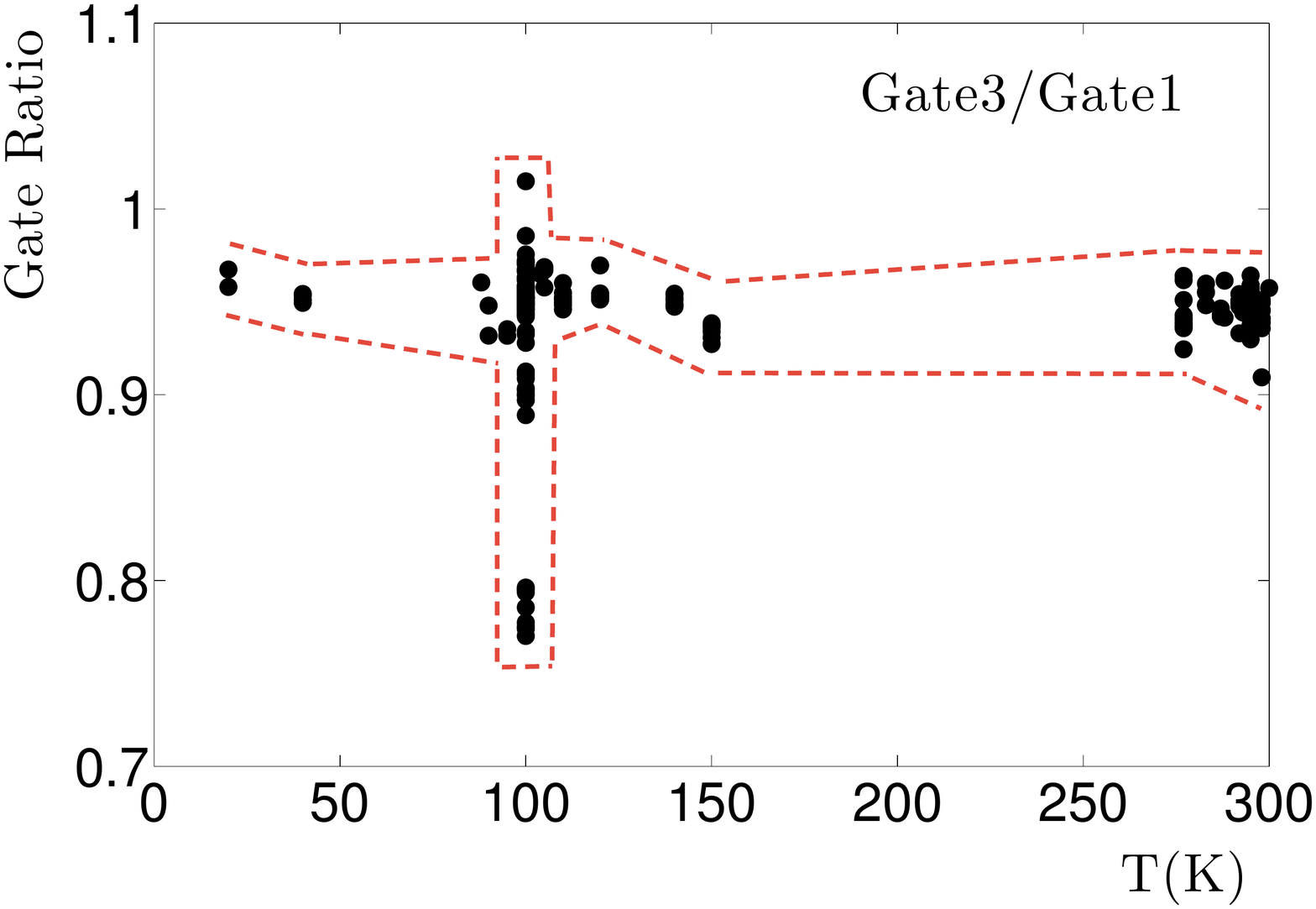}}
  \end{center}
\end{figure}

\vfill

\hfill{\large\sf Figure  \ref{fig20}}

\clearpage

\begin{figure}
  \begin{center}
    \resizebox{16.5cm}{!}{\includegraphics[]{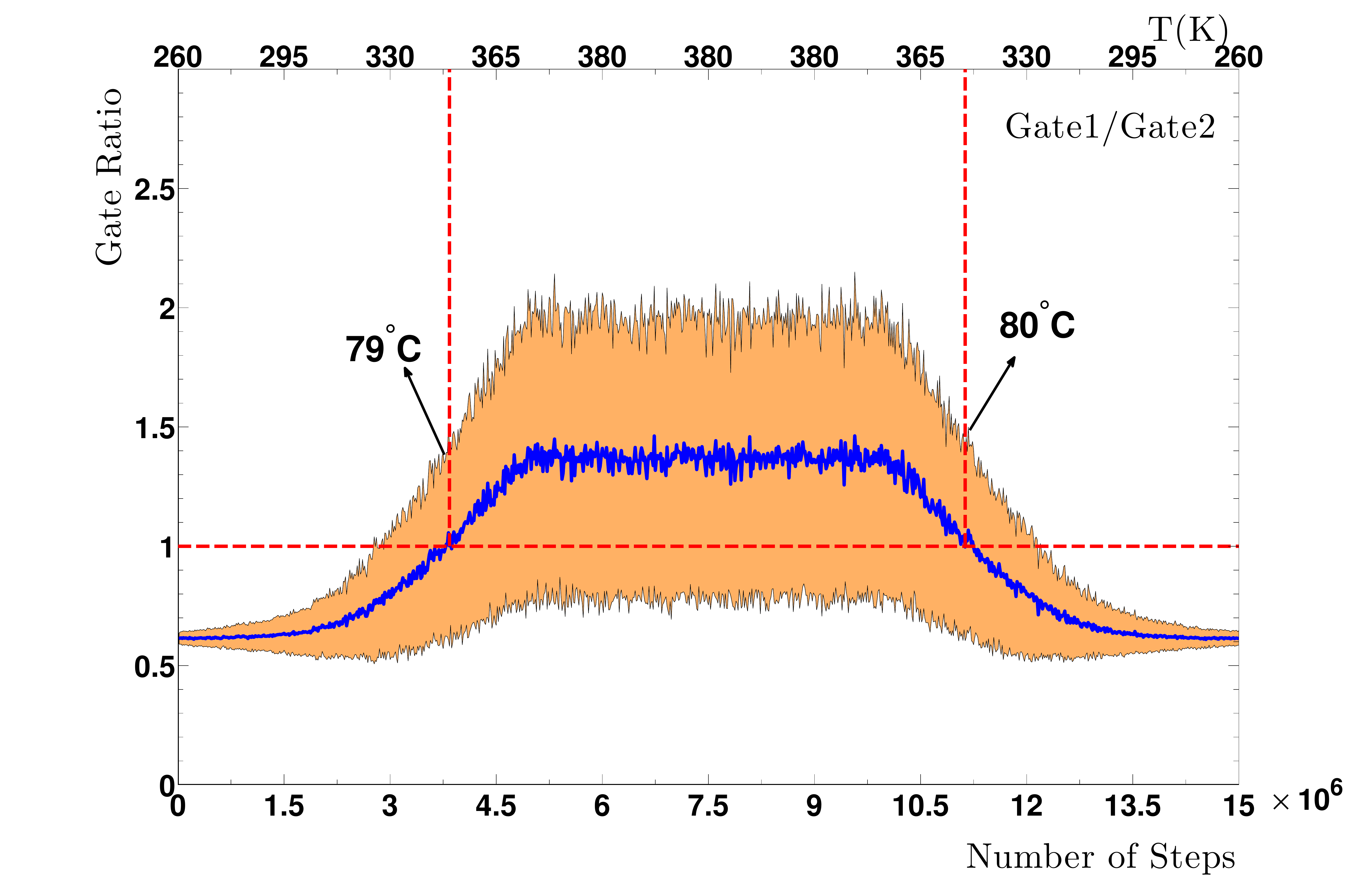}}
  \end{center}
\end{figure}

\vfill

\hfill{\large\sf Figure  \ref{fig21}}

\clearpage

\begin{figure}
  \begin{center}
    \resizebox{16.5cm}{!}{\includegraphics[]{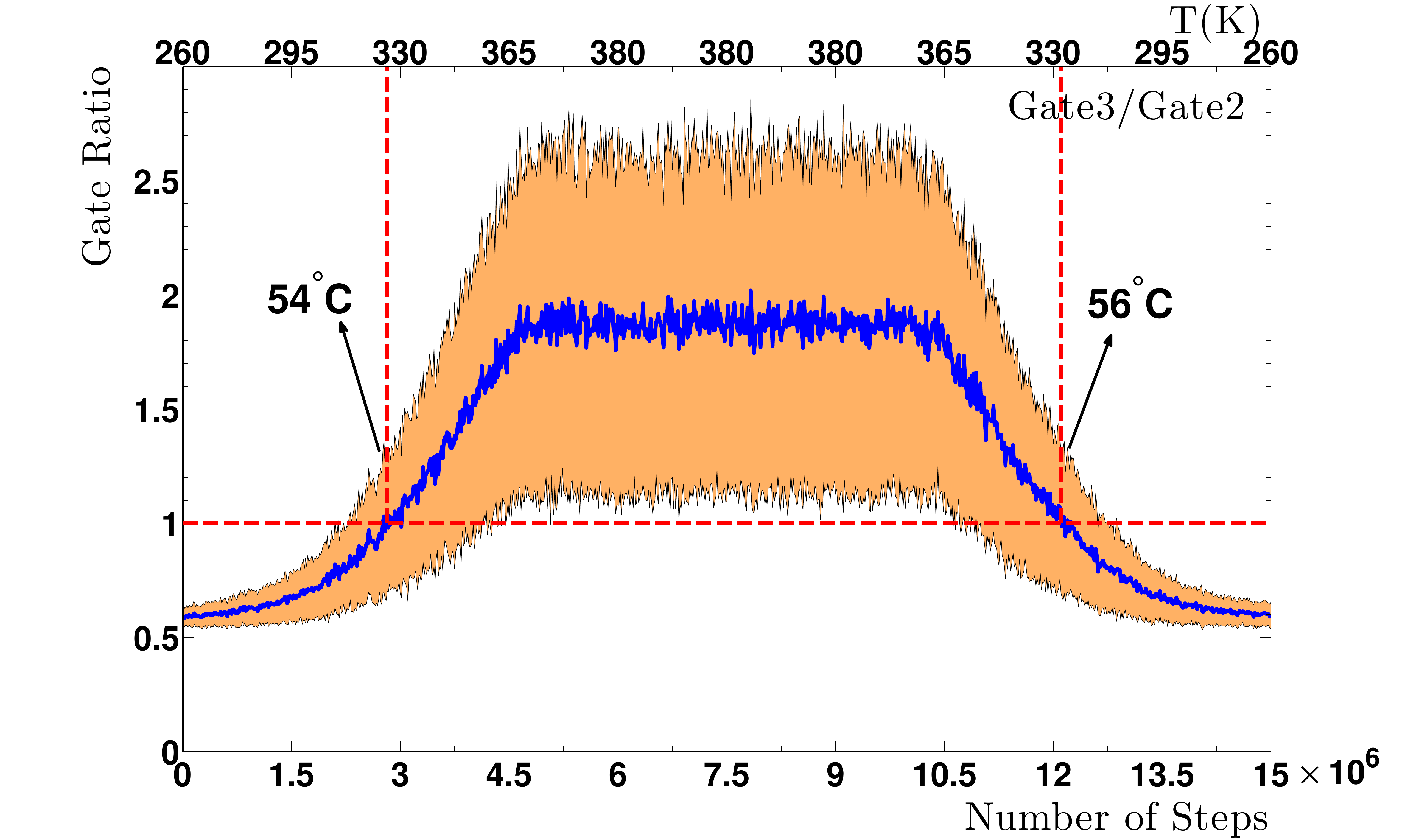}}
  \end{center}
\end{figure}

\vfill

\hfill{\large\sf Figure  \ref{fig22}}

\clearpage

\begin{figure}
  \begin{center}
    \resizebox{16.5cm}{!}{\includegraphics[]{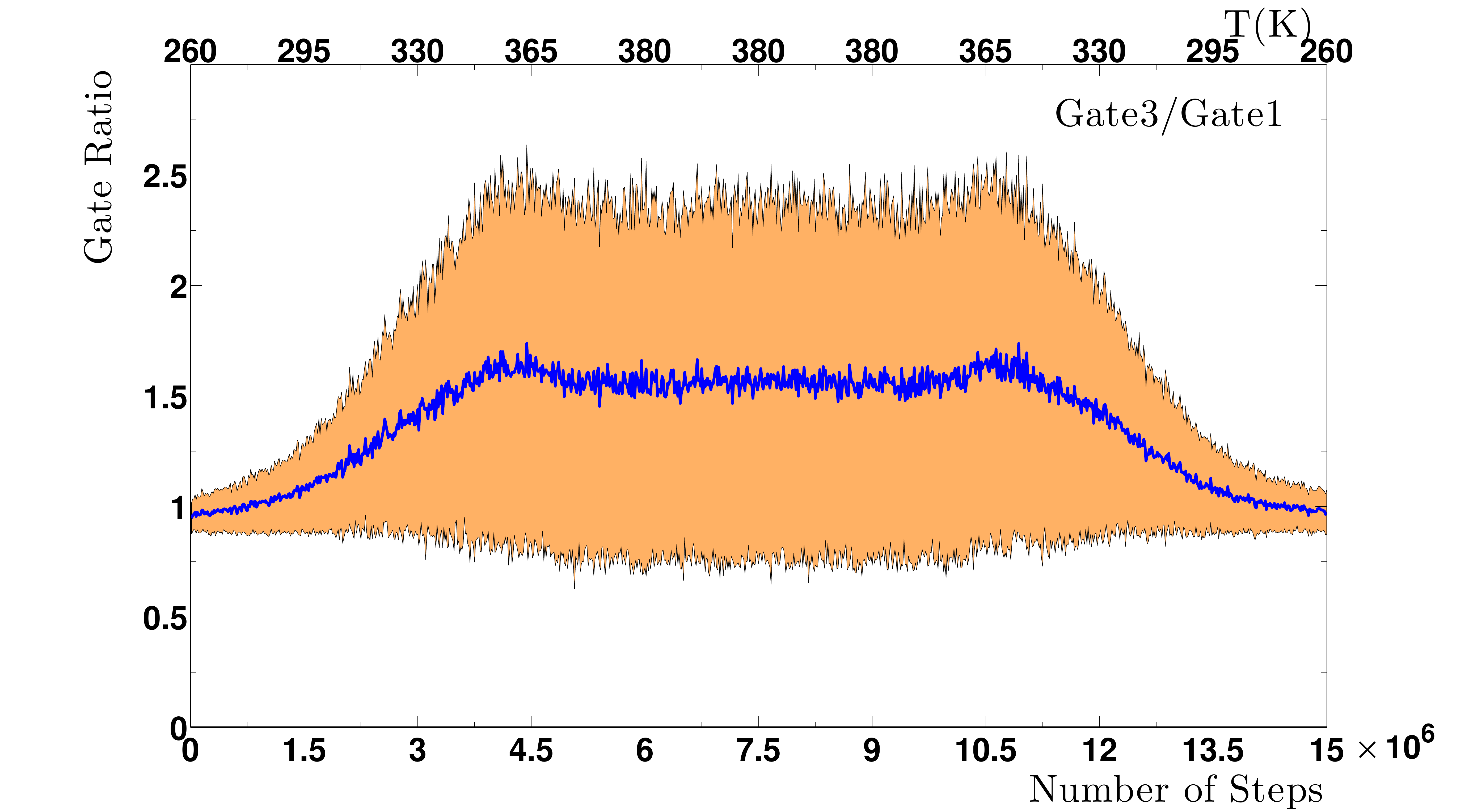}}
  \end{center}
\end{figure}

\vfill

\hfill{\large\sf Figure  \ref{fig23}}

\clearpage

\begin{figure}
  \begin{center}
    \resizebox{16.5cm}{!}{\includegraphics[]{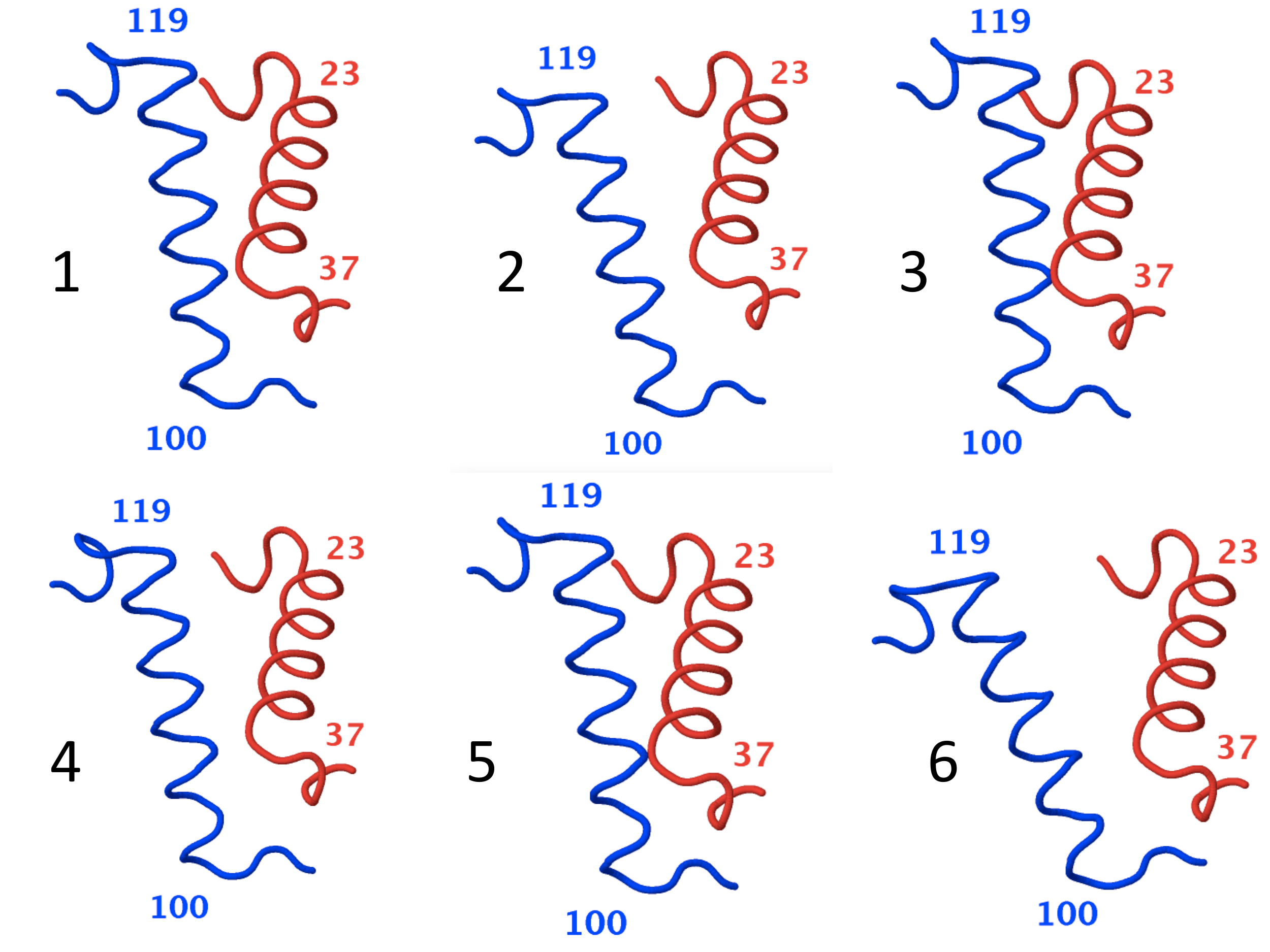}}
  \end{center}
\end{figure}

\vfill

\hfill{\large\sf Figure  \ref{fig24}}

\clearpage

{
\begin{table}[htb]
\caption{The solitons along the 1ABS  $C_\alpha$-backbone,
with indexing starting from the $N$ terminus. We have left out  the end sites that correspond
to monotonous helices, and the N and C termini segments. The type  identifies whether the soliton corresponds to a loop 
that connects $\alpha$-helices and (or) $3/10$-helices. }\vskip 0.5cm
\begin{tabular}{|c|c|c|c|c|c|}  
\hline 
soliton 
 & 1 & 2  & 3  & 4  & 5
 \tabularnewline 
\hline 
sites &  15-27 & 30-41 & 39-49  & 47-57 & 54-66 
 \tabularnewline \hline
type & $ \alpha$-$\alpha$ & $\alpha $-3/10   & 3/10-3/10 & 3/10-3/10  & 3/10-$\alpha $ 
 \tabularnewline \hline  \hline  
soliton & 6 & 7 & 8 & 9 & 10
 \tabularnewline \hline
site & 72-87 & 83-92 & 94-106 & 110-123 & 121-135 
\tabularnewline \hline
type &  $\alpha$-$\alpha  $ & $\alpha$-$\alpha$ & $\alpha$-$\alpha$ & $\alpha$-$\alpha$ & $\alpha$-$\alpha$
 \tabularnewline  
 \hline 
\end{tabular} 
\\ \vskip 0.2cm { 
 }\label{tab:loops}
\end{table}
}

\clearpage

 {
\begin{table}[htb]
\caption{The parameters in the Ansatz (\ref{bond2}), (\ref{ti}) for the solitons in 1ABS and  their root mean square 
distance to the corresponding
PDB loops. The backbone sites of the solitons are given in Table I.}\vskip 0.5cm
\begin{tabular}{|c|c|c|c|c|c|c|c||c|}
\hline 
soliton 
& $b_r  $  &
 $c_{r1}  $  &
 $c_{r2} $ &
$d_{r} $ & 
$m_{r1}$ &
$m_{r2}$ &
$s_r$ & ~
rmsd ~
\tabularnewline\hline
1   & ~ -9080236 ~ & ~ 2.83434 ~ & ~ 2.82279 ~  & ~ -3.8e{-6} ~ & ~ 83.227~  & ~ 83.258 ~ & ~ 18.497 ~ & ~ 0.26~ 
\tabularnewline\hline
2  & -1621461 & 2.96740 & 2.95728  & -2.4e-8 & 158.651 & 158.601 & 34.466 & 0.18
\tabularnewline\hline
3 & -9769758  & 2.20192  & 2.21558  & -1.3e-7 & 127.213 & 127.250 & 43.499 & 0.43
\tabularnewline\hline
4   & -1311756 & 2.90014 & 2.89872 & -1.1e-4 & 76.962 & 76.918 & 49.062 & 0.28
\tabularnewline\hline
5  & -248.174 & 3.22529 & 3.24216 & -1.1e-3 & 14.125 & 14.111 & 57.028 & 0.19
\tabularnewline\hline
6 &  -32318.8 & 2.57417 & 2.55045  & -9.3e-5 & 76.908 & 76.957 & 79.616 & 0.46
\tabularnewline\hline
7  &  -4166384 & 3.85257 & 3.84220 & -2.1e-6  & 20.358 & 20.359 & 85.104 & 0.12
\tabularnewline\hline
8  & -619147 & 3.92542 & 3.91562 & -1.4e-5 & 171.263 & 170.962 & 98.170 & 0.27
\tabularnewline\hline
9 & -409720 & 3.07447 & 3.07262 & -2.4e-4 & 51.778 & 51.789 & 117.571 & 0.36
\tabularnewline\hline
10 &  -3930247 &  2.59273  & 2.60829  & -2.7e-5  & 114.65 & 114.543 & 121.824 & 0.30
 \tabularnewline \hline   
\end{tabular} 
\\ \vskip 0.2cm { 
 }
	\label{tab:para2}
\end{table}
}


\clearpage

{
\begin{table}[htb]
\caption{Parameter values in energy (\ref{E1}) for the multi-soliton solution that describes 1ABS.  }\vskip 0.5cm
\begin{tabular}{|c|c|c|c|c|c|c|c|c|}
\hline
soliton &  $q_1$ & $q_2$ & $m_1$ & $m_2$  & $a$ & $b$ & $c$ & $d$ 
\tabularnewline \hline \hline
1  &  9.923 & 2.232 & 1.54097 & 1.54548 & -5.62412 e-08 & -4.13459 e-07 & 1.81044 e-08 & 4.273 e-09
\tabularnewline\hline
2  &  6.48516 & 0.9955 & 1.58013 & 1.54058 & -6.25287 e-11& -1.68598 e-05 & 1.47093 e-07 & 2.82807 e-07
\tabularnewline\hline
3  &  2.05153 & 0.657 & 1.66032 & 1.60224 & -9.05135 e-08 & 1.20232 e-06 & 5.10166 e-11 & 5.75389 e-09
\tabularnewline\hline
4  & 0.89676 & 6.74235 & 1.3563& 1.5232 & -2.33413 e-07 & -3.3991 e-07 & 2.36516 e-08 & 7.98841 e-09
\tabularnewline\hline
5 & 9.26118 & 0.83376 & 1.55206 & 1.5386 & -9.73035 e-08 & 4.78674 e-07 & 1.03189 e-10 & 4.88194 e-09
\tabularnewline\hline
6 &  0.98018 & 2.1337 & 1.45791 & 1.54653 & -7.25906 e-09 & 3.76092 e-09 & 6.82624 e-10 & 1.87212 e-14
\tabularnewline\hline
7 & 1.37667 & 3.16891 & 1.47151 & 1.04128 & -1.39052 e-13 & 5.97719 e-13 & 3.77897 e-14 & 5.81911 e-14
\tabularnewline\hline
8 &  10.3168 & 4.2801 & 1.18192 & 1.61334 & -1.27193 e-07 & 1.41736 e-06 & 1.07182 e-10 & 1.26295 e-08
\tabularnewline\hline
9 &  0.80042 & 1.28973 & 1.5154 & 1.60278 & -2.03487 e-07 & 1.13574 e-06 & 1.46007 e-11 & 7.82707 e-08
\tabularnewline\hline
10 & 3.15255 & 0.91475 & 1.55827 & 1.55151 & -1.07811 e-07 & 1.02768 e-06  & 7.49571 e-11 & 7.73639 e-09
\tabularnewline\hline
\end{tabular}
\\ \vskip 0.2cm { 
 }
	\label{tab:para2}
\end{table}
}

\clearpage

{
\begin{table}[htb]
\caption{RMSD distances in \.Angstr\"om between 1ABS and 1MBC, over the
backbone sites corresponding to our ten solitons. }\vskip 0.5cm
%
%
%
\begin{tabular}{|c||c|c|c|c|c|c|c|c|c|c|}
\hline
~ soliton ~ & 1 & 2 & 3 & 4 & 5 & 6 & 7 & 8 & 9 & 10
\tabularnewline\hline
~rmsd~  & ~0.25~ & ~0.17~ & ~0.30~ & ~0.23~ & ~0.19~ & ~0.31~ & ~0.30~ & ~0.34~ & ~0.44~ & ~0.33~
\tabularnewline\hline
\end{tabular}
\\ \vskip 0.2cm { 
 }
	\label{tab:para2}
\end{table}
}

\clearpage

{
\begin{table}[htb]
\caption{Parameter values in the fits (\ref{Rg1}), (\ref{rmsd1}) for the two ranges 0 - 7.5 and 7.5 - 15 (in million) of
iteration steps
 }\vskip 0.5cm
\begin{tabular}{|ccccc|cccc|}
\multicolumn{1}{c}{}  & \multicolumn{4}{c}{$R_g$} &  \multicolumn{4}{c}{$R_{rmsd}$}  \\
\cline{1-9} 
\multicolumn{1}{|c|}{range}  & \multicolumn{4}{c|}{ \hskip 0.2cm a  \hskip 1.3cm b \hskip 1.3cm c  \hskip 1.0cm d }
& 
 \multicolumn{4}{c|}{ \hskip 0.2cm a  \hskip 1.3cm b \hskip 1.3cm c  \hskip 1.0cm d }
\\
\cline{1-9}
\multicolumn{1}{|c|}{0 - 7.5 \hskip 0.3cm }  
& 
\multicolumn{4}{c|}{\hskip 0.65cm 3.519  \hskip 0.3cm 0.9047   \hskip 0.3cm 3.6855   \hskip 0.3cm
18.29 ~ } 
&  
\multicolumn{4}{c|}{ \hskip 0.3cm  7.9   \hskip 0.5cm 0.8318  \hskip 0.3cm 3.5715  \hskip 0.3cm 9.291}  
\\
\multicolumn{1}{|c|}{~7.5 - 15\hskip 0.2cm~}  
& 
\multicolumn{4}{c|}{  \hskip 0.25cm -3.486 \hskip 0.3cm  0.9193   \hskip 0.3cm 11.2965   \hskip 0.3cm 18.28 }
&  
\multicolumn{4}{c|}{  \hskip 0.1cm -7.872  \hskip 0.35cm 0.8327 \hskip 0.3cm 11.4255  \hskip 0.3cm 9.298 \hskip 0.1cm~} \\
\cline{1-9} 
\hline
\end{tabular}
\end{table}

\end{document}